\documentclass[twocolumn]{aastex631}

\usepackage[T1]{fontenc} 
\usepackage{amsmath} %
\usepackage{bm}
\usepackage{comment}
\usepackage{amsmath, amsthm, amssymb, amsfonts}
\usepackage{hyperref}
\usepackage{upgreek}
\usepackage{booktabs}
\usepackage{graphicx, nicefrac}
\usepackage{xcolor}

\newcommand{\Msun}{\mathrm{M}_{\odot}}

\defcitealias{Ricarte+2018a}{\scshape RN18}

\begin{document}

\title{Tracking the assembly of supermassive black holes: a comparison of diverse models across cosmic time}

\correspondingauthor{Antonio J. Porras-Valverde}
\email{antonio.porras@yale.edu}

\author[0000-0002-1996-0445]{Antonio J. Porras-Valverde}
\affiliation{Department of Astronomy, Yale University, P.O. Box 208101, New Haven, CT 06520, USA}

\author[0000-0001-5287-0452]{Angelo Ricarte}
\affiliation{Black Hole Initiative, Harvard University, Cambridge, MA 02138, USA}

\author[0000-0002-5554-8896]{Priyamvada Natarajan}
\affiliation{Department of Astronomy, Yale University, P.O. Box 208101, New Haven, CT 06520, USA}
\affiliation{Department of Physics, Yale University, P.O. Box 208121, New Haven, CT 06520, USA}
\affiliation{Black Hole Initiative, Harvard University, Cambridge, MA 02138, USA}

\author[0000-0002-6748-6821]{Rachel S. Somerville}
\affiliation{Center for Computational Astrophysics, Flatiron Institute, 162 5th Ave, New York, NY 10010, USA}

\author[0000-0003-4295-3793]{Austen Gabrielpillai}
\affiliation{The City University of New York, 365 5th Ave, New York, NY 10016, USA}

\author[0000-0003-3466-035X]{{L. Y. Aaron} {Yung}}
\affiliation{Space Telescope Science Institute, 3700 San Martin Drive, Baltimore, MD 21218, USA}

\begin{abstract}

Galaxies grow alongside their central supermassive black holes (SMBHs), linked through fueling and feedback. However, the origins and details of this co-evolution remain unclear and differ significantly amongst modeling frameworks. Using a suite of semi-analytic models (SAMs), we trace SMBH mass assembly across $M_{\rm BH} \sim 10^{6-10}, \mathrm{M}_{\odot}$. We find significant discrepancies between observations and physics-based models of the local black hole mass function (BHMF), likely due to differences in the underlying stellar mass function and the scaling relations therefrom used to infer the BHMF. However, most physics-based models agree at $z \sim 1-4$ and align reasonably well with broad-line AGN BHMF from JWST observations at $z=4-5$. Most physics-based models reproduce the bolometric AGN luminosity evolution, except {\sc Dark Sage}, which predicts an excess deviating from models and observations. Interestingly, this pronounced ``knee' in the bolometric AGN luminosity function predicted by {\sc Dark Sage} around $L_{\rm bol} \sim 10^{46} \, \mathrm{erg \, s^{-1}}$ is consistent with the inferred luminosity of ``Little Red Dots'' at $z=5-6$, assuming that their entire emission originates from AGN activity. We analyze black hole mass build-up and accretion histories in {\sc Dark Sage}, which, unlike other models, allows for super-Eddington accretion. We report that on average, SMBHs in {\sc Dark Sage} primarily grow through secular disk instabilities and merger-driven cold gas accretion, while black hole mergers contribute 60\% of the total mass budget only for the most massive SMBHs by $z=0$.

\end{abstract}


\section{Introduction}\label{introduction}

Observations of nearby galaxies have shaped the consensus that most massive galaxies host at least one central supermassive black hole ($10^6-10^{10} \, \Msun$) \citep{1995ARA&A..33..581K, 2013ARA&A..51..511K}. Several observational studies have established correlations between SMBH mass and the properties of their host elliptical galaxies such as the stellar velocity dispersion (using the $M_{\rm BH}-\sigma$ relation) \citep{2000ApJ...539L..13G, 2001ApJ...547..140M, 2002ApJ...574..740T}, luminosity \citep{2001MNRAS.327..199M, 2002MNRAS.331..795M}, Sersic index \citep{2001ApJ...563L..11G, 2007MNRAS.378..198G}, bulge mass \citep{1998AJ....115.2285M, 2003ApJ...589L..21M, 2004ApJ...604L..89H}, and binding energy \citep{2007ApJ...669...45H, 2007ApJ...665..120A}. These scaling relations suggest that the coevolution between SMBHs and elliptical galaxies is closely linked, primarily regulated by various factors such as active galactic nuclei (AGN) feedback. It is hypothesized that black holes grow by accreting gas during ``active" phases, possibly fueled by a major merger of gas-rich galaxies, until the energy from the SMBH feedback expels the gas and halts the accretion process \citep{1998A&A...331L...1S, 1999MNRAS.308L..39F, 2005MNRAS.361.1387B, 2005ApJ...618..569M}.

Many models have connected SMBH growth, the quasar phase, and galaxy evolution \citep{1998MNRAS.300..817H, 2000MNRAS.318L..35H, 2000MNRAS.311..576K, 2003ApJ...595..614W, 2003ApJ...582..559V, 2005MNRAS.364..407C, 2008MNRAS.391..481S,  2008ApJ...676...33D, 2009MNRAS.398...53B}. Although it appears that AGN feedback may play a critical role in balancing the growth of both SMBHs and galaxies, the mechanisms that fuel SMBH growth are plausibly diverse. Gas-rich galaxies that undergo major mergers may drive quasar activity at high redshifts, leading to the growth of the most massive SMBHs. However, only accounting for major mergers does not yield the observed number of faint X-ray AGN \citep{2007MNRAS.375..649M}. Instead, secular accretion of gas through internal galactic processes may fuel these dimmer, lower-mass AGN at lower redshifts \citep{2001ApJ...551..131C, 2006ApJS..166....1H}. Observational evidence points to AGN being located in late-type galaxies up to $z\approx1$ \citep{2006ApJS..166...89G, 2009ApJ...691..705G}. The X-ray luminosity function from the AGN in these galaxies suggests that internal instabilities may contribute significantly to the overall SMBH accretion history \citep{2009MNRAS.397..623G}.

Given the observed coevolution between SMBHs and galaxies, the BHMF offers valuable information about the primary fueling mechanisms responsible for black hole growth. Unfortunately, two overarching complications make observational measurements of the BHMF difficult: survey incompleteness, which relies on a flux limit, and SMBH mass uncertainty. Currently, obtaining reliable mass estimates for SMBHs beyond the local population through dynamical modeling of stellar or gaseous components is not feasible, so scaling relations are used instead.

Black hole mass estimates can be obtained by employing scaling relationships between $M_{\rm BH}$ and the properties of the host galaxy's bulge \citep{1999MNRAS.307..637S, 2007ApJ...663...53T}. Additionally, black hole mass can be estimated using the luminosity and width of broad emission lines in AGN \citep{1999ApJ...526..579W, 2006ApJ...641..689V}, as well as their X-ray variability \citep{2004MNRAS.350L..26N, 2010ApJ...710...16Z, 2011ApJ...730...52K}. These mass estimates would in turn yield statistically significant samples necessary to determine the BHMF. Comparing BHMFs derived from different black hole mass estimation methods is important to highlight the level of agreement or disagreement between them. Additionally, BHMFs derived from physical models can help determine the primary mechanisms driving black hole growth.

Both \textit{empirical} and \textit{physics-based} models have been used to investigate the multivariate distributions of properties of SMBHs, host galaxies, and their parent halos. We refer to as \textit{empirical} models those which attempt to constrain the connection between dark matter halos and galaxies by assigning galaxies to halos using diverse techniques such as abundance matching \citep{1999ApJ...523...32C}; halo occupation distributions (HOD) \citep{2004ApJ...609...35K}; conditional luminosity functions \citep{2003MNRAS.339.1057Y}, and models that flexibly link galaxies to dark matter halo growth histories over time \citep{2009ApJ...696..620C, 2010ApJ...710..903M, Hearin2013TheColour, 2018ARA&A..56..435W, Behroozi2019Universemachine:010, 2023MNRAS.518.2123Z}. We refer to as \textit{physics-based} models those which evolve galaxies and gas, incorporating an explicit parameterization of the physics involved in galaxy formation. \textit{Physics-based} models can be categorized into \textit{semi-analytic} models \citep[e.g.][]{Croton2006, 2008MNRAS.391..481S, Croton2016, Somerville2001TheGalaxies, Benson2012, 2015MNRAS.453.4337S, Henriques2015, 2015ARA&A..53...51S, Stevens2016, Ricarte+2018a}, and \textit{numerical hydrodynamics} simulations \citep[e.g.][]{2015MNRAS.450.1349K, 2016A&C....15...72M, 2017ARA&A..55...59N, Naiman2018FirstEuropium}. \textit{Semi-analytic} models use prescriptions motivated by observed trends to make physical predictions and track flows between different reservoirs. \textit{Numerical} simulations solve partial differential equations for hydrodynamics, thermodynamics, and gravity, while self-consistently evolving galaxies alongside the large-scale structure in simulated boxes. Although SAMs and numerical hydrodynamics simulations incorporate black hole growth prescriptions, both explicitly rely on the observationally determined local $M_{\rm BH}-\sigma$ or $M_{\rm BH}-M_{\rm bulge}$ relations for calibration. These models contain a diverse set of black hole growth channels that can effectively reproduce observed local black hole scaling relations \citep{2015ARA&A..53...51S,2022MNRAS.511.3751H}.

There is considerable disagreement among different large-scale numerical hydrodynamics models about how SMBHs grow as a function of redshift and stellar or halo mass \citep[e.g.,][]{Natarajan+2021,2022MNRAS.511.3751H, 2024arXiv241013958D}. First, the spatial and particle resolution are insufficient to resolve black hole accretion disks. In these models, black hole seeds, growth, and feedback are handled using sub-grid physics. Models vary in their sub-grid implementations, including the location, mass, and dynamics of black hole seeds, methods for modeling black hole accretion, and different approaches and efficiencies for injecting AGN energy and releasing it. Some simulations explicitly link AGN feedback channels to black hole mass, while others assume a uniform feedback mechanism. For a more comprehensive comparison between numerical hydrodynamics models, please see \citep[Table 1 from][]{2022MNRAS.509.3015H}.

In this work, we look at the properties of black hole populations and how they evolve over time for {\sc Dark Sage}, {\sc Ricarte \& Natarajan}'s SAM, the {\sc Santa Cruz} SAM, the {\sc TRINITY} empirical model, and the {\sc IllustrisTNG300} simulation. Then, we focus on deriving predictions for SMBH mass assembly histories from {\sc Dark Sage}. We parse the {\sc Dark Sage} black hole mass growth channels over cosmic time and quantify the relative contribution of accretion processes versus mergers. Our goal is to understand the physical processes responsible for SMBH growth. In particular, how do physics-based models compare with observations and what are the growth channels responsible for the bulk of black hole growth?

This paper is organized as follows. In Section \ref{Methodology}, we present an overview of the models used in this paper. Section \ref{model_comparisons} compares black hole population properties between {\sc Dark Sage}, the {\sc Santa Cruz} SAM, \citetalias{Ricarte+2018a}, {\sc IllustrisTNG}, and {\sc TRINITY}. In Section \ref{DS_BHchannels}, we focus on the black hole mass assembly in {\sc Dark Sage} through its various physical implementations of black hole growth. We present the discussion and conclusions of this study in Section \ref{conclusion}. The appendix \ref{app:calibration} provides more details about model calibration methods. Appendix \ref{BHgrowth_histdist} includes the Eddington ratio distributions divided into halo mass bins for {\sc Dark Sage}, the {\sc Santa Cruz} SAM, \citetalias{Ricarte+2018a}, and {\sc IllustrisTNG}.

\section{Black Hole Growth Models}
\label{Methodology}

In this section, we briefly summarize black hole seeding and growth prescriptions adopted in {\sc Dark Sage}, \citetalias{Ricarte+2018a}, the {\sc Santa Cruz} SAM, {\sc IllustrisTNG}, and {\sc TRINITY}. Section \ref{Methodology_DS} describes the {\sc Dark Sage} \textit{semi-analytic} model. Sections \ref{Methodology_Angelo} and \ref{Methodology_SC} present a brief summary of the \citetalias{Ricarte+2018a} models and the {\sc Santa Cruz} SAM, respectively. In Section \ref{Methodology_TNG}, we summarize the numerical hydrodynamics simulation {\sc IllustrisTNG300-1}, and in Section \ref{describing_trinity}, we describe the semi-empirical model {\sc TRINITY}. Table \ref{Table: sims_details} provides a summary of the details of the simulations used in our analysis. Note that the minimum halo mass in simulations using N-body simulations assumes that each halo contains at least 200 particles at $z=0$.

\begin{table*}
\centering
\caption{Summary of details pertaining to black hole properties from our model comparisons in figure \ref{model_comparisons}}
\begin{tabular*}{\textwidth}{@{\extracolsep{\stretch{1}}}*{4}{l}@{}}
\toprule
 Model name & Model type & Minimum Halo Mass [$\mathrm{M}_{\odot}$] & Black hole seeding  \\
\midrule
 {\sc Dark Sage} & SAM & $10^{11.2}$
   & $0\; \mathrm{M}_{\odot}$ \\
 \citetalias{Ricarte+2018a} & SAM & $5\times10^6$ & Light (common): $30\, \mathrm{M}_{\odot} \leq M_\bullet \leq 100\, \mathrm{M}_{\odot}$. Heavy (rare): $\sim10^5 \ \mathrm{M}_{\odot}$ \\
 {\sc Santa Cruz} & SAM & $10^{10}$ & $\sim10^4\, \mathrm{M}_{\odot}$  \\
 {\sc TNG300-1} & Hydro sim & $10^{10}$ & $1.1\times10^6\,\mathrm{M}_{\odot}$ \\
 {\sc TRINITY} & Empirical & $10^{11}$ & Functional black hole occupation fraction \citep[see equation 29 in][]{2023MNRAS.518.2123Z}   \\
\bottomrule   

\end{tabular*}\label{Table: sims_details}
\end{table*}

\subsection{\textit{Semi-analytic} Model: {\sc Dark Sage}}\label{Methodology_DS}

{\sc Dark Sage} \citep{Stevens2016} is a \textit{semi-analytic} model of galaxy formation that couples \textit{empirical} and phenomenological analytic prescriptions to describe the underlying physics of galaxy formation and evolution. This framework takes as an input a set of dark matter merger trees from the Millennium N-body simulation \citep{Springel2005Nat} and populates it with the predetermined physical conditions. The simulation uses a periodic box of length 684.9 Mpc with particle mass resolution of $1.1 \times 10^9\, \mathrm{M}_{\odot}$ and cosmological parameters from the Wilkinson Microwave Anisotropy Probe data \citep{Spergel2003}, where $\Omega_M = 0.25$, $\Omega_{\Lambda} = 0.75$, $\Omega_b = 0.045$, $\sigma_8 = 0.9$, and $h = 0.73$. The merger trees are constructed with {\sc L-HALOTREE} \citep{Springel2005} and the halos and subhalos are found using the {\sc SUBFIND} algorithm \citep{Springel2001}. 

{\sc Dark Sage} is the predecessor of {\sc SAGE }\citep{Croton2016}, in which the modeling starts with hot gas reservoirs modeled as an isothermal sphere within every dark matter halo. Through radiative cooling and condensation of hot gas, galaxies start to form and grow \citep{White1978}. As hot gas cools, it collapses gravitationally to form galactic disks \citep{Mo1998}. {\sc Dark Sage} uniquely evolves this one-dimensional disk structure, which is divided into 30 equally-spaced logarithmic bins of fixed specific angular momentum. The model uses two separate annular disk structures, one for stars and one for the gas, to allow convenient computation of relevant physical processes for the two species. {\sc Dark Sage} tracks Toomre instabilities \citep{Toomre1964} via a process of redistributing unstable gas or stars to adjacent rings while conserving angular momentum. When a particular ring becomes unstable, it can catalyze a burst of star formation and damp these instabilities. When these instabilities propagate to the innermost ring, unstable stars contribute to the instability-driven bulge, while unstable gas is channeled directly to feed the central black hole.  When a galaxy merger occurs, the angular momentum vectors of disks are summed, and misalignment between these vectors results in an overall gain or loss of specific angular momentum.  

In {\sc Dark Sage}, black hole seeds with zero mass are planted into every well-resolved halo that forms in the Millennium N-body simulation. Thereafter, black holes grow via five main channels: 

\begin{itemize}
    \item Accretion from hot halo gas, referred to as the (\textit{hot--mode})
    \item Cold gas accretion from galaxy mergers, referred to as the 
    (\textit{cold--mode})
    \item Ex-situ gas accretion from unstable cold gas with low angular momentum driven by galaxy mergers, referred to as the (\textit{Merger-driven instability})
    \item In-situ gas accretion from unstable cold gas with low angular momentum, referred to as the (\textit{Secular instability})   
    \item Mergers with other black holes (BH--BH merger mode)
\end{itemize}

We now describe each of these pathways in more detail. During major galaxy mergers, the two black hole masses are simply summed together, yielding the BH--BH merger mass growth channel. No black hole dynamics is implemented for black hole merger growth. Due to the absence of implementation of an explicit dynamical friction model, the mass produced from BH--BH mergers can be considered an upper limit to what a more realistic contribution from this channel might be as mergers might take a finite time and not always lead to completion. As galaxies evolve, black holes accumulate mass through cold gas accretion, which occurs as a result of the Toomre disk instabilities noted above and galaxy mergers. This mode may be episodically dependent on the \textit{in situ} cold gas availability; merger rate of galaxies; and their star formation history. In the case of \textit{hot--mode} gas accretion, it involves direct funneling of gas from the circum-galactic or the intra-halo medium. The accretion of gas from the hot halo is modeled using the Bondi-Hoyle-Lyttleton formula \citep{Bondi1952}. In the case of \textit{cold--mode} accretion, when a merger occurs, as described above, a fraction of gas denoted as $f_{\rm BH}$ is used to directly feed the central black hole on short time-scales, which results in rapid black hole growth following equation \ref{eq:BH_quasar}, adapted from \citet{2000MNRAS.311..576K}. This is applied individually to each annulus:

\begin{equation}
\label{eq:BH_quasar}
\begin{aligned}
\Delta m_{\mathrm{cold}} = f_{\mathrm{BH}} \left[ 1 + \left( \frac{280 \ \mathrm{km s^{-1}}}{V_{\mathrm{vir}}} \right)^{2} \right]^{-1} \\ \sum_{i=1}^{30} \left( m_{\mathrm{i,cen}} + m_{\mathrm{i,sat}} \right) \min \left( \frac{m_{\mathrm{i,sat}}}{m_{\mathrm{i,cen}}}, \frac{m_{\mathrm{i,cen}}}{m_{\mathrm{i,sat}}}  \right).
\end{aligned}
\end{equation}
The quantity $\Delta m_{\mathrm{cold}}$ represents the cold gas mass that is fed into the black hole from the disk. The parameter $f_{\mathrm{BH}}$ determines the efficiency of this accretion process and governs the proportion of cold gas intake attributed to major and minor mergers. Currently, the {\sc Dark Sage} fiducial model sets $f_{\mathrm{BH}}=20\%$. We discuss the implications of our results when lowering this value in section \ref{conclusion}. Additionally, $m_{\mathrm{i}}$ denotes the gas mass within the \textit{i}-th annular region, whether it pertains to the central galaxy or a satellite galaxy. It's worth noting that the majority of the gas that black holes accrete have low specific angular momentum, akin to what is commonly observed in \citet{2018ApJ...860...20S}. 

When a major galaxy merger occurs, after gas is funneled directly into the central black hole, the rest of the cold gas in each annulus is subject to a merger-driven starburst phase following the prescription in \citet{Somerville2001TheGalaxies}. Any remaining cold gas that exhibits gravitational instability undergoes the standard procedure implemented in {\sc Dark Sage}, which was briefly outlined in the preceding section, for the fate of unstable gas. The redistribution of this unstable gas might incite instabilities in other regions of the disk, potentially leading to a chain reaction that persists until the unstable gas reaches the innermost annulus. At this point, a portion of it is subsequently channeled directly into the black hole. We note that {\sc Dark Sage} limits the \textit{hot--mode} accretion to the Eddington limit. No other black hole growth channel follows such restrictions, therefore super-Eddington accretion is permitted for the other growth modes.

In this paper, we use the 2018 version of {\sc Dark Sage} \citep{Stevens2018ConnectingSAGE}. In this version, {\sc Dark Sage} updates the way in which the cooling scale radius and the velocity dispersion support in disks are calculated. Although a new version of {\sc Dark Sage} is now public \citep{2024PASA...41...53S}, for more information about our specific version, see \citet{Stevens2016}.

\subsection{The \textit{Semi-analytic} Model of RN18}\label{Methodology_Angelo}

Another SAM that we compare with in this work is the RN18 model, which employs a different set of recipes for BH seeding, growth and feedback. The \citetalias{Ricarte+2018a} model takes a hybrid \textit{empirical} and analytic approach to model the growth of SMBHs from $z=20$ to $z=0$ deeply embedded in the cold dark matter paradigm. The \citetalias{Ricarte+2018a} model uses \citet{2016A&A...594A..13P} cosmological parameters. Analytic dark matter merger trees generated using the \citet{Parkinson+2008} algorithm \citep[based on the formalism of][]{Press&Schechter1974} form the backbone of the model. This technique has the advantage of enabling much higher dynamic range than N-body based merger trees (down to a minimum halo mass of $5 \times 10^6\, \mathrm{M}_{\odot}$ in these models) at the cost of spatial information.  \citetalias{Ricarte+2018a} assumes \textit{empirical} relationships between galaxies and their host dark matter halos, then models SMBH-galaxy co-evolution semi-analytically. The model does not directly track gas evolution and star formation, instead relying on only one property: the galaxy's stellar velocity dispersion $\sigma$, which is estimated from the stellar mass and size \citepalias[see Fig.~4 of][]{Ricarte+2018a}. Recent James Webb Space Telescope (JWST) data indicate that the local $M_\bullet-\sigma$ relation appears to hold even for samples at $4 \lesssim z \lesssim 11$ although the $M_\bullet-M_*$ relation arguably varies with redshift \citep{Maiolino+2023}.

The \citetalias{Ricarte+2018a} model includes several variations that explore different physical assumptions.  

\begin{itemize}
\item {\bf Seeding:} First, SMBHs are initialized as either ``light'' or ``heavy'' seeds in the redshift range $15 \leq z \leq 20$.  Light seeds are assumed to be common but low-mass ($30\; \mathrm{M}_{\odot} \leq M_{\bullet} \leq 100\; \mathrm{M}_{\odot}$), while heavy seeds are rare but high-mass \citep[$\sim10^5\, \mathrm{M}_{\odot}$ following the formalism of][which is based solely on dark matter halo mass, redshift, and angular momentum, neglecting the radiation field]{Lodato&Natarajan2006,Lodato&Natarajan2007}.  
\item {\bf Growth:} When a major halo merger (mass ratio of 1:10 or greater) occurs, SMBHs grow via the ``burst'' mode at the Eddington limit until they reach the $M_\bullet-\sigma$ relation.  If they are not growing in this mode, they instead grow through a ``steady'' mode.  In the ``power law'' models included in this paper, a universal Eddington ratio distribution function is assumed, which is sub-dominant in terms of SMBH growth, but important for reproducing observed luminosity functions.  
\item {\bf Dynamics:} When a major halo merger occurs, SMBHs may merge with probability $p_\mathrm{merge}$ at the same time the burst mode is triggered, after a dynamical friction time delay estimated by the \citet{Boylan-Kolchin+2008} formulae. No further black hole dynamics modeling is implemented after the galaxies merge. 
\citetalias{Ricarte+2018a} found that models with $p_\mathrm{merge}=1$ would overshoot the local $M_\bullet-\sigma$ relationship because SMBH-SMBH mergers become the dominant growth mode in the most massive halos.
\end{itemize}

\subsection{The {\sc Santa Cruz} \textit{semi-analytic} model}\label{Methodology_SC}

The {\sc Santa Cruz} SAM \citep{1999MNRAS.310.1087S, Somerville2001TheGalaxies, 2008MNRAS.391..481S, 2014MNRAS.444..942P,2015MNRAS.453.4337S} is based within the backbone of cosmological merger trees, and uses simple, physically motivated recipes to track the flows of gas from the intergalactic medium (IGM) into the circumgalactic medium (CGM) and then into the interstellar medium (ISM), where it can form stars. Massive stars and supernovae drive outflows that eject gas from the ISM, back into the CGM and IGM. Cold ISM gas is partitioned into different phases (atomic, molecular, and ionized) using a prescription based on hydrodynamic simulations, and the transformation of molecular gas into stars is modeled using an empirical relationship based on observations of nearby galaxies \citep{2015MNRAS.453.4337S}. The model predicts a wide range of galaxy properties over cosmic time, like stellar mass, star formation rate, and the metallicity of gas and stellar phases, and has been used extensively to create forecasts and mock catalogs for comparison with past, present and future astronomical surveys \citep{2021MNRAS.502.4858S, 2022MNRAS.515.5416Y, 2023MNRAS.519.1578Y}. These works and others have shown that the {\sc Santa Cruz} SAM reproduces many key observational constraints on galaxy properties from $0 \lesssim z \lesssim 10$.

In the {\sc Santa Cruz} model, each dark matter halo with no progenitor is seeded with a black hole of fixed mass, typically $\sim 10^{4}\, \mathrm{M}_{\odot}$ \citep{2008MNRAS.391..481S,2012MNRAS.426..237H}. Black holes grow through Bondi accretion of gas from the hot halo, and inflows of cold gas from the interstellar medium (ISM), driven either by internal gravitational instabilities or mergers triggering the \textit{radiatively efficient mode} (which is analogous to \textit{cold--mode}). Similar to {\sc Dark Sage}, the {\sc Santa Cruz} SAM provides a channel for bulge and black hole growth through disk instabilities. The main difference between {\sc Dark Sage} and the {\sc Santa Cruz} SAM in this regard is the details of how the instabilities are calculated. The {\sc Santa Cruz} SAM employs a metric from \citet{1982MNRAS.199.1069E} to determine when disks become unstable. Using N-body simulations, \citet{1982MNRAS.199.1069E} showed that disks become unstable when the ratio between dark matter mass and disk mass drops below a specific threshold. They provide a parameterization defining the initiation of disk instabilities as follows:

\begin{eqnarray}   
     M_{\mathrm{disk,crit}}= \frac{v_{\mathrm{max}}^2\ R_{\mathrm{disk}}}{G\
      \epsilon},
\end{eqnarray}
where $M_{\mathrm{disk,crit}}$ represents the threshold disk mass, beyond which the disk is expected to become unstable. $v_{\mathrm{max}}$ denotes the peak circular velocity, $R_{\mathrm{disk}}$ stands for the exponential disk scale length, and $\epsilon$ is the stability parameter. Here, the {\sc Santa Cruz} SAM uses an $\epsilon = 0.75$.

For disk-instability triggered black hole growth, the {\sc Santa Cruz} SAM assumes a constant Eddington ratio of $f_{\mathrm{edd}}=0.01$ with a Gaussian distributed scatter of $0.02\ \mathrm{dex}$, motivated by observations \citep{2012MNRAS.426..237H}. Black holes continue to accrete as long as there is remaining gas fuel in the ``bulge reservoir'', with the accretion rate given by:

\begin{eqnarray}
      \dot{M}_{\bullet,\mathrm{disk}} = 1.26 \times
10^{38}\, \mathrm{erg}\,\mathrm{s}^{-1}\ \frac{1-\epsilon}{\epsilon}
\frac{f_{\mathrm{edd}}}{c^2} M_{\bullet}
\end{eqnarray}

The model also includes rapid black hole growth via gas accretion triggered by galaxy-galaxy mergers, based on findings from cosmological binary merger simulations \citep{2005ApJ...625L..71H, 2006MNRAS.373.1013C, 2006ApJ...641...21R, 2008MNRAS.384..386C}. During this phase, the black hole accretes matter at the Eddington rate until it reaches a critical mass, at which point the energy emitted becomes sufficient to stop further accretion. Following this, the accretion rate gradually decreases in a power-law ``blow-out" phase until the black hole \textit{bright mode} is turned off \citep{2005ApJ...630..716H}. For further details, refer to Section 2.9 from \citet{2008MNRAS.391..481S} and \citet{2012MNRAS.426..237H}. Additional model updates are described in \citet{2021MNRAS.508.2706Y}. 

The Santa Cruz SAM estimates the merger rates of satellite galaxies within their host dark matter halos using the modified Chandrasekhar formula of \citet{Boylan-Kolchin+2008}, as described in detail in \citet{2008MNRAS.391..481S}. The SMBH within each galaxy are assumed to merge instantaneously when the galaxies merge, with the sum of the SMBH remnant being equal to the sum of the progenitor masses.

We use central and satellite galaxies only from the {\sc Santa Cruz} SAM. In this study, we use the version of the {\sc Santa Cruz} SAM recently presented by \citet[2025 \textit{in prep}]{2022MNRAS.517.6091G}. Here, the SAM is run within dark matter merger trees generated using the {\sc ROCKSTAR} halo finder \citep{2013ApJ...762..109B} and {\sc Consistent-Trees} merger tree codes \citep{2013ApJ...763...18B} run on the {\sc IllustrisTNG 300-1} dark matter only simulation (more on {\sc IllustrisTNG} are provided below in section \ref{Methodology_TNG}). The cosmological parameters adopted in this run of the {\sc Santa Cruz} SAM are the same as those adopted by {\sc IllustrisTNG}, and are consistent with Planck results \citep{2016A&A...594A..13P}. The values of the free parameters in the SAM have been adjusted slightly relative to previous work to account for the change in the merger trees (see \citet{2022MNRAS.517.6091G} for details).

\subsection{\textit{The {\sc IllustrisTNG300-1} simulation}}\label{Methodology_TNG}

The Next Generation {\sc Illustris} simulations ({\sc IllustrisTNG}) come in a suite of numerical hydrodynamics simulations with sub-grid prescriptions for SMBH evolution and star formation \citep{Pillepich2018FirstGalaxies, Springel2018FirstClustering, 2019MNRAS.490.3234N, Naiman2018FirstEuropium, Nelson2018FirstBimodality, Marinacci2018FirstFields}. This simulation is an upgraded version derived from the original {\sc Illustris} model \citep{2014Natur.509..177V, 2014MNRAS.444.1518V, 2014MNRAS.445..175G, 2015MNRAS.452..575S, Nelson2015}, incorporating a comprehensive array of physical mechanisms including gas dynamics processes like cooling and heating, star formation, feedback from stars, black hole growth, and the feedback from AGN. Analysis of simulation slices have been used to examine the impact of feedback mechanisms in shaping galactic properties \citep{10.1093/mnras/sty1733, 2020MNRAS.493.1888T}, establishing links between galaxy properties and the dark matter halos enveloping them \citep{2019MNRAS.490.5693B, 2020MNRAS.491.5747M, 2020MNRAS.492.1671Z}, exploring the traits of galaxy clusters and their constituent galaxies \citep{2019ApJ...876...82N, 2020MNRAS.494.1848S, 2020MNRAS.496.2673J}, and predicting high redshift JWST results \citep{2020MNRAS.492.5167V, 2020MNRAS.495.4747S, 2022MNRAS.510.5560S}. 

There are three main boxes in the suite with side lengths of approximately 50, 100, and 300 cMpc, tracing the formation and evolution of structures influenced by gravity and magnetohydrodynamics, spanning from the Universe's early stages to $z=0$. In this paper, we compare our SAMs to the 300 cMpc box with highest resolution (the {\sc TNG300-1} is referred hereafter to as {\sc TNG300-1}), which has a particle mass resolution of 5.9 x $10^7\, \mathrm{M}_{\odot}$ and uses cosmological parameters derived from Planck \citep{2016A&A...594A..13P}, specifically $\Omega_M = 0.3089$, $\Omega_{\Lambda} = 0.6911$, $\Omega_b = 0.0486$, and $h = 0.6774$. As stated before, We use central and satellite galaxies in our sample. For the purpose of our comparison, we note that both {\sc TNG300-1} and Millennium (which is used to run {\sc Dark Sage}) use FoF and {\sc SUBFIND}. We consistently use $M_{200crit}$, which is the total mass in the FoF halo enclosed in a sphere whose mean density is 200 times the critical density of the Universe at the time the halo is recorded to set halo masses. 

In {\sc TNG300-1}, black holes with a mass of $1.1\times10^{6}\, \mathrm{M}_{\odot}$ are seeded in every halo that reaches a mass of $ 7\times10^{10}\, \mathrm{M}_{\odot}$ \citep{2017MNRAS.465.3291W}. Soon after being seeded, these black holes start growing through Bondi accretion (equation \ref{eq:BH_B}) capped at the Eddington rate (equation \ref{eq:BH_Edd}). The resulting accretion rate is determined by the lesser of the two rates, denoted as $\dot{m}_{\mathrm{BH}}$ in Equation \ref{eq:BH_Mdot}, 

\begin{equation}
    \dot{m}_{\mathrm{Bondi}}= \frac{4\uppi G^2 m_{\mathrm{BH}}^2 \rho}{c_{s}^3},
\label{eq:BH_B}
\end{equation}
\begin{equation}
    \dot{m}_{\mathrm{Edd}}= \frac{4\uppi G m_{\mathrm{BH}} m_p}{\epsilon_r \sigma_T} \ c,
\label{eq:BH_Edd}
\end{equation}
\begin{equation}
\dot{m}_{\mathrm{BH}}= {\rm min}(\dot{m}_{\mathrm{Bondi}},\,\dot{m}_{\mathrm{Edd}}),
\label{eq:BH_Mdot}
\end{equation}

\noindent
where $c_{s}$ is the speed of sound of the local interstellar medium, $\rho$ is the density of the nearby heated gas, $\epsilon_r$ is the efficiency of radiative accretion assumed to be 0.2, and $\sigma_T$ is the Thompson cross-section. 

Once equation \ref{eq:BH_Mdot} is calculated, one can determine if the black hole accretion comes from the \textit{thermal-mode} or \textit{kinetic-mode}. To determine which {\sc TNG300-1} black holes accrete in each of these modes, a black hole mass-dependent threshold is applied using the following equation:

\begin{equation}
  \chi = \min\left[ 0.002 \left(\frac{M_\text{BH}}{10^8 \,\mathrm{M}_{\odot}} \right)^2, 0.1\right].
\end{equation}
\noindent 
If the Eddington ratio exceeds this threshold, the black hole is deemed to be accreting in the \textit{thermal-mode}, whereas if the Eddington ratio is lower than the threshold, the black hole grows through \textit{kinetic-mode} accretion. Other than these two modes, black holes may also grow by merging with other black holes when galaxies merge. Like in {\sc Dark Sage} and the  {\sc Santa Cruz} SAM, no black hole dynamics are included in the simulation. Once merged and at any point in the history of the halo that hosts a black hole, the model enforces the SMBH position to be at the potential minimum of its host halo at every global integration timestep \citep{10.1093/mnras/sty1733}.

\subsection{The {\sc TRINITY} \textit{Empirical} Model }\label{describing_trinity}

{\sc TRINITY} connects galaxy properties to SMBHs \citep{2023MNRAS.518.2123Z} using the {\sc UNIVERSEMACHINE} \citep{Behroozi2019Universemachine:010} machinery as its foundation. First, the model defines the galactic star formation rate (SFR) as a function that depends both on the halo mass and redshift. Within this parameter space, considering the average mass growth history of halos, an integration of the resulting SFRs is performed. {\sc TRINITY} is constructed to reproduce the observed stellar mass functions, galaxy quenched fractions, cosmic star formation rates, specific star formation rates, galaxy UV luminosities, quasar luminosity functions, quasar probability distribution function, active SMBH mass function, SMBH mass-bulge mass relation, and the SMBH mass distribution of bright quasars within $z=0-10$. {\sc TRINITY} has the largest number of free parameters of the models in our study (56; see \autoref{app:calibration}). As a result, the model outputs a stellar mass-to-halo mass relation. This calculated stellar mass is used to obtain bulge masses using observational scaling relations \citep{2004ApJ...604L..89H, 2012MNRAS.419.2497B, 2013ARA&A..51..511K, 2013ApJ...764..184M, 2016ApJ...817...21S}. The chosen SMBH mass-bulge mass relation determines the average accretion rates of SMBHs, because the growth history of galaxies on average is determined by the relationship between SFR and halo mass. The distributions of Eddington ratios and the radiative efficiency are parametrized, and they dictate how the growth of SMBHs is connected to the observed distribution of SMBH luminosities. Lastly, the model compares these predictions to calculate a likelihood function and employs a Markov Chain Monte Carlo (MCMC) algorithm to derive the posterior distribution of model parameters that align with the observed data. Like {\sc TNG300-1}, {\sc TRINITY} uses a flat $\Lambda$ cold dark matter cosmology consistent with Planck results \citep{2016A&A...594A..13P}.

\section{Comparison of Models: A Diversity of SMBH Assembly Histories} \label{model_comparisons}

In this section, we present the results of comparing \textit{empirical}, \textit{semi-analytic} models, and \textit{numerical hydrodynamics} simulations. Section \ref{local_BHMF} discusses the local black hole mass function. The evolution of black hole mass is highlighted in section \ref{BHMF_z0_z6}. Section \ref{Eddratios_models} shows the time evolution of the Eddington ratio distributions. Lastly, section \ref{cosmicSMBHfunc_z0_6} compares the predicted and observationally inferred cosmic SMBH mass density functions.

\subsection{Local black hole mass function} \label{local_BHMF}
 
\begin{figure*}[t]
\centering
\includegraphics[width=1.5\columnwidth, clip]{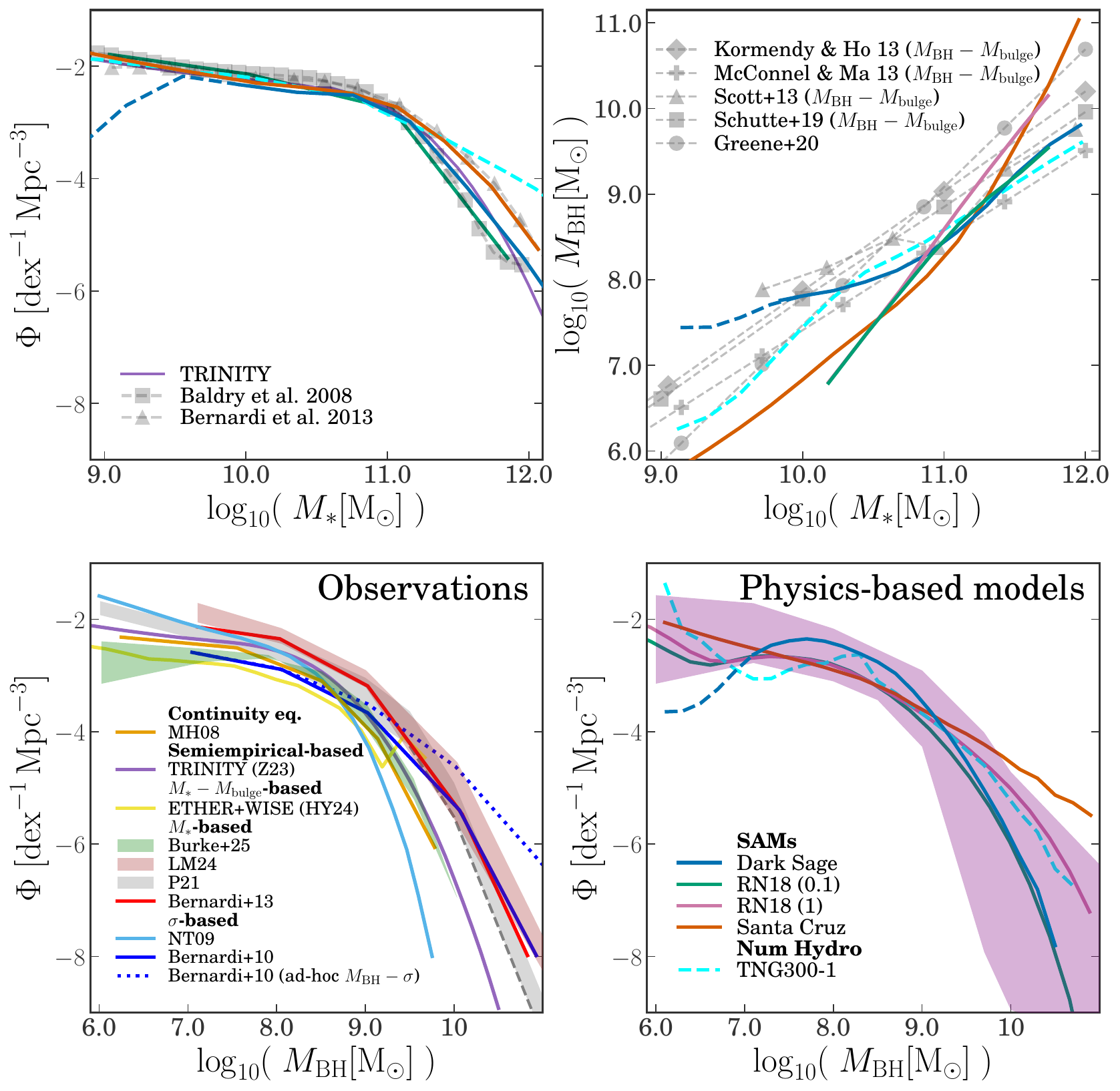}
\caption{The stellar mass function, black hole mass-stellar mass relation, and the black hole mass function at $z=0$. The top left panel shows the \textit{semi-analytic} models {\sc Dark Sage} (solid dark blue line), the {\sc Santa Cruz} SAM (solid brown line), and \citetalias{Ricarte+2018a} (solid green line: $p_{merge}$=0.1). We include {\sc TRINITY} (solid purple line) and {\sc TNG300-1} (dashed cyan line). The observed SMF from \citet{Baldry2008} and \citet{2013MNRAS.436..697B} (used to calibrate {\sc Dark Sage}, the {\sc Santa Cruz} SAM, and {\sc TNG300-1}) are included, while \citetalias{Ricarte+2018a} models use the stellar mass function from \citet{2013MNRAS.428.3121M}. The top right panel presents the mean of these models alongside observational data from \citet{Greene2020} and \citet{Reines&Volonteri2015}. The bottom left panel categorizes observational constraints based on inferred BHMFs, showing: the continuity equation from \citet{2008MNRAS.388.1011M} (solid orange line), semiempirical methods from \citet{2023MNRAS.518.2123Z}, stellar mass-bulge mass relation from \citet{2023Galax..11...15R} and \citet{Hernandez-Yevenes+2024} (ETHER+Wise shown in solid yellow-green line), stellar mass constraints from \citet{Burke2025}, \citet{2024ApJ...971L..29L}, \citet{2021ApJ...923..260P}, and \citet{2013MNRAS.436..697B} (shown as green, dark red, grey shaded regions, and solid red line, respectively), and the velocity dispersion function from \citet{2009MNRAS.393..838N} (solid light blue line), \citet{2010MNRAS.404.2087B} (solid blue line), and the $M_{\rm BH}-\sigma$ scaling relation from \citet{Sato-Polito+2023} with \citet{2010MNRAS.404.2087B}. The bottom right panel compares our physics-based models, with the violet shaded area representing the range of observational results from the bottom left panel. Note that the dashed line for {\sc Dark Sage} represents galaxies in halos with fewer than 200 particles. The grey dashed line in the bottom left panel from \citet{2021ApJ...923..260P} denotes extrapolated BHMFs for $M_{\rm BH} < 10^{9.5}\, \mathrm{M}_{\odot}$. Our results for the {\sc Santa Cruz} SAM on {\sc TNG300} dark matter trees are consistent with \citet{2022MNRAS.517.6091G}, who used the {\sc Santa Cruz} SAM model on dark matter trees from the {\sc TNG100} simulation. We find significant discrepancies in the observed BHMF. The physics-based models fall within the observed range.}\label{fig:BHMF_z0}
\end{figure*}

The SMBH mass function offers a valuable overview of the statistics of the local population of black holes. The inferred BHMF varies significantly depending on the choice of SMF, black hole mass scaling relation, or AGN accretion history. The top panels of Figure \ref{fig:BHMF_z0} compare the SMF and black hole mass-stellar mass relation at $z=0$. While \citet{Baldry2008} and \citet{2013MNRAS.436..697B} agree up to $M_{*} \sim 10^{11}\, \mathrm{M}_{\odot}$, the latter predicts a higher number density at larger masses. As a result, models calibrated to \citet{2013MNRAS.436..697B}, such as the {\sc Santa Cruz} SAM and {\sc TNG300-1}, produce a higher number density of massive black holes. In contrast, {\sc TRINITY} and \citetalias{Ricarte+2018a} rely on SMFs from \citet{Baldry2012}, \citet{2013ApJ...767...50M}, and \citet{2013MNRAS.428.3121M}, which are more consistent with \citet{Baldry2008}.

Studies have shown that the number density of massive galaxies has been underestimated by factors of 3-10 for $M_{\rm *} \sim 10^{11-11.6}\, \mathrm{M}_{\odot}$ and by up to 100 times for even higher stellar masses \citep{2017MNRAS.467.2217B, 2017MNRAS.468.2569B}. \citet{2018MNRAS.475..757B} suggests that a varying IMF, likely bottom-heavy in the most massive galaxies with high velocity dispersion, implies even more massive galaxies than predicted by the Santa Cruz SAM. Therefore, models calibrated to \citet{Baldry2008}, \citet{Baldry2012}, and \citet{2013ApJ...767...50M} likely underestimate the number of massive black holes, affecting even the \citet{2021ApJ...923..260P} ``upper limit'' based on the {\sc UniverseMachine} which uses the SMF from \citet{2013ApJ...767...50M}. Unless scaling relations become steeper at high masses, their inferred BHMFs likely underrepresent the high-mass end. Older empirical models based on pre-SDSS or early SDSS-EDR galaxy mass functions and velocity dispersion functions have been superseded.

For the black hole mass-stellar mass relation, we see overlap between the {\sc Santa Cruz} SAM and \citetalias{Ricarte+2018a} models (solid green line: $p_{merge}$=0.1, solid violet line: $p_{merge}$=1.0). {\sc Dark Sage} and {\sc TNG300-1} closely follow each other's mean relation for $M_{\rm BH} > 10^{10.2}\, \mathrm{M}_{\odot}$. All of our physics-based models fall within \citet{Greene2020} and \citet{Reines&Volonteri2015} observations. For $M_{\rm BH} > 10^{11}\, \mathrm{M}_{\odot}$, even though both the {\sc Santa Cruz} SAM and {\sc TNG300-1} are calibrated to match \citet{2013MNRAS.436..697B}, they have different $M_{\rm BH}-M_*$ normalizations. This is because the {\sc Santa Cruz} SAM is tuned to match the local black hole mass-bulge mass relation from \citet{2013ApJ...764..184M}. The local black hole mass–bulge mass relation in {\sc Dark Sage} has a lower normalization, leading to a shallower $M_{\rm BH}-M_*$ relation. Interestingly, the {\sc Santa Cruz} SAM and the \citetalias{Ricarte+2018a} $p_{\rm merge} = 1.0$ model converge at high masses, despite the latter not modeling bulges. This suggests that a higher black hole merger rate could also raise the normalization of the $M_{\rm BH}-M_*$ relation.

The normalization of the black hole mass–bulge mass relation varies by a factor of 2–3 \citep{2013ARA&A..51..511K}, with more recent estimates favoring higher black hole-to-bulge mass ratios at the high-mass end. For example, {\sc TRINITY} uses datasets that equally weighs older samples like \citet{2004ApJ...604L..89H}, which report a lower normalization, compared to newer scaling relations like those in \citet{2013ARA&A..51..511K}. ETHER+WISE use a modified version of \citet{2019ApJ...887..245S}, which has a comparable black hole mass-bulge mass normalization to \citet{2013ARA&A..51..511K}. There is a general agreement that the normalization of these scaling relations is on the higher side, which in turn leads to a more prominent black hole mass function at the massive end.

Bottom two panels show the local BHMFs for observations (left) and the physics-based models (right). In general, observations agree within $\lesssim 0.8\ \mathrm{dex}$ for $M_{\rm BH} \sim 10^{7-8}\, \mathrm{M}_{\odot}$ SMBHs. For $M_{\rm BH} < 10^{7}\, \mathrm{M}_{\odot}$, the differences between the models increase to $\lesssim 1.5\ \mathrm{dex}$. By $M_{\rm BH} \sim 10^{10}\, \mathrm{M}_{\odot}$, we show the largest disagreement between observations, with a scatter of $\lesssim 4.6\ \mathrm{dex}$. First, \citet{2008MNRAS.388.1011M} derive the BHMF by converting AGN X-ray luminosity to black hole accretion rates, which depend on the Eddington ratio, and applying the continuity equation \citep{1971ApJ...170..223C, 1992MNRAS.259..725S}. While their results align with the {\sc TRINITY} model, uncertainties remain, particularly in the bolometric correction of the local BHMF. Their method primarily accounts for low-accretion rate objects ($L/L_{edd} < 10^{-2}$) and extrapolates to higher rates under the assumption that radio luminosity is weak in the radiatively efficient mode. However, AGN in this mode significantly contribute to the X-ray luminosity function. For this reason, there is potential for substantial systematic errors in their BHMF estimate.

The {\sc TRINITY} model may decline exponentially at the high-mass end because it relies on the SMF from \citet{Baldry2012} and \citet{2013ApJ...767...50M}, along with their relatively shallow $M_{\rm BH}-M_*$ and $M_{\rm BH}-M_{\rm bulge}$ relations in this regime. For the bulge mass-stellar mass-inferred BHMF, we include the Event Horizon and Environs (ETHER) catalog, which currently compiles 1.6 million black hole mass estimates, approximately 15,500 milliarcsecond-scale radio fluxes, around 14,000 hard X-ray fluxes, and SED information derived from existing catalogs, database queries, and literature \citep{2023Galax..11...15R,Hernandez-Yevenes+2024}. This is combined with the Wide-field Infrared Survey Explorer (WISE) catalog, a comprehensive database with infrared wavelengths
at 3.4 $\mu$m, 4.6 $\mu$m, 12 $\mu$m, and 22 $\mu$m collected by the Wide-field Infrared Survey Explorer \citep{2010AJ....140.1868W}. \citet{Hernandez-Yevenes+2024} estimate stellar masses and bulge fractions using WISE photometry, calibrated to the Galaxy and Mass Assembly (GAMA) optical survey \citep{2011MNRAS.413..971D}. They adopted a modified \citet{2019ApJ...887..245S} black hole mass–bulge mass relation, which is closely aligned with \citet{2013ARA&A..51..511K}. We see that between $M_{\rm BH} \sim 10^{7-9}\, \mathrm{M}_{\odot}$ ETHER+WISE show the lowest number density of all observational samples. Above this range, there is a raise in the number density that comes from not considering the intrinsic scattter of the black hole mass-bulge mass relation \citep{Hernandez-Yevenes+2024}. At the high mass end, ETHER+WISE agree with \citet{2013MNRAS.436..697B}, \citet{2021ApJ...923..260P}, and \citet{2024ApJ...971L..29L}.

\citet{Burke2025} developed a Bayesian model to infer the BHMF by analyzing X-ray, radio, and optical variability observations. Here, they use the data release 4 of the GAMA survey \citep{2022MNRAS.513..439D}, convolved with the $M_{\rm BH}-M_*$ relations from \citet{Greene2020}. Below $M_{\rm BH} < 10^{9}\, \mathrm{M}_{\odot}$, \citet{Burke2025} is consistent with the ETHER+WISE sample, which agrees with results from tidal disruption event estimates \citep{2023ApJ...955L...6Y}. At the lowest black hole masses, \citet{2008MNRAS.388.1011M}, {\sc TRINITY}, \citet{2021ApJ...923..260P}, and \citet{2009MNRAS.393..838N} predict a number density roughly an order of magnitude higher than \citet{Burke2025}. Above $M_{\rm BH} > 10^{9}\, \mathrm{M}_{\odot}$, \citet{Burke2025}, \citet{2008MNRAS.388.1011M}, and {\sc TRINITY} converge, while \citet{2021ApJ...923..260P} and \citet{2024ApJ...971L..29L} lie with a higher number density at roughly all masses. The SMF used in \citet{2024ApJ...971L..29L} includes the results of the MASSIVE survey \citet{2014ApJ...795..158M, 2022ApJ...932..103G} (for $M_{*} > 10^{11.5}\, \mathrm{M}_{\odot}$) and \citet{2020ApJ...893..111L} (for $M_{*} < 10^{11.3}\, \mathrm{M}_{\odot}$), with an assumed $M_{\rm BH}-M_*$ relation from \citet{2013ARA&A..51..511K}. Between $M_{\rm BH} \sim 10^{8-10.2}\, \mathrm{M}_{\odot}$, \citet{2024ApJ...971L..29L} overlaps with BHMF limits from \citet{2021ApJ...923..260P}. Outside of this range, \citet{2024ApJ...971L..29L} is above most observations, with the exception of \citet{2010MNRAS.404.2087B} and \citet{Sato-Polito+2023} (ad-hoc $M_{\rm BH}-\sigma$ at the high mass end. The BHMF inferred from \citet{2013MNRAS.436..697B} is consistent with \citet{2024ApJ...971L..29L} shaded region. \citet{2021ApJ...923..260P} considers a wide range of possible BHMFs and presents a ``lower" and ``upper" limit from $z=0-6$. The ``lower" BHMF limit comes from \citet{2009ApJ...690...20S} derived from the continuity equation. The ``upper" BHMF limit is inferred empirically using the {\sc UNIVERSEMACHINE} SMF \citep{Behroozi2019Universemachine:010} combined with the black hole mass-stellar mass scaling relation from \citet{2013ARA&A..51..511K}. It is possible that other black hole mass-stellar mass relations may change this ``upper" limit slightly.

\citet{2009MNRAS.393..838N} derived the BHMF using the number density of black holes from the Sloan Digital Sky Survey (SDSS) \citep{2000AJ....120.1579Y}. The BHMF at the high-mass end decreases at $M_{\rm BH} < 10^{10}\, \mathrm{M}_{\odot}$ because, at that time, SDSS had not surveyed as large of a volume as it has in more recent observations. \citet{2010MNRAS.404.2087B} uses the velocity dispersion function and the $M_{\rm BH}-\sigma$ scaling relation from \citet{Sato-Polito+2023}. \citet{2010MNRAS.404.2087B} overlaps with \citet{2024ApJ...971L..29L} and partially with \citet{2021ApJ...923..260P} above $M_{\rm BH} > 3\times10^{9}\, \mathrm{M}_{\odot}$. At lower black hole masses, it shows a lower number density compared to most $M_*$-based BHMFs. \citet{Sato-Polito+2023} introduce a steeper second power law in their ad-hoc $M_{\rm BH}-\sigma$ relation to match the amplitude of the stochastic gravitational-wave background observed by PTA teams. This BHMF predicts significantly more SMBHs above $M_{\rm BH} > 10^{9}\, \mathrm{M}_{\odot}$ than a standard single power-law $M_{\rm BH}-\sigma$ relation.

If the high-mass end of the local BHMF is dictated by scaling relations, then tuning the {\sc Santa Cruz} SAM to match both \citet{2013MNRAS.436..697B} and \citet{2013ARA&A..51..511K} would likely produce the highest number density of massive black holes among all models. Calibrating to only one of these constraints is not enough. For example, tuning {\sc TNG300-1} to \citet{2013MNRAS.436..697B} increases the BHMF by at most $\lesssim 1.3\ \mathrm{dex}$ at $M_{\rm BH} \sim 10^{10.5}\, \mathrm{M}_{\odot}$, compared to a model matching the \citet{Baldry2008} local SMF. Similarly, using a steeper black hole mass-bulge mass relation also results in a $\lesssim 1.3\ \mathrm{dex}$ increase in number density at the same mass. Interestingly, \citet{2024ApJ...971L..29L} derived their stellar mass-inferred BHMF by convolving the SMF with a probability distribution function linked to the $M_{\rm BH}-M_*$ relation. They used \citet{2013ARA&A..51..511K} and stated that adopting \citet{Saglia+2016} would not significantly alter the results. However, \citet{Saglia+2016} presents a shallower black hole mass-bulge mass relation than \citet{2013ARA&A..51..511K}, suggesting that steepness alone does not fully explain the high number density of massive black holes.

As seen, the estimates diverge most strongly from each other at the highest and lowest masses. At the lowest masses, number densities are essentially extrapolations from more massive SMBH populations where SMBH-galaxy relationships are known. However, relationships found in high-mass (typically bulge-dominated) galaxies may not all agree for lower-mass galaxies \citep{Reines&Volonteri2015,Sturm&Reines2024}. We find a very large degree of disagreement in the number density of ultra-massive black holes (UMBHs), those with $M_\mathrm{BH}>10^9\, \mathrm{M}_{\odot}$, which have recently become an observationally exciting mass range due to the detection of the stochastic nHz gravitational wave background \citep{Agazie+2023}. At these masses, it has been argued that number densities of UMBHs could plausibly be suppressed due to self-regulated growth imposing de-facto an upper limit on SMBH masses \citep{2009MNRAS.393..838N}. Recently, several authors have argued that SMBHs in the most massive galaxies may be systematically {\it over-}massive with respect to local scaling relations at lower masses \citep{Mezcua+2018, 2023ApJ...957L...3P, 2024ApJ...966L..30M}.  A larger number density of detected UMBHs than anticipated by most relationships may also help explain the large amplitude of the nHz gravitational wave background \citep{Sato-Polito+2023,Izquierdo-Villalba+2024}.

In the \textit{semi-analytic} model and \textit{numerical hydrodynamics} simulation considered in this paper, for $M_{\rm BH} \sim 10^{8-9}\, \mathrm{M}_{\odot}$, all models agree within $\lesssim 0.4\ \mathrm{dex}$. At  $M_{\rm BH} \sim 10^{7.1-9}\, \mathrm{M}_{\odot}$, {\sc Dark Sage} overpredicts the number of black holes compared to all models. For $M_{\rm BH} < 10^{7}\, \mathrm{M}_{\odot}$, {\sc Dark Sage} predicts a lower number density of BH than the {\sc Santa Cruz} and \citetalias{Ricarte+2018a} SAMs by about $1.3\ \mathrm{dex}$. This difference decreases to $0.7\ \mathrm{dex}$ if we remove our {\sc Dark Sage} adopted halo mass cut. By $M_{\rm BH} \sim 10^{10}\, \mathrm{M}_{\odot}$, {\sc TNG300-1} and the {\sc Santa Cruz} SAM agree in having $0.9\ \mathrm{dex}$ higher densities than {\sc Dark Sage} and \citetalias{Ricarte+2018a} $p_{merge}$=0.1 model. This difference increases with increasing black hole mass. The {\sc Santa Cruz} SAM and TNG models are quite similar up to $M_{\rm BH} \sim 10^{10}\, \mathrm{M}_{\odot}$, as demonstrated by \citet{2022MNRAS.517.6091G}. However, for black holes with masses exceeding $M_{\rm BH} \sim 10^{10}\, \mathrm{M}_{\odot}$, the {\sc Santa Cruz} SAM model predicts a higher number density. Interestingly, the two models from the \citetalias{Ricarte+2018a} SAM, which differ in their predicted probabilities of black hole mergers, provide varying abundances that lie squarely within these mass ranges. The model with the higher black hole merger rate predicts a greater number density of black holes, since SMBH mergers come to dominate the growth of the most massive SMBHs. Overall, all models fall within the range of observations discussed above. 

The discrepancies found in our model comparison can be attributed to differences in calibration procedures (see appendix \ref{app:calibration}), which in turn rely on observational data sets that often do not agree, and, as we show in later results, the existence of multiple methodologies to derive the same final outcome. Although all these models are calibrated to match the observed stellar mass function in addition to the $M_{\rm BH}-\sigma$, $M_{\rm BH}-M_{*}$, or $M_{\rm BH}-M_{\rm bulge}$ relation at $z=0$, differences in the set of SMF and steepness in the scaling relations used are the most likely cause of the disagreements seen.

\subsection{The evolution of black hole mass} \label{BHMF_z0_z6}

\begin{figure*}[t]
\centering
\includegraphics[width=2.1\columnwidth, clip]{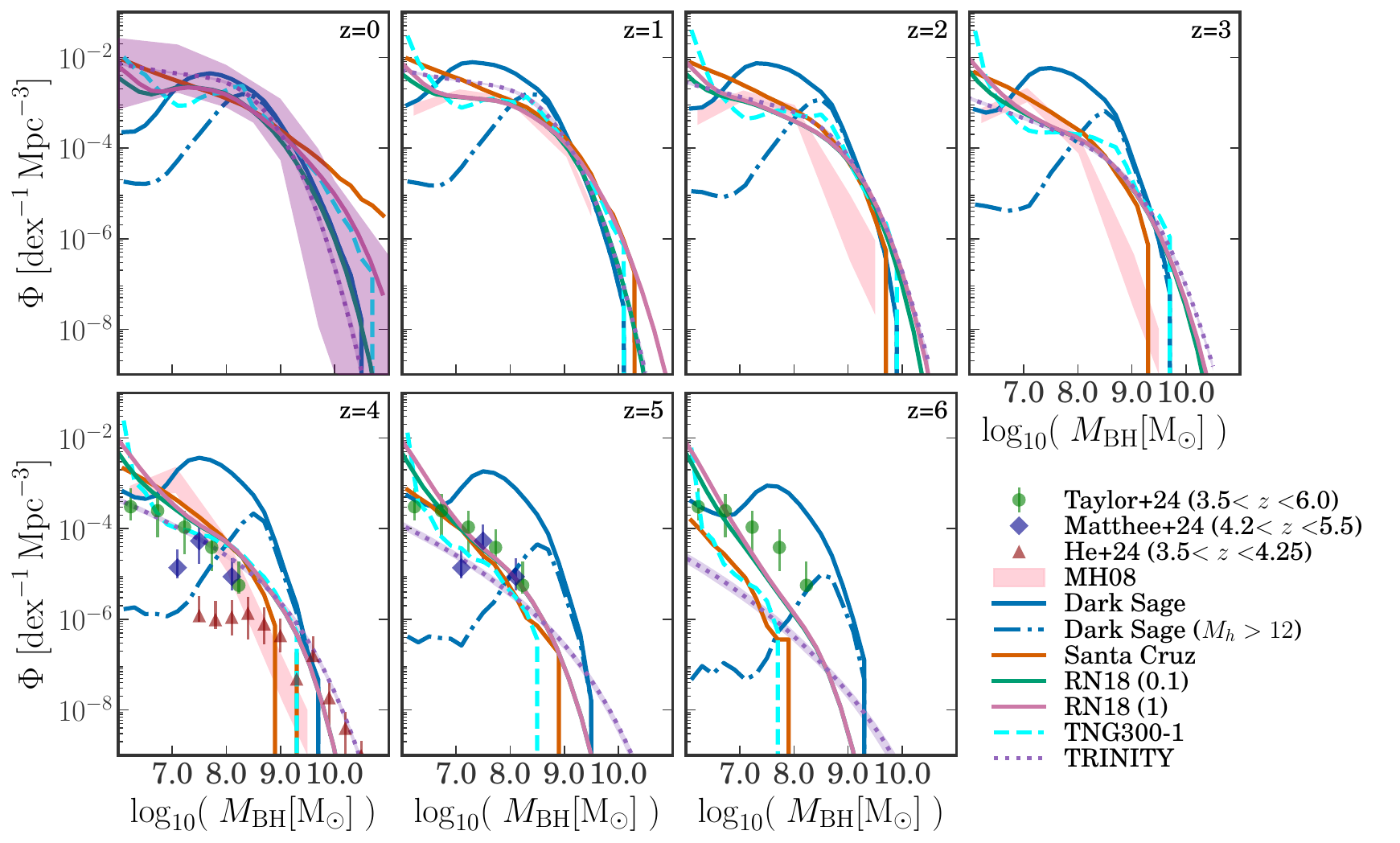}
\caption{Black hole mass function from $z=0$ to $z=6$ for {\sc Dark Sage} (solid dark blue line), {\sc Dark Sage} black holes in $M_h > 10^{12}\, \mathrm{M}_{\odot}$ (dashed dotted dark blue line), the {\sc Santa Cruz} SAM (solid brown line), and \citetalias{Ricarte+2018a} (solid green line: $p_{merge}$=0.1, solid violet line: $p_{merge}$=1.0). We include {\sc TNG300-1} (dashed cyan line) and {\sc TRINITY} (dotted purple line) in this comparison. The BHMF derived from the continuity equation in \citet{2008MNRAS.388.1011M} for $1 < z < 4$ is shown in pink. We include the broad-line AGN BHMF for $3.5 < z < 6$ in green circles  from \citet{2024arXiv240906772T}. NIRCam WFSS from \citet{2024ApJ...963..129M} for $4.2 < z < 5.5$ are shown as purple diamonds. Ground-based SDSS+HSC data from \citet{2024ApJ...962..152H} for $3.5 < z < 4.25$ are in red triangles. Note that the lowest mass datapoint from \citet{2024arXiv240906772T} and \citet{2024ApJ...963..129M} suffer from incompleteness. Models that show significant discrepancies at $z=0$ align more closely at $z\sim1-4$, except for the distinct bump in {\sc Dark Sage}.}\label{fig:BHMF_z0_z6}

\end{figure*}

We now turn to the evolution of the SMBH mass function. The BHMF from $z=0$ to $z=6$ is plotted in Figure \ref{fig:BHMF_z0_z6}. Physics-based models that disagree at $z=0$ converge between $z\approx 1-4$, with the exception of {\sc Dark Sage} which shows a distinct ``knee'' near $M_{\rm BH} \sim 10^{8}\, \mathrm{M}_{\odot}$ that remains fixed across time. This ``knee'' becomes progressively narrower with increasing redshift, leading to a sharper drop at the high mass end. We compare our results to \citet{2008MNRAS.388.1011M}, who use the X-ray AGN luminosity functions and an assumed Eddington ratio distribution. Because they only capture actively accreting black holes, they likely underestimate the number of massive, quiescent black holes. At $z=1$, the BHMF from \citet{2008MNRAS.388.1011M} is roughly in agreement with most physics-based models. At $z=2-3$, the disagreement between \citet{2008MNRAS.388.1011M} and all models is more apparent at the high mass end, likely due to the lack of obscured AGN sources and SMBH mergers included in the inferred BHMF. Interestingly, by $z=4$, \citet{2008MNRAS.388.1011M} closely traces the BHMFs derived from JWST \citep{2024arXiv240906772T, 2024ApJ...963..129M, 2024ApJ...962..152H}. While {\sc TNG300-1} and {\sc TRINITY} align closely with JWST sources at $z=4$, by $z=5$, the {\sc Santa Cruz} SAM and \citetalias{Ricarte+2018a} models provide a better match. At $z=6$, for $M_{\rm BH} > 10^{7}\, \mathrm{M}_{\odot}$, all models except {\sc Dark Sage} underestimate number densities relative to \citet{2024arXiv240906772T}.

By construction, the BHMFs from \citetalias{Ricarte+2018a} have an exponential drop-off that becomes steeper with increasing redshift. Both $p_{merge}$ models have consistent BHMFs with each other, except for deviations at the high-mass end below $z < 2$. At fixed mass in this range, the $p_{merge}$=1.0 model shows a higher number density of black holes than the $p_{merge}$=0.1 model, indicating a higher rate of BH--BH mergers. Interestingly, by $z=1$, $p_{merge}$=1.0 model agrees better with the {\sc Santa Cruz} SAM, while the $p_{merge}$=0.1 model aligns better with the {\sc Dark Sage} prediction. 

The {\sc Santa Cruz} SAM aligns closely with \citetalias{Ricarte+2018a}, {\sc TRINITY}, and {\sc TNG300-1} between $z = 0$ and $2$. Between $z = 3$ and $6$, the {\sc Santa Cruz} SAM and {\sc TNG300-1} exhibit a sharp drop in number density around $M_{\rm BH} \sim 10^{8-9}\, \mathrm{M}_{\odot}$, likely due to the relatively small volume of the simulation box. By $z = 6$, the {\sc Santa Cruz} SAM and {\sc TNG300-1} BHMFs closely agree. In {\sc TRINITY}, the ``knee'' in the BHMF becomes less pronounced at higher redshifts. This is because, at high redshift, the $M_{*}-M_h$ and $M_{BH}-M_h$ relationships can be approximated by a single power law \citep{2023MNRAS.518.2123Z}. There is also strong evolution in the BHMF above $z = 5$ driven by universally high Eddington ratios at high redshifts. In fact, {\sc TRINITY} allows super-Eddington accretion. In general, the high-growth period ends earlier for more massive black holes \citep{2023MNRAS.518.2123Z}. This is commonly referred to as ``AGN downsizing,'' whereby more massive active black holes exhibit their peak activity at earlier cosmic epochs \citep{2004MNRAS.353.1035M, 2005AJ....129..578B, 2012MNRAS.426..237H}. Below $z = 3$, AGN downsizing slows the evolution of the BHMF at the massive end, while $M_{\rm BH} \sim 10^{8-9}\, \mathrm{M}_{\odot}$ grow significantly. This continued growth establishes the ``knee'' in the BHMF at low redshift.

The number density in the low-mass end in {\sc Dark Sage} is underestimated due to the resolution of the underlying dark matter simulation. In general, the BHMF of {\sc Dark Sage} stands out as significantly distinct from those of the other models. The distinct bump in {\sc Dark Sage} around $M_{\rm BH} \approx 10^{8}\, \mathrm{M}_{\odot}$ comes from the large contribution of \textit{cold-mode} accretion during galaxy mergers, that is, \textit{ex-situ} cold gas added directly into the black hole after galaxies merge (see Fig. \ref{fig:fracSMBH_channels}). Even if we only account for black holes in $M_{h} > 10^{12}\, \mathrm{M}_{\odot}$ (dashed dotted dark blue line), we see that the bump is preserved with a smaller normalization in the BHMF, indicating the importance of \textit{cold-mode} accretion to all black hole masses.

\begin{figure*}[t]
\centering
\includegraphics[width=2.1\columnwidth, clip]{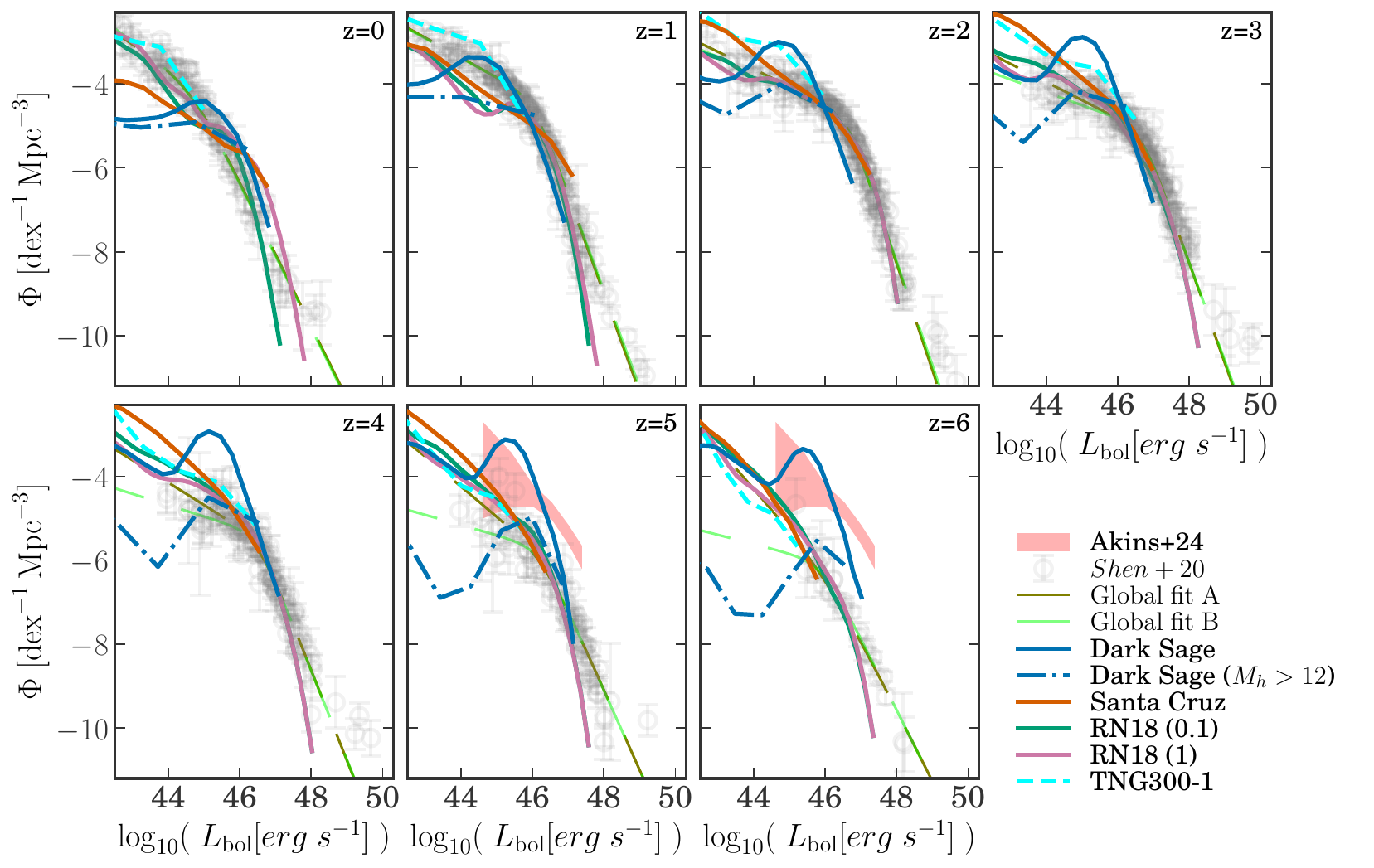}
\caption{The bolometric AGN luminosity function at $z=0-6$ for all models described in figure \ref{fig:BHMF_z0_z6}. Here, we include observations from \citet{2020MNRAS.495.3252S} in grey, which include the rest-frame IR, B band, UV, soft, and hard X-ray as well as bolometric, dust, and extinction corrections.  We add \citet{2020MNRAS.495.3252S} Global fit A and B in dashed olive and lime lines, respectively. The shaded red region at $z=5-6$ is the inferred bolometric luminosity if we assume that the ``little red dots'' discovered by JWST are AGNs \citep{2024arXiv240610341A}. Overall, the {\sc TNG300-1} and the \citetalias{Ricarte+2018a} models agree with observations across time. The {\sc Santa Cruz} SAM deviates from observations at $z=0$ in the bright end of the bolometric AGN luminosity function but aligns with observations at higher redshifts. {\sc Dark Sage}'s bump surprisingly falls within the inferred bolometric AGN luminosity function from \citet{2024arXiv240610341A}.}\label{fig:QLF_z0z6}

\end{figure*}

Figure \ref{fig:QLF_z0z6} presents the bolometric AGN luminosity function across redshift. We compare predictions from physics-based models to observations from \citet{2020MNRAS.495.3252S} and \citet{2024arXiv240610341A}. At all redshifts, {\sc TNG300-1} roughly matches observations, generally sitting slightly above the Global fit A line. The faint-end of the luminosity function steepens with increasing redshift, consistent with \citet{2022MNRAS.509.3015H}. Note that Global fits A and B differ in their slopes for $L_{\rm bol} \lesssim 10^{46}\, erg \ s^{-1}$, with Global fit B flattening with increasing redshift. The {\sc Santa Cruz} SAM reproduces the observed AGN luminosity function at $z=1-6$, in agreement with \citet{2021MNRAS.508.2706Y}. We find that the bright end of the quasar luminosity function becomes steeper at $z\gtrsim2$, while the faint end steepens at $z\gtrsim3$ and continues to do so at higher redshifts \citep{2020MNRAS.495.3252S}. We notice that at $z=2-3$, the faint end slope in the {\sc Santa Cruz} SAM is overpredicting the number density. Its slope stays consistently steep up to $z=6$ when it finally matches the observations. At $z=0$, the {\sc Santa Cruz} SAM\footnote{For Figures \ref{fig:QLF_z0z6} and \ref{cosmicSMBHfunc_z0_6}, we use only the ``radiatively efficient'' (merger and disk instability triggered) accretion mode from the {\sc Santa Cruz} SAM  model, as we expect accretion associated with the ``jet mode'' to be radiatively inefficient.} matches the observed bright end slope. Note that none of these models, with the exception of \citetalias{Ricarte+2018a}, are tuned to match the observed bolometric quasar luminosity function, so it is surprising to see the level of agreement between some physics-based models and observations. \citetalias{Ricarte+2018a} is tuned to reproduce the $z=0.1$ AGN bolometric luminosity function from \citet{Hopkins+2007} and \citet{Ueda+2014}. Yet, at $z=1-6$, it naturally matches the observed luminosity functions. It appears more discrepant at the high-luminosity end for redshifts $z=1-6$, but \citetalias{Ricarte+2018a} determined that this was most likely due to a lack of massive halos in the simulation.

Between $z=0-1$, although {\sc Dark Sage} closely matches the observed bolometric AGN luminosity function at the high-luminosity end, at low-luminosities, it underpredicts the number density by about two orders of magnitude. The low-luminosity end in {\sc Dark Sage} starts to converge with other models at $z\gtrsim3$. By $z=2$, the distinct ``knee'' near $M_{\rm BH} \sim 10^{8}\, \mathrm{M}_{\odot}$ (Figure \ref{fig:BHMF_z0_z6}) becomes more pronounced, diverging from the observed single power-law distribution with increasing redshift. By $z=5-6$, this feature is about two orders of magnitude above the observations from \citet{2020MNRAS.495.3252S}, yet it remains consistent with the inferred bolometric AGN luminosity function from ``little red dots'' \citep{2024arXiv240610341A}. Note that ``little red dots'' are assumed to be actively accreting black holes with AGNs emitting broad lines. The current literature remains uncertain whether ``little red dots'' are galaxies or AGN \citep{2024ApJ...961L..39S}. If they are indeed AGN, there is little evidence on how well the scaling relations hold for these sources.

\begin{figure*}[t]
\centering
\includegraphics[width=2.1\columnwidth, clip]{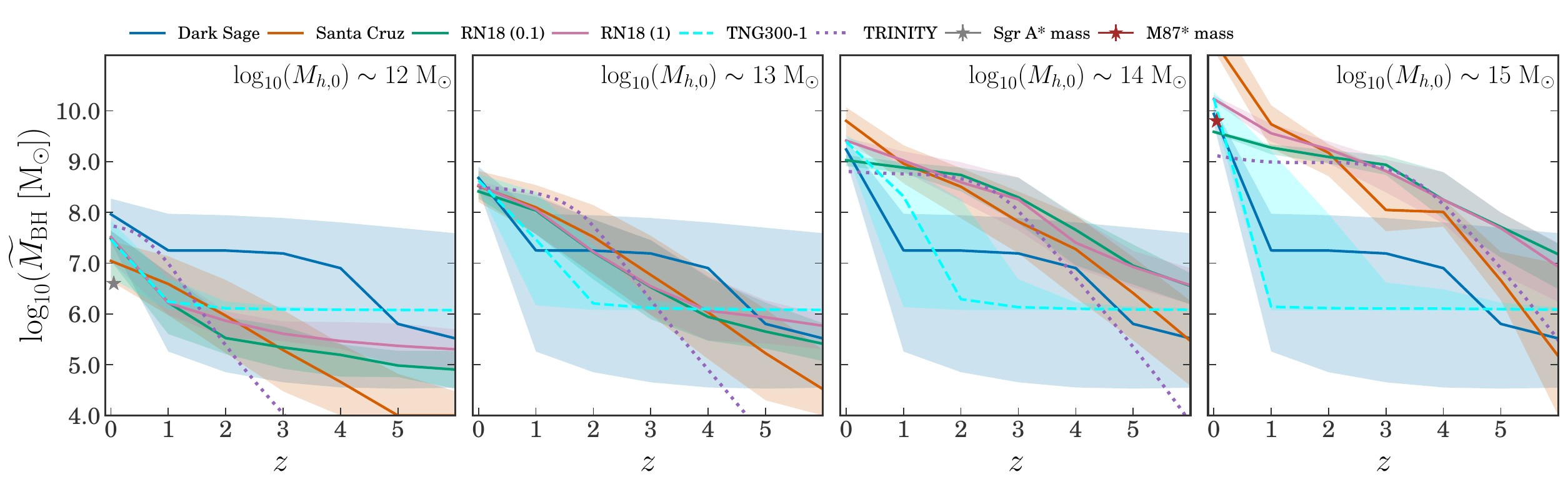}
\caption{Median black hole mass growth history as a function of redshift. Each panel shows a fixed halo mass from $M_h \sim 10^{12-15}\, \mathrm{M}_{\odot}$ taken at $z=0$ for \textit{semi-analytic} models {\sc Dark Sage} (solid dark blue line), the {\sc Santa Cruz} SAM (solid brown line), and \citetalias{Ricarte+2018a} (solid green line: $p_{merge}$=0.1, solid violet line: $p_{merge}$=1.0). The shaded regions enclose 68 percent of the data. We compare those models to {\sc TNG300-1} (dashed cyan line) and {\sc TRINITY} (dotted purple line). 
The grey star represents Sgr A*, which is hosted in a halo with $M_h \sim 10^{12}\, \mathrm{M}{\odot}$, while the dark red star represents M87*, hosted in a halo with $M_h \sim 10^{15}\, \mathrm{M}{\odot}$. For a more complete representation of this figure, note the black hole seeding mass denoted in Table \ref{Table: sims_details}. In most cases, all models converge within an order of magnitude in the final black hole mass at $z=0$. However, in the highest halo mass bin, the difference extends to two orders of magnitude.}\label{fig:meanSMBH_total}

\end{figure*}

We now explore the assembly of SMBHs as a function of the final host dark matter halo mass over redshift, from which we gain insights into SMBH-galaxy co-evolution. We plot the median black hole mass of the largest progenitor galaxy for a root $M_h \sim 10^{12-15}\, \mathrm{M}_{\odot}$ halo at $z=0$ in Figure \ref{fig:meanSMBH_total}. Note that, at $z=0$, our Milky Way galaxy's dark matter halo is $M_h \sim 10^{12}\, \mathrm{M}_{\odot}$ \citep{1987ApJ...320..493L, 1992MNRAS.255..105K, 2002ApJ...573..597K, 2013ApJ...768..140B, 2022ApJ...925....1S} and M87's dark matter halo mass from the Virgo cluster is $M_h \sim 10^{15}\, \mathrm{M}_{\odot}$ \citep{1978ApJ...219..413M, 2001A&A...375..770F, 2017ApJ...850..207S, 2020A&A...635A.135K}. The manner in which black hole masses assemble in these halos is determined by the timing difference in how the halos themselves assemble. Although models generally converge within an order of magnitude in their predicted final SMBH masses across all halo mass bins except the highest, their respective growth histories differ significantly.

Firstly, we note that in {\sc Dark Sage}, black hole growth over the range $0 \leq z \leq 6$ is slower but rapidly increases with halo mass, similar to the {\sc Santa Cruz} SAM and \citetalias{Ricarte+2018a}. In contrast, black holes in {\sc TNG300-1} stall at $M_{\rm BH} \sim 10^{6}\, \mathrm{M}_{\odot}$ until $z<2$. After this point, they gain between one to four orders of magnitude in mass, depending on their host halo mass. In most cases, {\sc Dark Sage}, the {\sc Santa Cruz} SAM, and \citetalias{Ricarte+2018a} show a sharp increase in the median black hole mass growth at $z < 1$. In contrast, {\sc TRINITY} displays a flattening curve in this redshift range, indicating a slowdown in growth. SMBHs grow rapidly, with the growth rate depending on the halo mass they reside in. For example, in more massive halos, the SMBH growth phase ends above $z>3$.

In Milky Way-like halos, most models produce SMBHs that are slightly more massive than Sgr A*. This is likely because Sgr A* is undermassive compared to black holes in the global $M_{\rm BH}-M_*$ or $M_{\rm BH}-M_{\rm bulge}$ relation. {\sc Dark Sage}, has the highest median curve due to its super Eddington accretion at high-z and its orders of magnitude drop to sub-Eddington at lower-z (see figure \ref{fig:Eddington_ratio}). In the case of {\sc TNG300-1}, its initial black hole seeding mass is already close to the Sgr A* mass (see Table \ref{Table: sims_details}), so we see less growth across time. In \citetalias{Ricarte+2018a}, we see a flattened median line that slightly increases with redshift until it converges at $z=1$. In the {\sc Santa Cruz} SAM, black holes gain three orders of magnitude of steady growth from $z=5$ up to $z=0$.

For halos with $M_h \sim 10^{14}\, \mathrm{M}_{\odot}$, the median black hole mass in the {\sc Dark Sage} model increases by $\approx 4.5\ \mathrm{dex}$ from $z=6$ to $z=0$ similar to results from the {\sc Santa Cruz} SAM. Both \citetalias{Ricarte+2018a} models gain about $\approx 2.5\ \mathrm{dex}$ within this halo mass and redshift range. Below $z=2$, {\sc Dark Sage}, \citetalias{Ricarte+2018a}, and the {\sc Santa Cruz} SAM start to diverge for $M_h \sim 10^{14-15}\, \mathrm{M}_{\odot}$, indicating a higher rate of BH--BH mergers in this regime. This shows more clearly for $M_h \sim 10^{15}\, \mathrm{M}_{\odot}$, where the \citetalias{Ricarte+2018a} $p_{merge}$=1 model clearly produces higher mass black holes compared to $p_{merge}$=0.1. In galaxy clusters, black hole mass growth flattens at low redshift for \citetalias{Ricarte+2018a} but steepens in {\sc Dark Sage}, the {\sc Santa Cruz} SAM, and {\sc TNG300-1}. In \citetalias{Ricarte+2018a}, this flattening follows strict adherence to the M-sigma relation, while other models steepen due to frequent black hole mergers and the absence of dynamics delaying mergers after galaxy mergers. The contribution of BH--BH mergers to the total black hole mass growth in {\sc Dark Sage} is clearly shown in Figure \ref{fig:medianSMBH_channels}. Within the highest halo mass bin, the {\sc Santa Cruz} SAM predicts a median black hole mass that is $\sim 1.6\ \mathrm{dex}$ higher than M87* mass, while {\sc TRINITY} predicts a black hole mass an order of magnitude lower. At all halo masses, {\sc Dark Sage} predicts the largest amount of scatter in black hole mass at fixed halo mass, around $\sim 3\ \mathrm{dex}$ above $z=1$.

\subsection{Eddington Ratio distributions }\label{Eddratios_models}

\begin{figure*}[t]
\centering
\includegraphics[width=2.1\columnwidth, clip]{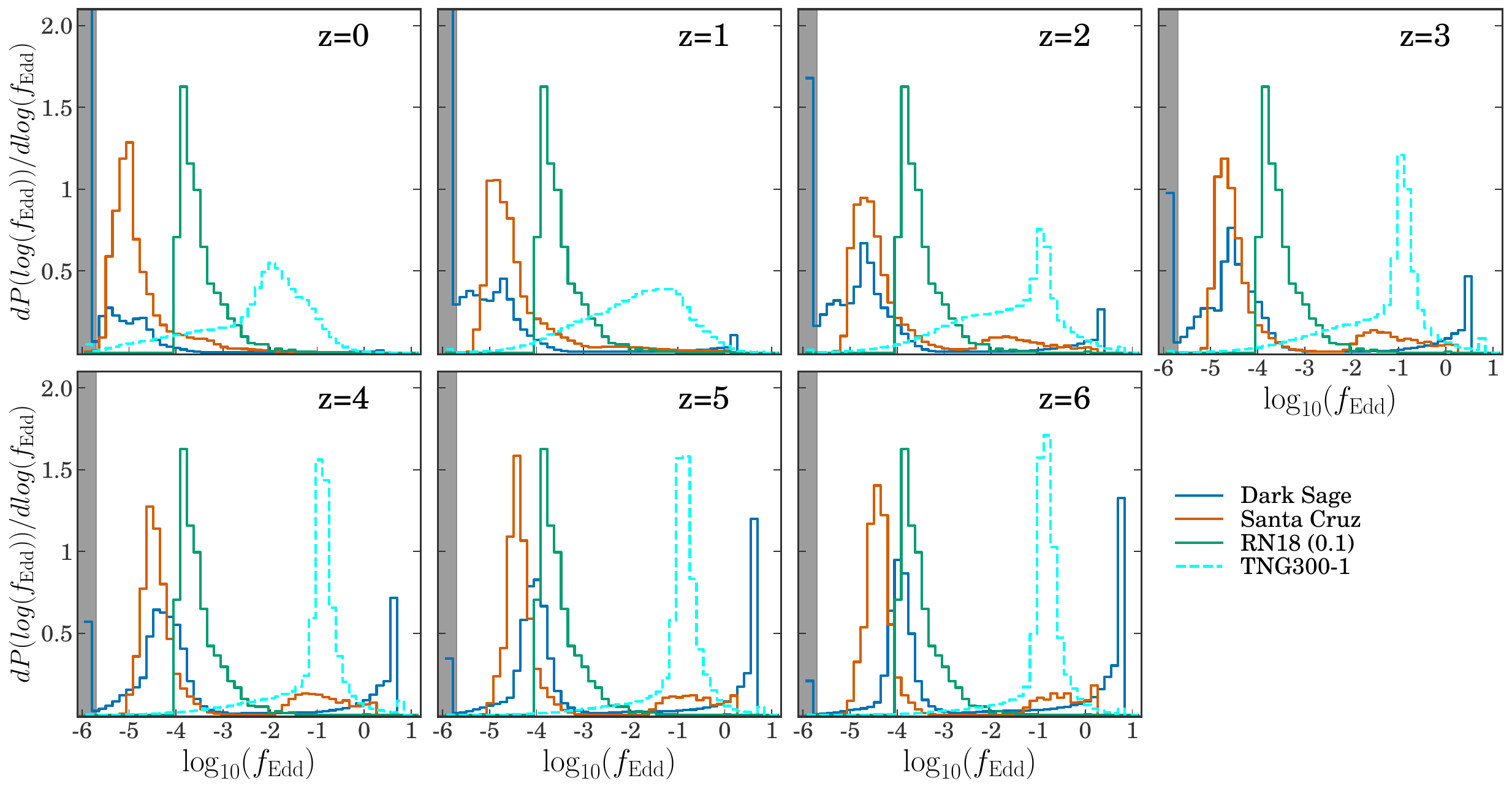}
\caption{Histograms of the Eddington ratio distributions from $z=0$ to $z=6$ for {\sc Dark Sage} (solid dark blue line), the {\sc Santa Cruz} SAM (solid brown line), and \citetalias{Ricarte+2018a} (solid green line: $p_{merge}$=0.1. Note that the \citetalias{Ricarte+2018a} model with $p_{merge}=1.0$ has Eddington ratio distributions similar to the model with $p_{merge}=0.1$. Therefore, we have chosen to present only one of these models. We compare those models to {\sc TNG300-1} (dashed cyan line). Not surprisingly, we see high variation in the Eddington ratio distributions between models at all redshifts.}\label{fig:Eddington_ratio}

\end{figure*}

Figure \ref{fig:Eddington_ratio} highlights a diverse set of $\log_{10}(f_\mathrm{Edd})$ distributions amongst models. We show the Eddington ratio distributions for all galaxies as a function of cosmic epoch. We find strikingly different distributions amongst models and across redshifts. At $z=0$, {\sc Dark Sage} shows a bimodal peak between $\log_{10}(f_\mathrm{Edd}) \sim -5.5$ and $-4.5$. This distribution overlaps with the {\sc Santa Cruz} SAM, which shows a skewed distribution peaking at $\log_{10}(f_\mathrm{Edd}) \sim -5$. \citetalias{Ricarte+2018a} shows a peak about one and half orders of magnitude higher than the one {\sc Santa Cruz} SAM at $\log_{10}(f_\mathrm{Edd}) \sim -3.5$. {\sc TNG300-1} shows a broad distribution that peaks around $\log_{10}(f_\mathrm{Edd}) \sim -2$. As redshift increases, the {\sc TNG300-1} distribution narrows and shifts an order of magnitude higher in $\log_{10}(f_\mathrm{Edd})$ by $z=6$. The {\sc Santa Cruz} SAM distributions shift approximately $0.5\ \mathrm{dex}$ towards higher $\log_{10}(f_\mathrm{Edd})$ by $z=6$. By construction, the \citetalias{Ricarte+2018a} distributions maintain a peak at $\log_{10}(f_\mathrm{Edd}) \sim -4$ across all redshifts, while a new peak emerges around $\log_{10}(f_\mathrm{Edd}) \sim 0$ at $z=2$ and becomes more pronounced by $z=6$. Lastly, the {\sc Dark Sage} bimodal distribution shifts about $1\ \mathrm{dex}$ higher in $\log_{10}(f_\mathrm{Edd})$ by $z=6$, with an emergent pronounced peak at $\log_{10}(f_\mathrm{Edd}) \sim 1$ at this redshift. 

For the {\sc Santa Cruz} and {\sc Dark Sage} SAMs, the bimodal and trimodal distributions of the black hole growth appear to depend on halo mass. In the case of {\sc Dark Sage} (see Appendix \ref{BHgrowth_histdist}), at $z<3$, when breaking down the Eddington ratio distributions into halo mass bins, the distribution more tilted toward sub-Eddington values (the leftmost one) corresponds to black holes in high-mass halos around $M_h \sim 10^{14} \, \mathrm{M}_{\odot}$, whereas the other sub-Eddington distribution at $\log(f_{\rm Edd}) \approx -4$ arises from black holes in low-mass halos around $M_h \sim 10^{12} \, \mathrm{M}_{\odot}$. Similarly, in the {\sc Santa Cruz} model, high-mass halos $M_h > 10^{13} \, \mathrm{M}_{\odot}$ display higher Eddington ratio distributions compared to halos around $M_h \sim 10^{12} \, \mathrm{M}_{\odot}$. The observed bimodality in \citetalias{Ricarte+2018a}, the {\sc Santa Cruz} SAM, and {\sc Dark Sage} is due to a rapid accretion mode triggered by galaxy-galaxy mergers.

\subsection{The integrated cosmic SMBH mass density} \label{cosmicSMBHfunc_z0_6}

\begin{figure*}[t]
\centering
\includegraphics[width=2.0\columnwidth, clip]{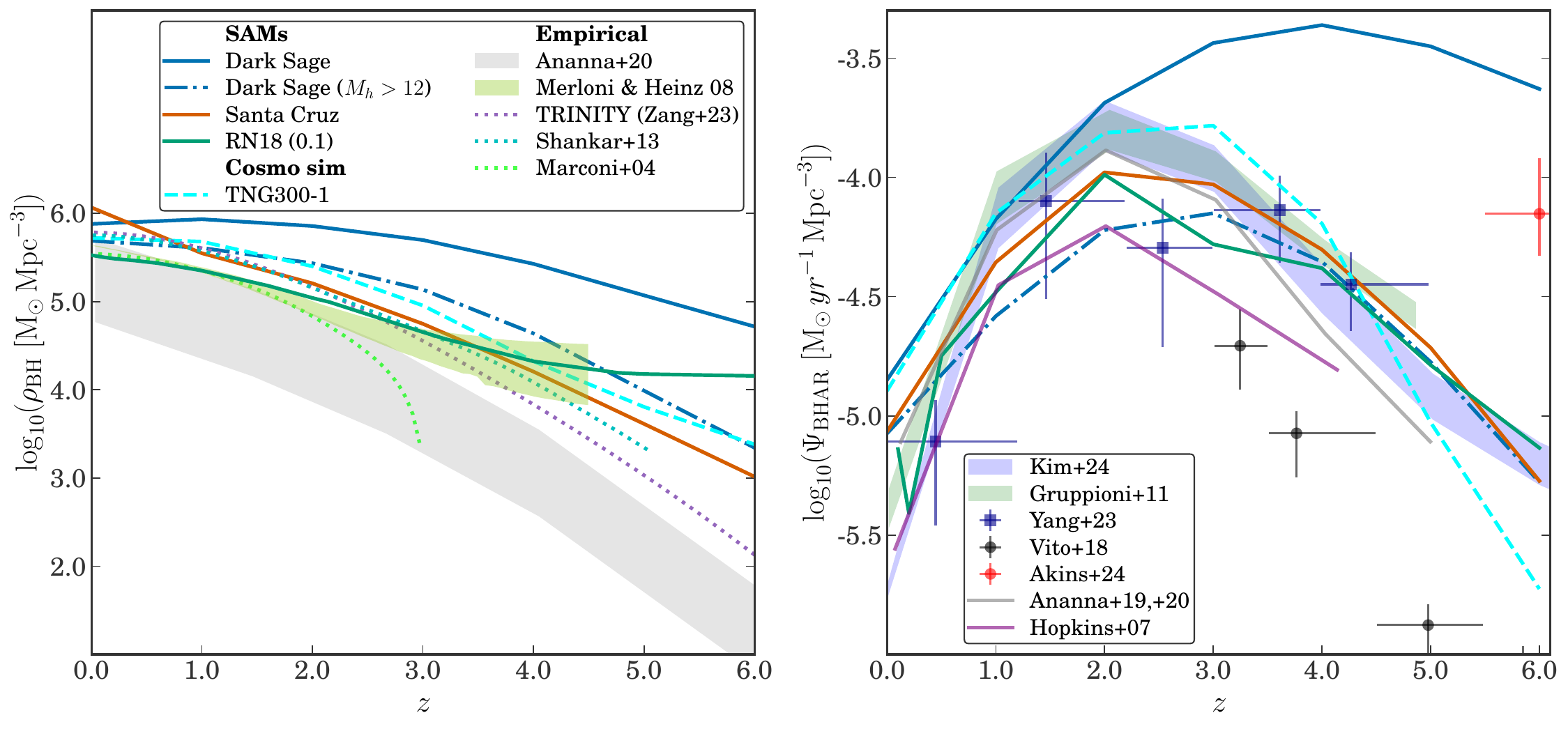}
\caption{The cosmic SMBH mass and accretion density function for physics-based models and observations. The left panel includes \citet{2023MNRAS.518.2123Z} ({\sc TRINITY}) (dotted purple line), \citet{2020ApJ...903...85A} (shaded yellowgreen region)
\citet{2013MNRAS.428..421S} (dotted cyan line), \citet{2008MNRAS.388.1011M} (shaded aquamarine region) and \citet{2004MNRAS.351..169M} (dotted lime line) \textit{empirical} models compared to \textit{semi-analytic} models {\sc Dark Sage} (solid dark blue line), {\sc Dark Sage} black holes in $M_h > 10^{12}\, \mathrm{M}_{\odot}$ (dashed dotted dark blue line), the {\sc Santa Cruz} SAM (solid brown line), and \citetalias{Ricarte+2018a} (solid green line: $p_{merge}$=0.1. The dashed cyan line shows {\sc TNG300-1}. The right panel shows same physics-based models, while including observations from \citet{Hopkins+2007}, \citet{2011MNRAS.416...70G}, \citet{2018MNRAS.473.2378V}, 
\citet{2019ApJ...871..240A, 2020ApJ...903...85A}, \citet{2023ApJ...950L...5Y}, \citet{2024MNRAS.527.5525K}, and \citet{2024arXiv240610341A}. For the cosmic SMBH mass density, models with different $\log_{10}(\rho_{BH})$ histories converge at $z=0$. Most models agree with observations for the black hole accretion rate density, except for {\sc Santa Cruz} at $z=0$ and {\sc Dark Sage} at $z>2$. In both panels, {\sc Dark Sage} recovers the observed normalization by only including black holes in $M_h > 10^{12}\, \mathrm{M}_{\odot}$.
}\label{fig:cosmicSMBHfunc}

\end{figure*}

The integrated mass density of SMBHs at a particular redshift is computed by summing over mass growth via accretion processes that have occurred over preceding cosmic times. More specifically, for a given bolometric AGN luminosity function $\Phi (L, z)$, the cumulative mass density can be derived by integrating over a range of bolometric luminosities and redshifts, taking into account parameters such as the radiative efficiency, kinetic efficiency, and time elapsed in the Universe. 

Figure \ref{fig:cosmicSMBHfunc} shows the cosmic SMBH mass and accretion density as a function of time for various physics-based and \textit{empirical} models. The left panel includes \citet{2013MNRAS.428..421S}, \citet{2008MNRAS.388.1011M}, and \citet{2004MNRAS.351..169M}, which solves the continuity equation. The \textit{empirical} cosmic SMBH mass density is obtained using the intrinsic X-ray luminosity function in the rest-frame 2-10 keV band with a bolometric correction following equation \ref{eqn:rho}: 

\begin{equation}\label{eqn:rho}
\begin{split}
\rho_{\rm BH} (z) &= \int^{z_{S}}_{z} \int^{L_{max}}_{L_{min}}  
\frac{(1 - \eta - \eta_{\rm kin} )~L}{\eta~c^2} \Phi(L, z) d\log L  \\
&\quad \times \frac{dt}{dz} dz .
\end{split}
\end{equation}

\noindent
where $L$ is the bolometric luminosity, $z_S$ is the redshift upper limit at which significant black hole growth begins, $\eta$ is the radiative efficiency as mentioned in the above paragraphs, and $c$ is the speed of light. \citet{2020ApJ...903...85A} integrate over bolometric luminosities 10$^{42}-10^{50}\,$ erg/s in equation \ref{eqn:rho} to obtain the cosmic SMBH mass density results shown in the left panel of figure \ref{fig:cosmicSMBHfunc}. One can also change equation \ref{eqn:rho} to obtain the cosmic SMBH mass density with a dependence on the black hole accretion history, in the form of $\rho_{\rm BH}(z)$ using two parameters: the local black hole mass density $\rho_{\rm BH,0}$ and the (average) radiative efficiency $\epsilon_{\rm rad}$. This leads to an intriguing implication: without assuming anything about the distribution of the Eddington ratios in the AGN population, constraints can be obtained on the (mass-weighted) average radiative efficiency. The Soltan argument \citep{2008MNRAS.388.1011M} establishes a robust {\it lower} limit for $\epsilon_{\rm rad}$. Specifically, it is enough to identify the value of the radiative efficiency that renders the integral $\int_0^{z}\frac{\Psi_{\rm BHAR}(z')}{\rho_{\rm BH,0}}\frac{dt}{dz'}dz'$ greater than unity. \citet{2008MNRAS.388.1011M} obtain a lower limit of 0.065 with the chosen local SMBH mass function and AGN bolometric luminosity function.\footnote{In our simulations, we calculate the cosmic SMBH mass density by binning black holes by redshift, summing their mass and dividing by the volume of the corresponding simulation box. Empirical curves in figure \ref{fig:cosmicSMBHfunc} are retrieved from the literature.}

Current estimates for the \textit{empirical} local black hole mass density ($\rho_{\rm BH}$) converge at $z=0$ to within about $1\ \mathrm{dex}$ with estimates between $5-6\times 10^{10}\, \mathrm{M}_{\odot}$ Mpc$^{-3}$ \citep{2004MNRAS.351..169M, 2008MNRAS.388.1011M, 2013MNRAS.428..421S, 2023MNRAS.518.2123Z}. The cosmic SMBH density function obtained from \citet{2020ApJ...903...85A} is derived using a range of radiative efficiencies between 0.1-0.3. {\sc TRINITY} includes the contribution of wandering black holes in their results, which accounts for about 15\% of the total black hole mass density at $z=0$ \citep{2023MNRAS.518.2123Z}. Their results are stated to be consistent with \citet{2003ApJ...582..559V} and \citet{Ricarte+2021}. Below $z \sim 2$, discrepancies in mass density comparisons across various studies are likely to arise from differences in models with varying efficiencies in AGN energy, occupation fraction of unobscured AGNs, and/or the missing contribution of black holes in low-mass halos. Above $z \sim 2$, the consistent difference between \citet{2004MNRAS.351..169M} and other models becomes more noticeable. This discrepancy is possibly a result of their approach to modeling AGN evolution, assuming that SMBH growth took place before $z \sim 3$. However, these initial assumptions did not consider the mass growth history of SMBHs at higher redshifts, which explains the difference in SMBH mass distribution between \citet{2004MNRAS.351..169M} and other models like {\sc TRINITY}, where SMBHs start growing as early as redshift 15. The high-redshift plateau seen in the \citetalias{Ricarte+2018a} models occurs because the mass density at these early epochs is dominated by seeds. Mass locked up in seeds is invisible to mass density estimates inferred from radiation and is thus unaccounted for in \textit{empirical} models. Physics-based models either have light seeds or do not resolve the halos at epochs and masses where the seeds dominating the \citetalias{Ricarte+2018a} mass density reside. Our results are consistent with \citet{Natarajan+2021}, where even models which roughly reproduce the observed local SMBH mass density at $z \sim 0$ begin to diverge significantly by $z\sim 3$, differing by several orders of magnitude.

At $z=0$, {\sc Dark Sage}, the {\sc Santa Cruz} SAM and {\sc TNG300-1} closely agree to about $\log_{10}(\rho_{BH}) \approx 6$. {\sc Dark Sage}\footnote{In {\sc Dark Sage}, there is a $0.05\ \mathrm{dex}$ difference in the $z=0$ cosmic SMBH density if we remove our halo mass cut.} lies $0.2\ \mathrm{dex}$ lower than the {\sc Santa Cruz} SAM and {\sc TNG300-1}. \citet{2013MNRAS.428..421S} and \citet{2023MNRAS.518.2123Z} agree with $\log_{10}(\rho_{BH}) \approx 5.7$, whereas \citetalias{Ricarte+2018a} models predict $0.5\ \mathrm{dex}$ lower black hole mass density than the {\sc Santa Cruz} SAM and {\sc TNG300-1}, in agreement with \citet{2004MNRAS.351..169M, 2008MNRAS.388.1011M}, and \citet{2020ApJ...903...85A}. We note that the {\sc Santa Cruz} SAM results are consistent with Gabrielpillai et al. 2024 (in prep). By $z>3$, \citet{2020ApJ...903...85A} has the lowest number density of all. In this redshift range, the cosmic mass density normalization from  \citet{2013MNRAS.428..421S}, \citet{2020ApJ...903...85A}, and {\sc TRINITY} is slightly off by about $0.1\ \mathrm{dex}$

The right panel presents the cosmic SMBH accretion histories. \citet{2024MNRAS.527.5525K} used a model to interpret mid-infrared galaxy source counts across a wide flux range, using extremely faint galaxies \citep{2022MNRAS.517..853L, 2023MNRAS.522.1138L, 2023MNRAS.523.5187W}. By applying a backward evolution model of parametrized mid-infrared luminosity functions, following \citet{2011MNRAS.416...70G}, they fit this model to JWST/MIRI source counts alongside data from the Infrared Space Observatory (ISO), AKARI, and Spitzer \citep{1997MNRAS.289..471O, 2000MNRAS.316..768S, 2007A&A...475..801R, 2010A&A...514A...8P, 2014MNRAS.444..846P, 2017MNRAS.472.4259D}, providing constraints to the cosmic star formation history and black hole accretion history. \citet{2011MNRAS.416...70G} models the local galaxy luminosity function for mid-infrared bands for five different types of galaxies. Then, they approximated the bolometric luminosity to the infrared luminosity integrated between 1 and 1000$\mu$m. Because their sample have different galaxies, the AGN contribution to the luminosity is weighted. \citet{2023ApJ...950L...5Y} points are derived from MIRI-selected AGN. \citet{2018MNRAS.473.2378V} and \citet{Hopkins+2007} come from X-ray detections.

From $z=0$ to $z=1$, observations agree within $\approx0.4\ \mathrm{dex}$ from each other. By $z>1$, observations start to diverge, where \citet{2024MNRAS.527.5525K}, \citet{2020ApJ...903...85A}, and \citet{2011MNRAS.416...70G} show values that are higher than \citet{2018MNRAS.473.2378V} and
\citet{Hopkins+2007}. Such differences may arise from variations in methodology, wavelength coverage, and the datasets used. \citetalias{Ricarte+2018a} is the only model that spans the range of the observed black hole accretion density from $z=0-6$. While {\sc TNG300-1} is slightly higher than most observations at $z=0$, it matches \citet{2024MNRAS.527.5525K} and \citet{2011MNRAS.416...70G} between $z\approx3-5$. The {\sc Santa Cruz} SAM matches observations from \citet{2023ApJ...950L...5Y} black hole accretion density at $z>1$. By $z=0$, the model lies about 0.2 dex below {\sc Dark Sage} and {\sc TNG300-1}, whose models overpredict the number density of accreting black holes compared to observations. Between $z=1-2$, {\sc Dark Sage} agrees with \citet{2024MNRAS.527.5525K} and \citet{2011MNRAS.416...70G}. However, above $z=2$, a large number of black holes continue to accrete highly up until $z=3$. The black hole accretion density for ``little red dots'' lies orders of magnitude above all observations \citet{2024arXiv240610341A}, closer to {\sc Dark Sage} prediction. While most physics-based models derived accretion density peak at $z\approx$ like in the observations, {\sc Dark Sage} shows an earlier peak around $z=4$ due to its high number of low mass halos and black holes growing rapidly at earlier times.

The differences in results between observations and {\sc Dark Sage} are primarily due to sample selection. As illustrated in the dashed dotted dark blue line, where we take only black holes with $M_h > 10^{12}\, \mathrm{M}_{\odot}$, we find a closer match to the observations in the cosmic SMBH mass and accretion density, consistent with previous results from \citet{2023A&A...669A.127W} who took a sample of massive halos from the Horizon-AGN \citep{2016MNRAS.460.2979V} and Astrid \citep{2022MNRAS.513..670N} to match the observed black hole accretion rate density. General discrepancies between physics-based models likely arise from how each of the models account for co-evolutionary processes between black holes and their host galaxies, such as the black hole mass-stellar/bulge mass relation at $z=0$ and/or variation in the black hole merger rate and its contribution to black hole growth. Variances in the strength or timing of feedback processes in physics-based models also appear to contribute to differences in the predictions.

\section{Focusing on the SMBH Assembly in {\sc Dark Sage}}\label{DS_BHchannels}

\begin{figure*}[t]
\centering
\includegraphics[width=2.1\columnwidth, clip]{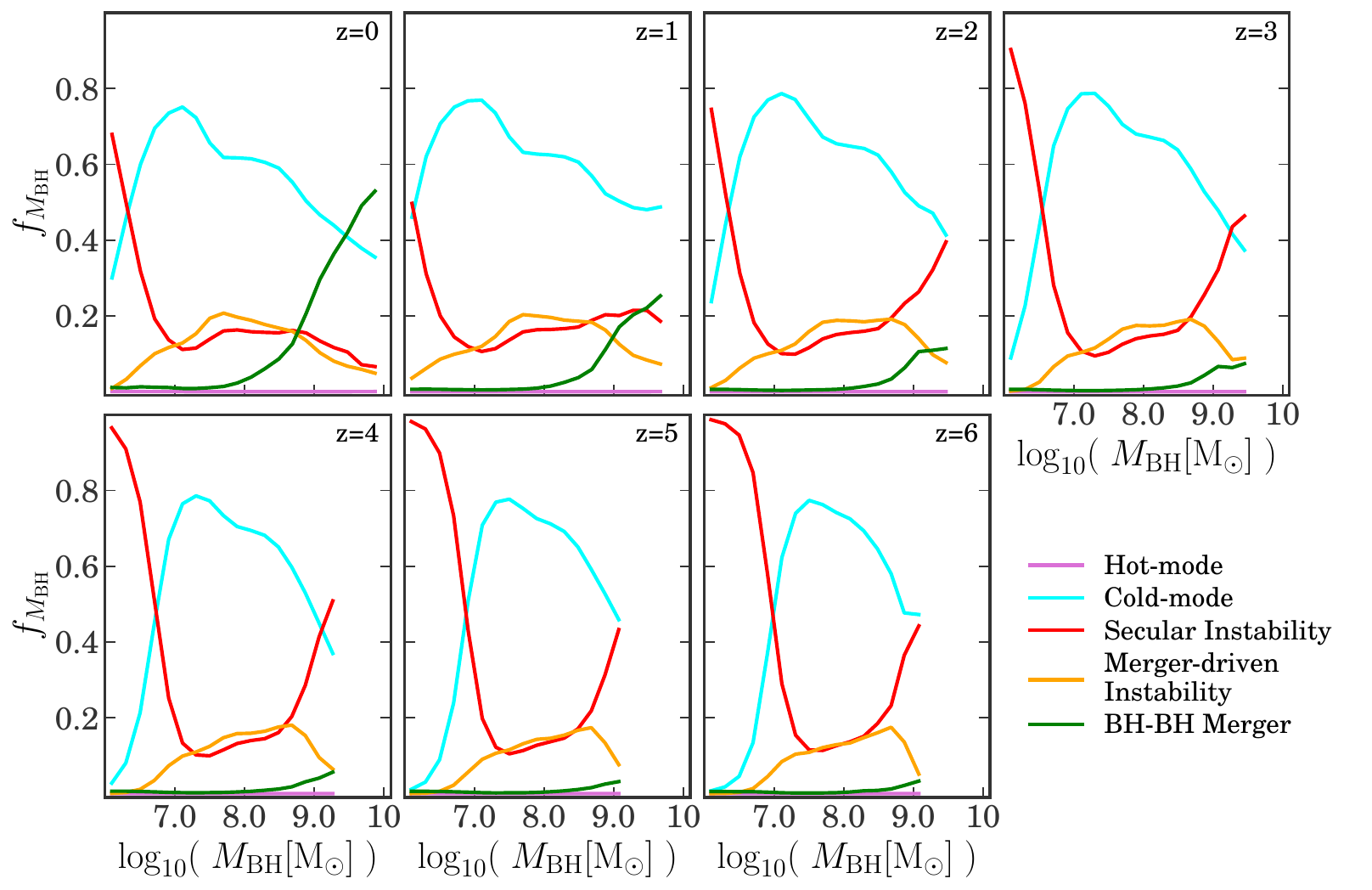}
\caption{The fraction of black hole mass contributed by the five, different  distinct black hole growth channels in {\sc Dark Sage}: \textit{hot--mode} (solid violet line), \textit{cold--mode} (solid cyan line), \textit{secular instability mode} (solid red line), \textit{merger-driven instability mode} (solid yellow line), and BH--BH merger mode (solid green line). These are binned by black hole mass for $z=0$ to $z=6$. At all redshifts for the fiducial model, black holes that are $\sim 10^{6}\, \mathrm{M}_{\odot}$ growth through \textit{secular instabilities}. At the high-mass end, the \textit{cold--mode}, BH--BH mergers, and \textit{secular instabilities} contribute the most to the final black hole mass budget across redshift.}\label{fig:fracSMBH_channels}

\end{figure*}

With {\sc Dark Sage}, we investigate how different growth channels contribute to final SMBH masses across redshift. In Figure \ref{fig:fracSMBH_channels}, we present the fraction of black hole mass divided by the contributions from five distinct growth channels in {\sc Dark Sage}: \textit{hot mode}, \textit{cold mode}, \textit{secular instability mode}, \textit{merger-driven instability mode}, and BH--BH merger mode. Regardless of redshift, black holes with masses around $\sim 10^{6}\, \mathrm{M}_{\odot}$ primarily grow through \textit{secular instabilities}. For $M_{\rm BH} >10^{6}\, \mathrm{M}_{\odot}$, {\sc Dark Sage} reveals a diversity of growth paths. At $z=0$, most of the growth in $M_{\rm BH} >10^{9.5}\, \mathrm{M}_{\odot}$ is driven by BH--BH mergers and \textit{cold--mode} accretion. The contribution from BH--BH mergers decreases with lower black hole mass and increasing redshift, becoming nearly insignificant by $z=3$. The large contribution of BH--BH mergers to the final black hole mass census at low redshift agrees with predictions from other \textit{physical} models, both \textit{numerical hydrodynamics} and \textit{semi-analytic} models \citep{10.1093/mnras/sty1733, Ricarte+2018a, 2020ApJ...895...95P}.

In all cases, the \textit{hot mode} growth channel contributes the least to the final mass. Across redshifts, the \textit{cold--mode} channel consistently adds to the black hole mass budget, contributing up to 70\% of the final mass at $z=6$ and as low as 18\% at $z=4$. Interestingly, just before cosmic noon at $z=4$, black holes with $M_{\rm BH} >10^{9}\, \mathrm{M}_{\odot}$ shift from being predominantly \textit{cold--mode} dominated to having their primary mass contribution from \textit{secular instabilities}. Recall that \textit{cold--mode} and accretion due to \textit{merger-driven instabilities} occurs only when there is a galaxy merger and/or there is unstable cold gas mass with a low Q Toomre parameter and low angular momentum that is feeding the black hole episodically. Even in the absence of galaxy mergers, {\sc Dark Sage} SMBHs may still grow through secular accretion of unstable cold gas. 

\begin{figure*}[t]
\centering
\includegraphics[width=2.1\columnwidth, clip]{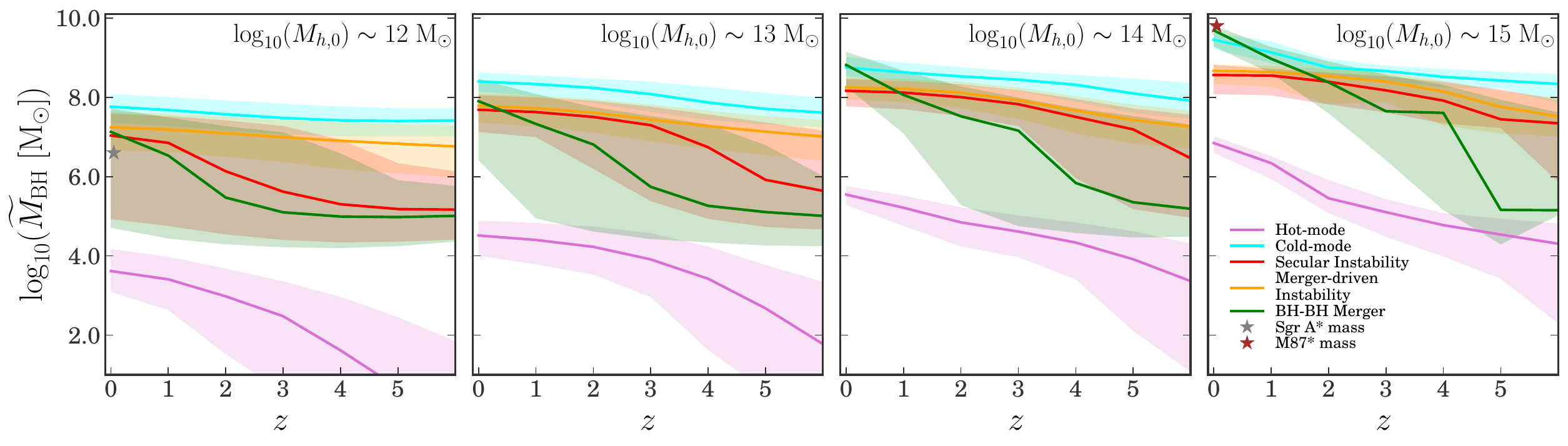}
\caption{Median black hole mass growth history separated by the five black hole growth channels described in figure \ref{fig:fracSMBH_channels}. The shaded regions enclose 68 percent of the data. The grey star represents the black hole mass of Sgr A* and the dark red star represents the black hole mass of M87*, each at their respective halo masses. Generally, both \textit{cold--mode} and \textit{instabilities} contribute the most mass to black hole mass across redshift.}\label{fig:medianSMBH_channels}

\end{figure*}

Figure \ref{fig:medianSMBH_channels} displays the median black hole mass as a function of halo mass and redshift for each of the black hole growth channels. For most SMBHs, we find that the \textit{cold--mode}, \textit{merger-driven} and \textit{secular instabilities} contribute most significantly to mass growth. Bondi accretion from the halo, \textit{hot--mode}, provides the least mass across redshift. Recall that the \textit{cold--mode} and \textit{merger-driven instabilities} occurs only when there is a galaxy merger. After a galaxy merger, a fraction of the total cold gas mass from the secondary galaxy, $f_{\rm BH}$, is added directly into the primary black hole. The remaining cold gas is added into the primary's cold gas annuli. {\sc Dark Sage} enters a starburst phase. Then, it calculates the Q Toomre parameter. Any unstable cold gas that is not resolved by mass transfer from neighboring annuli or by a starburst is transferred to the lowest angular momentum annuli. Thereafter, such unstable gas is added into the primary black hole. Even in the absence of galaxy mergers, black holes may still grow through \textit{secular instabilities} of unstable cold gas. Figure \ref{fig:medianSMBH_channels} shows that \textit{merger-driven} and \textit{secular instabilities} contribute almost equally for $M_h \approx 10^{15}\, \mathrm{M}_{\odot}$. Their contribution differs is more significant for the low-mass halos.

For $z\sim 0$ and $M_h \approx 10^{14}\, \mathrm{M}_{\odot}$, black hole mergers overtake all other growth channels. This agrees with other \textit{physical} models, both \textit{hydrodynamic} and \textit{semi-analytic} \citep{10.1093/mnras/sty1733, Ricarte+2018a, 2020ApJ...895...95P}.

\begin{figure*}[t]
\centering
\includegraphics[width=2.1\columnwidth, clip]{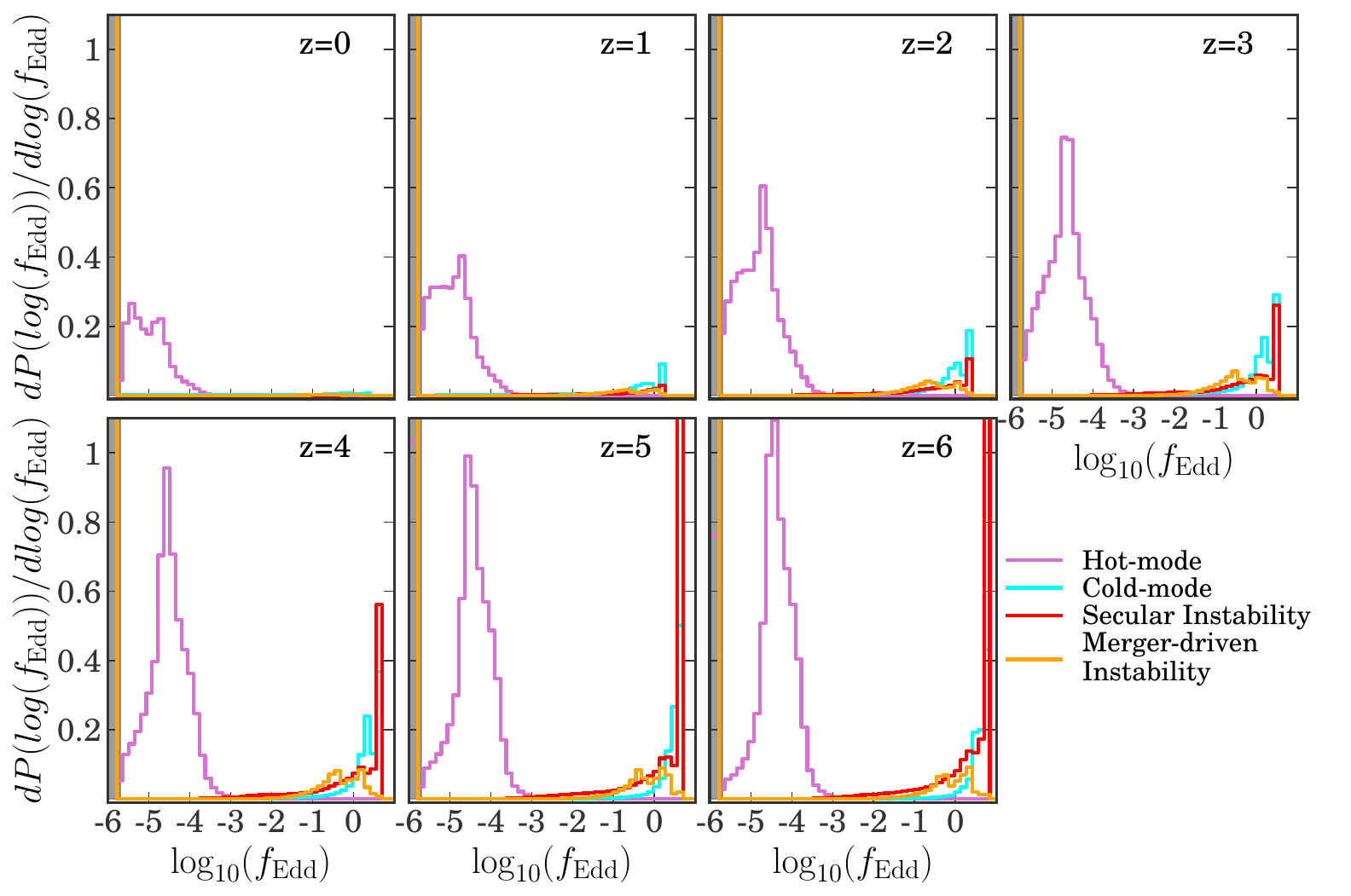}
\caption{Histograms of the Eddington ratio distributions from $z=0$ to $z=6$ separated by the same distinct black hole growth channels presented in figure \ref{fig:fracSMBH_channels}, except BH--BH mergers. The grey shaded region denote valus of $\log_{10}(f_\mathrm{Edd}) \leq -6$. We learn that \textit{secular instability} accretion dominate at high redshift.}\label{fig:EddratioSMBH_channels}

\end{figure*}

Keeping in mind the pronounced super-Eddington peak at high redshift illustrated in Figure \ref{fig:Eddington_ratio}, we now turn to dissecting its source. Figure \ref{fig:EddratioSMBH_channels} shows the instantaneous Eddington ratio distributions for the black hole growth channels in {\sc Dark Sage}. At all redshifts, sub-Eddington accretion comes from the \textit{hot--mode} channel, while super-Eddington accretion results primarily from \textit{secular instabilities}. Similar to Figure \ref{fig:Eddington_ratio}, a super-Eddington peak appears starting at $z=1$. Between $z=1$ and $z=3$, \textit{cold mode} accretion dominates over \textit{secular instabilities}. Beyond $z=3$, the peak is predominantly driven by \textit{secular instabilities}. Throughout, \textit{merger-driven instability} accretion remains a minor contributor.

\section{Discussion and Conclusions}
\label{conclusion}

We study the mass evolution of SMBHs using a variety of \textit{semi-analytic} models {\sc Dark Sage}, the {\sc Santa Cruz} SAM, and \citetalias{Ricarte+2018a}. We focus on the local and evolving BHMF, bolometric AGN luminosity, median black hole mass growth histories, Eddington ratio distributions, cosmic SMBH mass and accretion rate histories. We compare model predictions with a set of inferred BHMFs from scaling relations and luminosity from the AGN. We include the most up-to-date inferred JWST results for the BHMF, the bolometric AGN luminosity function, and the cosmic black hole accretion rate history. Then, using {\sc Dark Sage}, we dissect the contribution of the black hole mass from the various growth channels implemented in the model. We present black hole mass fractions from each growth channel, median black hole mass growth histories, and Eddington ratio distributions across time.

Our key results can be summarized as follows:
\begin{itemize}
    \item The variations in the observationally inferred local BHMF are due to a) fitting to SMFs with varying numbers of massive galaxies and b) tunning to scaling relations with varying steepness in their fitting slopes. Physics-based models lie within the range of observational values. Their differences are also affected by the choice of SMFs and scaling relations when calibrating (Figure \ref{fig:BHMF_z0}).
    \item Despite the large variations in the local BHMF, in general, physics-based models tend to converge between $z=1-4$, with the exception of {\sc Dark Sage}. By $z=4-5$, \citet{2008MNRAS.388.1011M}, {\sc TRINITY}, \citetalias{Ricarte+2018a}, the {\sc Santa Cruz} SAM, and {\sc TNG300-1} roughly agree with the BHMF derived from broad-line AGN detected by JWST \citep{2024arXiv240906772T, 2024ApJ...963..129M, 2024ApJ...962..152H} (Figure \ref{fig:BHMF_z0_z6}).
    \item Above $z=2$, {\sc Dark Sage} has a distinct ``knee'' near $M_{\rm BH} \sim 10^{8}\, \mathrm{M}_{\odot}$ that remains fixed over time, becoming progressively narrower with increasing redshift and resulting in a sharper high-mass drop-off. This bump is also apparent in the bolometric AGN luminosity around $L_{\rm bol} \sim 10^{46}\, erg \ s^{-1}$. Between $z=5-6$, the bump in {\sc Dark Sage} matches the inferred AGN emission from ``little red dots''\citep{2024arXiv240610341A}. Roughly all physics-based models, with the exception of the {\sc Santa Cruz} SAM at $z=0$ and {\sc Dark Sage} above $z=2$, match the bolometric AGN luminosity across redshift from \citet{2020MNRAS.495.3252S} (Figure \ref{fig:QLF_z0z6}).
    \item Physics-based models converge with \textit{empirical} estimates around $10^{5-6}\, \mathrm{M}_{\odot}$ Mpc$^{-3}$ in the local black hole mass density despite large differences in their cosmic mass density histories. Most models match the observed black hole accretion rate density, except {\sc Santa Cruz} at $z=0$ and {\sc Dark Sage} at $z>2$. The cosmic black hole accretion density in {\sc Dark Sage} peaks around $z=4$, whereas all other physics-based models and observations peak near $z=2$ (Figure \ref{fig:cosmicSMBHfunc}).
    \item Figure \ref{fig:fracSMBH_channels} dissects the black hole mass from different growth channels within {\sc Dark Sage}. At all redshifts, $M_{\rm BH} \sim 10^{6}\, \mathrm{M}_{\odot}$ grow through \textit{secular instabilities}. For the most massive black holes at $z=0$, BH--BH mergers contribute 60\% of the final mass, while the \textit{cold--mode} channel adds 38\%. In most redshifts, the \textit{cold--mode} channel contributes as much as 70\% and as little as 18\%.
    \item {\sc Dark Sage} displays an increasing peak in super-Eddington accretion with increasing redshift, which is revealed to originate from \textit{secular instabilities} (Figure \ref{fig:EddratioSMBH_channels}). 
    
\end{itemize}

Most physics-based models generally calibrate to the local $M_{\rm BH}-\sigma$, $M_{\rm BH}-M_*$, or $M_{\rm BH}-M_{\rm bulge}$ relation as discussed in Appendix \ref{app:calibration}. Despite this, we reveal significant differences in the high and low mass end of their black hole mass function, black hole mass growth histories, Eddington ratio distributions across redshift, and cosmic SMBH mass and accretion density. Several causes that likely account for these differences across models compared here are enumerated below.

\begin{itemize}
    \item Models are usually only calibrated loosely to reproduce observed local relations for ``typical'' SMBHs.  However, there may be deviations at the lowest masses (from e.g., supernova feedback) and the highest masses (from e.g., SMBH mergers and the details of AGN feedback implementation).
    \item Even observational and empirical inferences of local relationships vary significantly. Calibration becomes sensitive to which relationships are assumed between black hole mass and the properties of their hosts, and how the scatter in these scaling relations is treated.
    \item Physics-based models (especially numerical hydrodynamics simulations) are usually only calibrated at $z=0$, and there are many paths to reach the same final relationships. 
\end{itemize}

We find that exactly which scaling relationship is used to infer the BHMF matters. \citet{2007ApJ...663...53T}, \citet{2007ApJ...660..267B}, and \citet{2007ApJ...662..808L} have argued that the black hole mass function varies depending on whether the bulge mass or the velocity dispersion is used to relate the black hole mass to its host galaxy. Meanwhile, others claim that there are no significant differences in the observed scatter between $M_{\rm BH}-\sigma$ and $M_{\rm BH}-M_{\rm bulge}$ in massive ellipticals \citep{2013ApJ...764..184M, 2019ApJ...876..155S}. The local BHMF is estimated based on the distribution of host galaxy properties, assuming that black hole masses follow a constant log-normal scatter around a single-power law scaling relationship. Recent observations, however, suggest that this assumption may not hold true, indicating that the correlations break down at the highest and lowest black hole masses \citep{2007ApJ...662..808L, 2009ApJ...698..198G, 2012AdAst2012E...7K}. This could introduce biases in the BHMF derived from these scaling relationships, which subsequently affects the BHMF estimated from the continuity equation. Therefore, it is crucial to obtain more direct $M_{\rm BH}$ estimates through dynamical and kinematic modeling for a diverse range of galaxy types, particularly at the extremes of the $M_{BH}$ spectrum. However, in the current state of the field, statistically abundant derivations of black hole masses through direct dynamical and kinematic modeling is observationally expensive.

When constructing models of SMBH evolution, we often trade-off between physical realism and the match to observational data in each physical aspect tackled in the modeling process. Although empirical models like {\sc TRINITY} can reproduce all observational data considered by construction, physics-based models such as the {\sc Santa Cruz} SAM and {\sc IllustrisTNG} unexpectedly match observed high-redshift properties and trends, including the BHMF, bolometric AGN luminosity, and cosmic SMBH mass and accretion rate history, despite not being explicitly calibrated to them. Our study highlights that much of the discrepancy in, for example the BHMF, between observations and physics-based models are attributed to differences in SMFs and scaling relations adopted in calibrating the models.

Throughout this study, the outlier in the BHMF, bolometric AGN luminosity function, and the cosmic SMBH mass and accretion rate density at high-redshift has been {\sc Dark Sage}. The dump near $M_{\rm BH} \approx 10^{8}\, \mathrm{M}_{\odot}$ found in the BHMF above $z=2$ is imprinted in the bolometric AGN luminosity function. Surprisingly, by $z=5-6$, it matches the inferred bolometric AGN luminosity from ``little red dots'' \citep{2024arXiv240610341A}. The distinct shape of {\sc Dark Sage} BHMF likely arises from a high black hole occupation fraction (seeding a central black hole in every galaxy) and allowing super-Eddington accretion, enabling many black holes to grow into SMBHs. From dissecting the black hole mass growth channel contributions in {\sc Dark Sage}, we know that the \textit{cold mode} growth commonly dominates the SMBH mass budget in the high-redshift Universe, a result consistent with \citet{2019MNRAS.485.2694A}. For $M_{\rm BH} \approx 10^{8}\, \mathrm{M}_{\odot}$, the mass acquired from the \textit{cold mode} channel is about 80\% of the total mass at $z=5-6$. The amount of \textit{cold mode} growth can be manually adjusted using the free parameter $f_{\mathrm{BH}}$. Lowering $f_{\mathrm{BH}}$ reduces the pronounced bump at $L_{\rm bol} \sim 10^{46}\, erg \ s^{-1}$ above $z=2$, leading to a better match with the physics-based models and the bolometric AGN luminosity function from \citet{2020MNRAS.495.3252S}. Matching the AGN luminosity function from \citet{2024arXiv240610341A} suggests that if these ``little red dots'' are AGNs, their growth may be strongly tied to galaxy mergers.

As suggested above, a caveat to our results is that the mass contribution from \textit{cold-mode}, \textit{merger-driven}, and \textit{secular instabilities} in {\sc Dark Sage} depend on the value chosen for $f_{\mathrm{BH}}$. For example, reducing $f_{\mathrm{BH}}$ to 3\% increases the contribution from \textit{merger-driven instability} at all redshifts because more mass is added to the galactic disk, triggering starburst episodes and allowing the galaxy to grow in stellar mass. Therefore, any unstable gas with low angular momentum is added to the primary black hole. This also affects secular instabilities since galaxies that have regulated instabilities after a merger are likely to have larger cold gas reservoirs, which increases the likelihood of this gas becoming unstable over time. Lower values of $f_{\mathrm{BH}}$ for $z<2$ allow black holes with $M_{\rm BH} \sim 10^{6}\, \mathrm{M}_{\odot}$ to primarily gain their mass through the \textit{cold-mode} channel, rather than through \textit{secular instabilities}. This is because galaxies can regulate their instabilities and grow their stellar mass more efficiently relative to their black hole mass. For more massive black holes, contributions from secular and merger-driven instabilities become more significant.

Several observational studies support the idea that much of the rapid black hole accretion is triggered by cold-mode and secular instabilities rather than mergers, especially for late-type galaxies \citep{2006ApJS..166....1H, 2009MNRAS.397..623G, 2012ApJ...757...81B, 2019ApJ...882..141M, 2021ApJ...919..129L}. \citet{2024MNRAS.527.9461S} show that AGN activity is generally not driven by major mergers, except for a few gas-rich mergers at $z <$ 0.9. Major mergers do not contribute significantly to the growth of central SMBHs, resulting in these mergers not sustaining long-term accretion. Rapid black hole accretion driven by cold-mode and secular instabilities might explain the lack of a stronger observational link between mergers and AGN activity.

Although the division of black hole growth channels in {\sc Dark Sage} is similar to other semi-analytic models \citep{2008MNRAS.391..481S, Benson2012, Croton2016}, its main distinction lies in the treatment of instabilities. While instabilities do not physically happen in annuli, as calculated in {\sc Dark Sage}, the model behaves more like a global prescription. Because most unstable mass is transferred inward, if annulus \textit{i} becomes unstable, it increases the likelihood that annulus \textit{i-1} will also become unstable. Since we check annulus \textit{i-1} next, this instability can cascade all the way to the center of the galaxy, effectively resembling a global instability.

When tracing SMBHs at $z > 6$, two distinct populations of host galaxies emerge: those that have not experienced any mergers, resulting in both their galaxies and black holes growing through secular accretion, and those that have undergone minor mergers and a few major mergers. The latter population benefits from ex-situ accretion, allowing black holes to gain mass through the \textit{cold--mode} channel and \textit{merger-driven instabilities}. We limit our discussion of black holes and galaxies above $z>6$ due to the insufficient resolution of Millennium merger trees in capturing low-mass halos and galaxy mergers.


Our results highlight the diversity of black hole growth histories explored across models. We use the semi-analytic model {\sc Dark Sage} to separate the SMBH assembly into its multiple different growth channels across redshift. Low-mass black holes in {\sc Dark Sage} primarily grow via \textit{secular instabilities}. Most of the mass for black holes above $M_{\rm BH} \sim 10^{7}\, \mathrm{M}_{\odot}$ comes from the \textit{cold--mode} growth channel. At the lowest redshifts, black hole--black hole mergers contributes up to 60\% its current mass. We interpret the mass assembly of {\sc Dark Sage} black holes as initially undergoing a super-Eddington growth phase, followed by a gradual decrease in Eddington ratio as their mass increases. As \citet{2024MNRAS.530.1732P} highlights, while super-Eddington and Eddington-limited accretion models match the observed SMBHs and host galaxy properties at  $z=5-7$, they diverge at higher redshifts and in the intermediate mass range. In {\sc Dark Sage}, many black holes may have experienced stalled growth following their initial episode of super-Eddington accretion resulting in a population of undermassive black holes at later times. Several models incorporate super-Eddington accretion onto stellar-mass black holes, suggesting that such accretion is possible in certain conditions \citep{2009ApJS..183..171S, 2014MNRAS.439..503S, 2016MNRAS.456.3929S, 2016MNRAS.456.3915S}, though its feasibility and occurrence in reality remains uncertain.

Furthermore, observational evidence also supports super-Eddington accretion \citep{2018ApJ...856....6D}. In {\sc Dark Sage}, super-Eddington accretion results from Toomre instabilities, which are dependent on our assumption for the velocity dispersion of the gas in the galaxy disk. Lower sound speeds result in lower Toomre-Q values, making disks more prone to intense, centrally concentrated bursts of star formation. This compaction in galaxies, allowed by {\sc Dark Sage}, is affected by the angular momentum of infalling gas, determined by accretion from the halo and independent of disk dynamics. As a result, {\sc Dark Sage} galaxies experience a broad range of outcomes, with many undergoing compaction. Although these bursts might explain compact star-forming systems at $z\sim2$, models that incorporate higher velocity dispersions due to disk instability generally result in less erratic evolution by allowing for some degree of self-regulation.

\citet{Lodato&Natarajan2006} presents the formation of direct collapse black holes (DCBHs) using Toomre disk instabilities, a similar treatment to the one adopted in {\sc Dark Sage} with one key difference regarding outcome, i.e. disk fragmentation. Both models assume disk evolution is mainly driven by angular momentum redistribution induced by gravitational instabilities. While {\sc Dark Sage} grows $10^5 \, \mathrm{M}_{\odot}$ black holes above $z>10$ with metal-poor gas, our work does not account for disk fragmentation. In {\sc Dark Sage}, the gas within an annulus is evenly distributed, and its starburst phase, along with its Toomre-Q criterion, is considered across the entire annulus. Yet, our model presents a promising path towards DCBH formation with super-Eddington accreted gas. This will be explored further in future works.

The diversity of SMBH assembly histories in the models considered here may be distinguishable by their resultant spin distributions, which encode information about recent growth by mergers and accretion \citep[e.g.,][]{2008ApJ...684..822B}.  Present X-ray reflection spectroscopy constraints point towards high spins for most SMBHs with thin accretion disks \citep{Reynolds+2021}, and event horizon scale studies aim to constrain spins for up to tens of additional sources at (more typical) low Eddington ratios \citep{Ricarte+2023}.  We plan to investigate spin distributions in future work as it will likely provide strong discrimination between models.

\software{IPython \citep{Perez2007IPythonFor}, 
    Scipy \citep{2020SciPy-NMeth}, 
    matplotlib \citep{Hunter2007MatplotlibEnvironment}, 
    Astropy \citep{Robitaille2013Astropy:Astronomy}, 
    NumPy \citep{VanDerWalt2011TheComputation}
         }

\begin{acknowledgments}
AJPV gratefully acknowledges support of a post-doctoral fellowship from the Heising-Simons foundation. AJPV also thanks John C. Forbes, Colin Burke, Tonima Ananna, and Meg Urry for helpful comments and discussions that inspired this manuscript. We used computational facilities from the Vanderbilt Advanced Computing Center for Research and Education (ACCRE) and the Yale Center for Research Computing (YCRC). Literature reviews for this work was made using the NASA’s Astrophysics Data System. P.N. and A.R. acknowledge support from the Gordon and Betty Moore Foundation and the John Templeton Foundation that fund the Black Hole Initiative (BHI) at Harvard University. The Flatiron Institute is a division of the Simons Foundation.
\end{acknowledgments}

\clearpage

\bibliographystyle{aasjournal}
\bibliography{bibfile.bib}

\begin{thebibliography}{}
\expandafter\ifx\csname natexlab\endcsname\relax\def\natexlab#1{#1}\fi
\providecommand{\url}[1]{\href{#1}{#1}}
\providecommand{\dodoi}[1]{doi:~\href{http://doi.org/#1}{\nolinkurl{#1}}}
\providecommand{\doeprint}[1]{\href{http://ascl.net/#1}{\nolinkurl{http://ascl.net/#1}}}
\providecommand{\doarXiv}[1]{\href{https://arxiv.org/abs/#1}{\nolinkurl{https://arxiv.org/abs/#1}}}

\bibitem[{{Agazie} {et~al.}(2023){Agazie}, {Anumarlapudi}, {Archibald}, {Baker}, {B{\'e}csy}, {Blecha}, {Bonilla}, {Brazier}, {Brook}, {Burke-Spolaor}, {Burnette}, {Case}, {Casey-Clyde}, {Charisi}, {Chatterjee}, {Chatziioannou}, {Cheeseboro}, {Chen}, {Cohen}, {Cordes}, {Cornish}, {Crawford}, {Cromartie}, {Crowter}, {Cutler}, {D'Orazio}, {Decesar}, {Degan}, {Demorest}, {Deng}, {Dolch}, {Drachler}, {Ferrara}, {Fiore}, {Fonseca}, {Freedman}, {Gardiner}, {Garver-Daniels}, {Gentile}, {Gersbach}, {Glaser}, {Good}, {G{\"u}ltekin}, {Hazboun}, {Hourihane}, {Islo}, {Jennings}, {Johnson}, {Jones}, {Kaiser}, {Kaplan}, {Kelley}, {Kerr}, {Key}, {Laal}, {Lam}, {Lamb}, {Lazio}, {Lewandowska}, {Littenberg}, {Liu}, {Luo}, {Lynch}, {Ma}, {Madison}, {McEwen}, {McKee}, {McLaughlin}, {McMann}, {Meyers}, {Meyers}, {Mingarelli}, {Mitridate}, {Natarajan}, {Ng}, {Nice}, {Ocker}, {Olum}, {Pennucci}, {Perera}, {Petrov}, {Pol}, {Radovan}, {Ransom}, {Ray}, {Romano}, {Runnoe}, {Sardesai}, {Schmiedekamp}, {Schmiedekamp}, {Schmitz},
  {Schult}, {Shapiro-Albert}, {Siemens}, {Simon}, {Siwek}, {Stairs}, {Stinebring}, {Stovall}, {Sun}, {Susobhanan}, {Swiggum}, {Taylor}, {Taylor}, {Turner}, {Unal}, {Vallisneri}, {Vigeland}, {Wachter}, {Wahl}, {Wang}, {Witt}, {Wright}, {Young}, \& {Nanograv Collaboration}}]{Agazie+2023}
{Agazie}, G., {Anumarlapudi}, A., {Archibald}, A.~M., {et~al.} 2023, \apjl, 952, L37, \dodoi{10.3847/2041-8213/ace18b}

\bibitem[{{Akins} {et~al.}(2024){Akins}, {Casey}, {Lambrides}, {Allen}, {Andika}, {Brinch}, {Champagne}, {Cooper}, {Ding}, {Drakos}, {Faisst}, {Finkelstein}, {Franco}, {Fujimoto}, {Gentile}, {Gillman}, {Gozaliasl}, {Harish}, {Hayward}, {Hirschmann}, {Ilbert}, {Kartaltepe}, {Kocevski}, {Koekemoer}, {Kokorev}, {Liu}, {Long}, {McCracken}, {McKinney}, {Onoue}, {Paquereau}, {Renzini}, {Rhodes}, {Robertson}, {Shuntov}, {Silverman}, {Tanaka}, {Toft}, {Trakhtenbrot}, {Valentino}, \& {Zavala}}]{2024arXiv240610341A}
{Akins}, H.~B., {Casey}, C.~M., {Lambrides}, E., {et~al.} 2024, arXiv e-prints, arXiv:2406.10341, \dodoi{10.48550/arXiv.2406.10341}

\bibitem[{{Aller} \& {Richstone}(2007)}]{2007ApJ...665..120A}
{Aller}, M.~C., \& {Richstone}, D.~O. 2007, The Astrophysical Journal, 665, 120, \dodoi{10.1086/519298}

\bibitem[{{Amarantidis} {et~al.}(2019){Amarantidis}, {Afonso}, {Messias}, {Henriques}, {Griffin}, {Lacey}, {Lagos}, {Gonzalez-Perez}, {Dubois}, {Volonteri}, {Matute}, {Pappalardo}, {Qin}, {Chary}, \& {Norris}}]{2019MNRAS.485.2694A}
{Amarantidis}, S., {Afonso}, J., {Messias}, H., {et~al.} 2019, Monthly Notices of the Royal Astronomical Society, 485, 2694, \dodoi{10.1093/mnras/stz551}

\bibitem[{{Ananna} {et~al.}(2019){Ananna}, {Treister}, {Urry}, {Ricci}, {Kirkpatrick}, {LaMassa}, {Buchner}, {Civano}, {Tremmel}, \& {Marchesi}}]{2019ApJ...871..240A}
{Ananna}, T.~T., {Treister}, E., {Urry}, C.~M., {et~al.} 2019, The Astrophysical Journal, 871, 240, \dodoi{10.3847/1538-4357/aafb77}

\bibitem[{{Ananna} {et~al.}(2020){Ananna}, {Urry}, {Treister}, {Hickox}, {Shankar}, {Ricci}, {Cappelluti}, {Marchesi}, \& {Turner}}]{2020ApJ...903...85A}
{Ananna}, T.~T., {Urry}, C.~M., {Treister}, E., {et~al.} 2020, The Astrophysical Journal, 903, 85, \dodoi{10.3847/1538-4357/abb815}

\bibitem[{Baldry {et~al.}(2008)Baldry, Glazebrook, \& Driver}]{Baldry2008}
Baldry, I.~K., Glazebrook, K., \& Driver, S.~P. 2008, Monthly Notices of the Royal Astronomical Society, 959, 945, \dodoi{10.1111/j.1365-2966.2008.13348.x}

\bibitem[{{Baldry} {et~al.}(2012){Baldry}, {Driver}, {Loveday}, {Taylor}, {Kelvin}, {Liske}, {Norberg}, {Robotham}, {Brough}, {Hopkins}, {Bamford}, {Peacock}, {Bland-Hawthorn}, {Conselice}, {Croom}, {Jones}, {Parkinson}, {Popescu}, {Prescott}, {Sharp}, \& {Tuffs}}]{Baldry2012}
{Baldry}, I.~K., {Driver}, S.~P., {Loveday}, J., {et~al.} 2012, \mnras, 421, 621, \dodoi{10.1111/j.1365-2966.2012.20340.x}

\bibitem[{{Barger} {et~al.}(2005){Barger}, {Cowie}, {Mushotzky}, {Yang}, {Wang}, {Steffen}, \& {Capak}}]{2005AJ....129..578B}
{Barger}, A.~J., {Cowie}, L.~L., {Mushotzky}, R.~F., {et~al.} 2005, The Astronomical Journal, 129, 578, \dodoi{10.1086/426915}

\bibitem[{{Begelman} \& {Nath}(2005)}]{2005MNRAS.361.1387B}
{Begelman}, M.~C., \& {Nath}, B.~B. 2005, Monthly Notices of the Royal Astronomical Society, 361, 1387, \dodoi{10.1111/j.1365-2966.2005.09249.x}

\bibitem[{Behroozi {et~al.}(2019)Behroozi, Wechsler, Hearin, \& Conroy}]{Behroozi2019Universemachine:010}
Behroozi, P., Wechsler, R.~H., Hearin, A.~P., \& Conroy, C. 2019, Monthly Notices of the Royal Astronomical Society, 488, 3143, \dodoi{10.1093/mnras/stz1182}

\bibitem[{{Behroozi} {et~al.}(2013{\natexlab{a}}){Behroozi}, {Wechsler}, \& {Wu}}]{2013ApJ...762..109B}
{Behroozi}, P.~S., {Wechsler}, R.~H., \& {Wu}, H.-Y. 2013{\natexlab{a}}, The Astrophysical Journal, 762, 109, \dodoi{10.1088/0004-637X/762/2/109}

\bibitem[{{Behroozi} {et~al.}(2013{\natexlab{b}}){Behroozi}, {Wechsler}, {Wu}, {Busha}, {Klypin}, \& {Primack}}]{2013ApJ...763...18B}
{Behroozi}, P.~S., {Wechsler}, R.~H., {Wu}, H.-Y., {et~al.} 2013{\natexlab{b}}, The Astrophysical Journal, 763, 18, \dodoi{10.1088/0004-637X/763/1/18}

\bibitem[{{Beifiori} {et~al.}(2012){Beifiori}, {Courteau}, {Corsini}, \& {Zhu}}]{2012MNRAS.419.2497B}
{Beifiori}, A., {Courteau}, S., {Corsini}, E.~M., \& {Zhu}, Y. 2012, Monthly Notices of the Royal Astronomical Society, 419, 2497, \dodoi{10.1111/j.1365-2966.2011.19903.x}

\bibitem[{{Benson}(2012)}]{Benson2012}
{Benson}, A.~J. 2012, New Astronomy, 17, 175, \dodoi{10.1016/j.newast.2011.07.004}

\bibitem[{{Bernardi} {et~al.}(2017{\natexlab{a}}){Bernardi}, {Fischer}, {Sheth}, {Meert}, {Huertas-Company}, {Shankar}, \& {Vikram}}]{2017MNRAS.468.2569B}
{Bernardi}, M., {Fischer}, J.~L., {Sheth}, R.~K., {et~al.} 2017{\natexlab{a}}, \mnras, 468, 2569, \dodoi{10.1093/mnras/stx677}

\bibitem[{{Bernardi} {et~al.}(2017{\natexlab{b}}){Bernardi}, {Meert}, {Sheth}, {Fischer}, {Huertas-Company}, {Maraston}, {Shankar}, \& {Vikram}}]{2017MNRAS.467.2217B}
{Bernardi}, M., {Meert}, A., {Sheth}, R.~K., {et~al.} 2017{\natexlab{b}}, \mnras, 467, 2217, \dodoi{10.1093/mnras/stx176}

\bibitem[{{Bernardi} {et~al.}(2013){Bernardi}, {Meert}, {Sheth}, {Vikram}, {Huertas-Company}, {Mei}, \& {Shankar}}]{2013MNRAS.436..697B}
---. 2013, Monthly Notices of the Royal Astronomical Society, 436, 697, \dodoi{10.1093/mnras/stt1607}

\bibitem[{{Bernardi} {et~al.}(2010){Bernardi}, {Shankar}, {Hyde}, {Mei}, {Marulli}, \& {Sheth}}]{2010MNRAS.404.2087B}
{Bernardi}, M., {Shankar}, F., {Hyde}, J.~B., {et~al.} 2010, \mnras, 404, 2087, \dodoi{10.1111/j.1365-2966.2010.16425.x}

\bibitem[{{Bernardi} {et~al.}(2007){Bernardi}, {Sheth}, {Tundo}, \& {Hyde}}]{2007ApJ...660..267B}
{Bernardi}, M., {Sheth}, R.~K., {Tundo}, E., \& {Hyde}, J.~B. 2007, The Astrophysical Journal, 660, 267, \dodoi{10.1086/512719}

\bibitem[{{Bernardi} {et~al.}(2018){Bernardi}, {Sheth}, {Fischer}, {Meert}, {Chae}, {Dominguez-Sanchez}, {Huertas-Company}, {Shankar}, \& {Vikram}}]{2018MNRAS.475..757B}
{Bernardi}, M., {Sheth}, R.~K., {Fischer}, J.~L., {et~al.} 2018, \mnras, 475, 757, \dodoi{10.1093/mnras/stx3171}

\bibitem[{{Berti} \& {Volonteri}(2008)}]{2008ApJ...684..822B}
{Berti}, E., \& {Volonteri}, M. 2008, The Astrophysical Journal, 684, 822, \dodoi{10.1086/590379}

\bibitem[{Bondi(1952)}]{Bondi1952}
Bondi, H. 1952, Monthly Notices of the Royal Astronomical Society, 112, 195, \dodoi{10.1093/mnras/112.2.195}

\bibitem[{{Booth} \& {Schaye}(2009)}]{2009MNRAS.398...53B}
{Booth}, C.~M., \& {Schaye}, J. 2009, Monthly Notices of the Royal Astronomical Society, 398, 53, \dodoi{10.1111/j.1365-2966.2009.15043.x}

\bibitem[{{Bose} {et~al.}(2019){Bose}, {Eisenstein}, {Hernquist}, {Pillepich}, {Nelson}, {Marinacci}, {Springel}, \& {Vogelsberger}}]{2019MNRAS.490.5693B}
{Bose}, S., {Eisenstein}, D.~J., {Hernquist}, L., {et~al.} 2019, Monthly Notices of the Royal Astronomical Society, 490, 5693, \dodoi{10.1093/mnras/stz2546}

\bibitem[{{Bournaud} {et~al.}(2012){Bournaud}, {Juneau}, {Le Floc'h}, {Mullaney}, {Daddi}, {Dekel}, {Duc}, {Elbaz}, {Salmi}, \& {Dickinson}}]{2012ApJ...757...81B}
{Bournaud}, F., {Juneau}, S., {Le Floc'h}, E., {et~al.} 2012, \apj, 757, 81, \dodoi{10.1088/0004-637X/757/1/81}

\bibitem[{{Boylan-Kolchin} {et~al.}(2013){Boylan-Kolchin}, {Bullock}, {Sohn}, {Besla}, \& {van der Marel}}]{2013ApJ...768..140B}
{Boylan-Kolchin}, M., {Bullock}, J.~S., {Sohn}, S.~T., {Besla}, G., \& {van der Marel}, R.~P. 2013, The Astrophysical Journal, 768, 140, \dodoi{10.1088/0004-637X/768/2/140}

\bibitem[{{Boylan-Kolchin} {et~al.}(2008){Boylan-Kolchin}, {Ma}, \& {Quataert}}]{Boylan-Kolchin+2008}
{Boylan-Kolchin}, M., {Ma}, C.-P., \& {Quataert}, E. 2008, \mnras, 383, 93, \dodoi{10.1111/j.1365-2966.2007.12530.x}

\bibitem[{Brown {et~al.}(2015)Brown, Catinella, Cortese, Kilborn, Haynes, \& Giovanelli}]{Brown2015TheGalaxies}
Brown, T., Catinella, B., Cortese, L., {et~al.} 2015, Monthly Notices of the Royal Astronomical Society, 452, 2479, \dodoi{10.1093/mnras/stv1311}

\bibitem[{{Burke} {et~al.}(2025){Burke}, {Natarajan}, {Baldassare}, \& {Geha}}]{Burke2025}
{Burke}, C.~J., {Natarajan}, P., {Baldassare}, V.~F., \& {Geha}, M. 2025, \apj, 978, 77, \dodoi{10.3847/1538-4357/ad94d9}

\bibitem[{{Calette} {et~al.}(2018){Calette}, {Avila-Reese}, {Rodr{\'\i}guez-Puebla}, {Hern{\'a}ndez-Toledo}, \& {Papastergis}}]{2018RMxAA..54..443C}
{Calette}, A.~R., {Avila-Reese}, V., {Rodr{\'\i}guez-Puebla}, A., {Hern{\'a}ndez-Toledo}, H., \& {Papastergis}, E. 2018, Revista Mexicana de Astronomia y Astrofisica, 54, 443, \dodoi{10.48550/arXiv.1803.07692}

\bibitem[{{Cattaneo} {et~al.}(2005){Cattaneo}, {Blaizot}, {Devriendt}, \& {Guiderdoni}}]{2005MNRAS.364..407C}
{Cattaneo}, A., {Blaizot}, J., {Devriendt}, J., \& {Guiderdoni}, B. 2005, Monthly Notices of the Royal Astronomical Society, 364, 407, \dodoi{10.1111/j.1365-2966.2005.09608.x}

\bibitem[{{Cavaliere} {et~al.}(1971){Cavaliere}, {Morrison}, \& {Wood}}]{1971ApJ...170..223C}
{Cavaliere}, A., {Morrison}, P., \& {Wood}, K. 1971, The Astrophysical Journal, 170, 223, \dodoi{10.1086/151206}

\bibitem[{{Ciotti} \& {Ostriker}(2001)}]{2001ApJ...551..131C}
{Ciotti}, L., \& {Ostriker}, J.~P. 2001, The Astrophysical Journal, 551, 131, \dodoi{10.1086/320053}

\bibitem[{{Col{\'\i}n} {et~al.}(1999){Col{\'\i}n}, {Klypin}, {Kravtsov}, \& {Khokhlov}}]{1999ApJ...523...32C}
{Col{\'\i}n}, P., {Klypin}, A.~A., {Kravtsov}, A.~V., \& {Khokhlov}, A.~M. 1999, The Astrophysical Journal, 523, 32, \dodoi{10.1086/307710}

\bibitem[{{Conroy} \& {Wechsler}(2009)}]{2009ApJ...696..620C}
{Conroy}, C., \& {Wechsler}, R.~H. 2009, The Astrophysical Journal, 696, 620, \dodoi{10.1088/0004-637X/696/1/620}

\bibitem[{{Cox} {et~al.}(2006){Cox}, {Jonsson}, {Primack}, \& {Somerville}}]{2006MNRAS.373.1013C}
{Cox}, T.~J., {Jonsson}, P., {Primack}, J.~R., \& {Somerville}, R.~S. 2006, Monthly Notices of the Royal Astronomical Society, 373, 1013, \dodoi{10.1111/j.1365-2966.2006.11107.x}

\bibitem[{{Cox} {et~al.}(2008){Cox}, {Jonsson}, {Somerville}, {Primack}, \& {Dekel}}]{2008MNRAS.384..386C}
{Cox}, T.~J., {Jonsson}, P., {Somerville}, R.~S., {Primack}, J.~R., \& {Dekel}, A. 2008, Monthly Notices of the Royal Astronomical Society, 384, 386, \dodoi{10.1111/j.1365-2966.2007.12730.x}

\bibitem[{Croton {et~al.}(2006)Croton, Springel, White, De~Lucia, Frenk, Gao, Jenkins, Kauffmann, Navarro, \& Yoshida}]{Croton2006}
Croton, D.~J., Springel, V., White, S.~D., {et~al.} 2006, Monthly Notices of the Royal Astronomical Society, 365, 11, \dodoi{10.1111/j.1365-2966.2005.09675.x}

\bibitem[{{Croton} {et~al.}(2016){Croton}, {Stevens}, {Tonini}, {Garel}, {Bernyk}, {Bibiano}, {Hodkinson}, {Mutch}, {Poole}, \& {Shattow}}]{Croton2016}
{Croton}, D.~J., {Stevens}, A. R.~H., {Tonini}, C., {et~al.} 2016, The Astrophysical Journal Supplement Series, 222, 22, \dodoi{10.3847/0067-0049/222/2/22}

\bibitem[{{Dattathri} {et~al.}(2024){Dattathri}, {Natarajan}, {Porras-Valverde}, {Burke}, {Chen}, {Di Matteo}, \& {Ni}}]{2024arXiv241013958D}
{Dattathri}, S., {Natarajan}, P., {Porras-Valverde}, A.~J., {et~al.} 2024, arXiv e-prints, arXiv:2410.13958, \dodoi{10.48550/arXiv.2410.13958}

\bibitem[{{Davidge} {et~al.}(2017){Davidge}, {Serjeant}, {Pearson}, {Matsuhara}, {Wada}, {Dryer}, \& {Barrufet}}]{2017MNRAS.472.4259D}
{Davidge}, H., {Serjeant}, S., {Pearson}, C., {et~al.} 2017, \mnras, 472, 4259, \dodoi{10.1093/mnras/stx1935}

\bibitem[{{Di Matteo} {et~al.}(2008){Di Matteo}, {Colberg}, {Springel}, {Hernquist}, \& {Sijacki}}]{2008ApJ...676...33D}
{Di Matteo}, T., {Colberg}, J., {Springel}, V., {Hernquist}, L., \& {Sijacki}, D. 2008, The Astrophysical Journal, 676, 33, \dodoi{10.1086/524921}

\bibitem[{{Driver} {et~al.}(2011){Driver}, {Hill}, {Kelvin}, {Robotham}, {Liske}, {Norberg}, {Baldry}, {Bamford}, {Hopkins}, {Loveday}, {Peacock}, {Andrae}, {Bland-Hawthorn}, {Brough}, {Brown}, {Cameron}, {Ching}, {Colless}, {Conselice}, {Croom}, {Cross}, {de Propris}, {Dye}, {Drinkwater}, {Ellis}, {Graham}, {Grootes}, {Gunawardhana}, {Jones}, {van Kampen}, {Maraston}, {Nichol}, {Parkinson}, {Phillipps}, {Pimbblet}, {Popescu}, {Prescott}, {Roseboom}, {Sadler}, {Sansom}, {Sharp}, {Smith}, {Taylor}, {Thomas}, {Tuffs}, {Wijesinghe}, {Dunne}, {Frenk}, {Jarvis}, {Madore}, {Meyer}, {Seibert}, {Staveley-Smith}, {Sutherland}, \& {Warren}}]{2011MNRAS.413..971D}
{Driver}, S.~P., {Hill}, D.~T., {Kelvin}, L.~S., {et~al.} 2011, \mnras, 413, 971, \dodoi{10.1111/j.1365-2966.2010.18188.x}

\bibitem[{{Driver} {et~al.}(2022){Driver}, {Bellstedt}, {Robotham}, {Baldry}, {Davies}, {Liske}, {Obreschkow}, {Taylor}, {Wright}, {Alpaslan}, {Bamford}, {Bauer}, {Bland-Hawthorn}, {Bilicki}, {Bravo}, {Brough}, {Casura}, {Cluver}, {Colless}, {Conselice}, {Croom}, {de Jong}, {D'Eugenio}, {De Propris}, {Dogruel}, {Drinkwater}, {Dvornik}, {Farrow}, {Frenk}, {Giblin}, {Graham}, {Grootes}, {Gunawardhana}, {Hashemizadeh}, {H{\"a}u{\ss}ler}, {Heymans}, {Hildebrandt}, {Holwerda}, {Hopkins}, {Jarrett}, {Heath Jones}, {Kelvin}, {Koushan}, {Kuijken}, {Lara-L{\'o}pez}, {Lange}, {L{\'o}pez-S{\'a}nchez}, {Loveday}, {Mahajan}, {Meyer}, {Moffett}, {Napolitano}, {Norberg}, {Owers}, {Radovich}, {Raouf}, {Peacock}, {Phillipps}, {Pimbblet}, {Popescu}, {Said}, {Sansom}, {Seibert}, {Sutherland}, {Thorne}, {Tuffs}, {Turner}, {van der Wel}, {van Kampen}, \& {Wilkins}}]{2022MNRAS.513..439D}
{Driver}, S.~P., {Bellstedt}, S., {Robotham}, A. S.~G., {et~al.} 2022, \mnras, 513, 439, \dodoi{10.1093/mnras/stac472}

\bibitem[{{Du} {et~al.}(2018){Du}, {Zhang}, {Wang}, {Huang}, {Zhang}, {Lu}, {Hu}, {Li}, {Bai}, {Bian}, {Yuan}, {Ho}, {Wang}, \& {SEAMBH Collaboration}}]{2018ApJ...856....6D}
{Du}, P., {Zhang}, Z.-X., {Wang}, K., {et~al.} 2018, The Astrophysical Journal, 856, 6, \dodoi{10.3847/1538-4357/aaae6b}

\bibitem[{{Efstathiou} {et~al.}(1982){Efstathiou}, {Lake}, \& {Negroponte}}]{1982MNRAS.199.1069E}
{Efstathiou}, G., {Lake}, G., \& {Negroponte}, J. 1982, Monthly Notices of the Royal Astronomical Society, 199, 1069, \dodoi{10.1093/mnras/199.4.1069}

\bibitem[{{Fabian}(1999)}]{1999MNRAS.308L..39F}
{Fabian}, A.~C. 1999, Monthly Notices of the Royal Astronomical Society, 308, L39, \dodoi{10.1046/j.1365-8711.1999.03017.x}

\bibitem[{{Fouqu{\'e}} {et~al.}(2001){Fouqu{\'e}}, {Solanes}, {Sanchis}, \& {Balkowski}}]{2001A&A...375..770F}
{Fouqu{\'e}}, P., {Solanes}, J.~M., {Sanchis}, T., \& {Balkowski}, C. 2001, Astronomy \& Astrophysics, 375, 770, \dodoi{10.1051/0004-6361:20010833}

\bibitem[{{Gabor} {et~al.}(2009){Gabor}, {Impey}, {Jahnke}, {Simmons}, {Trump}, {Koekemoer}, {Brusa}, {Cappelluti}, {Schinnerer}, {Smol{\v{c}}i{\'c}}, {Salvato}, {Rhodes}, {Mobasher}, {Capak}, {Massey}, {Leauthaud}, \& {Scoville}}]{2009ApJ...691..705G}
{Gabor}, J.~M., {Impey}, C.~D., {Jahnke}, K., {et~al.} 2009, The Astrophysical Journal, 691, 705, \dodoi{10.1088/0004-637X/691/1/705}

\bibitem[{{Gabrielpillai} {et~al.}(2022){Gabrielpillai}, {Somerville}, {Genel}, {Rodriguez-Gomez}, {Pandya}, {Yung}, \& {Hernquist}}]{2022MNRAS.517.6091G}
{Gabrielpillai}, A., {Somerville}, R.~S., {Genel}, S., {et~al.} 2022, Monthly Notices of the Royal Astronomical Society, 517, 6091, \dodoi{10.1093/mnras/stac2297}

\bibitem[{{Gallazzi} {et~al.}(2005){Gallazzi}, {Charlot}, {Brinchmann}, {White}, \& {Tremonti}}]{2005MNRAS.362...41G}
{Gallazzi}, A., {Charlot}, S., {Brinchmann}, J., {White}, S. D.~M., \& {Tremonti}, C.~A. 2005, Monthly Notices of the Royal Astronomical Society, 362, 41, \dodoi{10.1111/j.1365-2966.2005.09321.x}

\bibitem[{{Gebhardt} {et~al.}(2000){Gebhardt}, {Bender}, {Bower}, {Dressler}, {Faber}, {Filippenko}, {Green}, {Grillmair}, {Ho}, {Kormendy}, {Lauer}, {Magorrian}, {Pinkney}, {Richstone}, \& {Tremaine}}]{2000ApJ...539L..13G}
{Gebhardt}, K., {Bender}, R., {Bower}, G., {et~al.} 2000, The Astrophysical Journal, Letters, 539, L13, \dodoi{10.1086/312840}

\bibitem[{{Genel} {et~al.}(2014){Genel}, {Vogelsberger}, {Springel}, {Sijacki}, {Nelson}, {Snyder}, {Rodriguez-Gomez}, {Torrey}, \& {Hernquist}}]{2014MNRAS.445..175G}
{Genel}, S., {Vogelsberger}, M., {Springel}, V., {et~al.} 2014, Monthly Notices of the Royal Astronomical Society, 445, 175, \dodoi{10.1093/mnras/stu1654}

\bibitem[{{Georgakakis} {et~al.}(2009){Georgakakis}, {Coil}, {Laird}, {Griffith}, {Nandra}, {Lotz}, {Pierce}, {Cooper}, {Newman}, \& {Koekemoer}}]{2009MNRAS.397..623G}
{Georgakakis}, A., {Coil}, A.~L., {Laird}, E.~S., {et~al.} 2009, Monthly Notices of the Royal Astronomical Society, 397, 623, \dodoi{10.1111/j.1365-2966.2009.14951.x}

\bibitem[{{Graham} {et~al.}(2007){Graham}, {Driver}, {Allen}, \& {Liske}}]{2007MNRAS.378..198G}
{Graham}, A.~W., {Driver}, S.~P., {Allen}, P.~D., \& {Liske}, J. 2007, Monthly Notices of the Royal Astronomical Society, 378, 198, \dodoi{10.1111/j.1365-2966.2007.11770.x}

\bibitem[{{Graham} {et~al.}(2001){Graham}, {Erwin}, {Caon}, \& {Trujillo}}]{2001ApJ...563L..11G}
{Graham}, A.~W., {Erwin}, P., {Caon}, N., \& {Trujillo}, I. 2001, The Astrophysical Journal, Letters, 563, L11, \dodoi{10.1086/338500}

\bibitem[{{Greene} {et~al.}(2020){Greene}, {Strader}, \& {Ho}}]{Greene2020}
{Greene}, J.~E., {Strader}, J., \& {Ho}, L.~C. 2020, \araa, 58, 257, \dodoi{10.1146/annurev-astro-032620-021835}

\bibitem[{{Gruppioni} {et~al.}(2011){Gruppioni}, {Pozzi}, {Zamorani}, \& {Vignali}}]{2011MNRAS.416...70G}
{Gruppioni}, C., {Pozzi}, F., {Zamorani}, G., \& {Vignali}, C. 2011, \mnras, 416, 70, \dodoi{10.1111/j.1365-2966.2011.19006.x}

\bibitem[{{Gu} {et~al.}(2022){Gu}, {Greene}, {Newman}, {Kreisch}, {Quenneville}, {Ma}, \& {Blakeslee}}]{2022ApJ...932..103G}
{Gu}, M., {Greene}, J.~E., {Newman}, A.~B., {et~al.} 2022, \apj, 932, 103, \dodoi{10.3847/1538-4357/ac69ea}

\bibitem[{{G{\"u}ltekin} {et~al.}(2009){G{\"u}ltekin}, {Richstone}, {Gebhardt}, {Lauer}, {Tremaine}, {Aller}, {Bender}, {Dressler}, {Faber}, {Filippenko}, {Green}, {Ho}, {Kormendy}, {Magorrian}, {Pinkney}, \& {Siopis}}]{2009ApJ...698..198G}
{G{\"u}ltekin}, K., {Richstone}, D.~O., {Gebhardt}, K., {et~al.} 2009, The Astrophysical Journal, 698, 198, \dodoi{10.1088/0004-637X/698/1/198}

\bibitem[{{Guyon} {et~al.}(2006){Guyon}, {Sanders}, \& {Stockton}}]{2006ApJS..166...89G}
{Guyon}, O., {Sanders}, D.~B., \& {Stockton}, A. 2006, The Astrophysical Journal, Supplement, 166, 89, \dodoi{10.1086/505030}

\bibitem[{{Habouzit} {et~al.}(2022{\natexlab{a}}){Habouzit}, {Onoue}, {Ba{\~n}ados}, {Neeleman}, {Angl{\'e}s-Alc{\'a}zar}, {Walter}, {Pillepich}, {Dav{\'e}}, {Jahnke}, \& {Dubois}}]{2022MNRAS.511.3751H}
{Habouzit}, M., {Onoue}, M., {Ba{\~n}ados}, E., {et~al.} 2022{\natexlab{a}}, Monthly Notices of the Royal Astronomical Society, 511, 3751, \dodoi{10.1093/mnras/stac225}

\bibitem[{{Habouzit} {et~al.}(2022{\natexlab{b}}){Habouzit}, {Somerville}, {Li}, {Genel}, {Aird}, {Angl{\'e}s-Alc{\'a}zar}, {Dav{\'e}}, {Georgiev}, {McAlpine}, {Rosas-Guevara}, {Dubois}, {Nelson}, {Banados}, {Hernquist}, {Peirani}, \& {Vogelsberger}}]{2022MNRAS.509.3015H}
{Habouzit}, M., {Somerville}, R.~S., {Li}, Y., {et~al.} 2022{\natexlab{b}}, Monthly Notices of the Royal Astronomical Society, 509, 3015, \dodoi{10.1093/mnras/stab3147}

\bibitem[{{Haehnelt} \& {Kauffmann}(2000)}]{2000MNRAS.318L..35H}
{Haehnelt}, M.~G., \& {Kauffmann}, G. 2000, Monthly Notices of the Royal Astronomical Society, 318, L35, \dodoi{10.1046/j.1365-8711.2000.03989.x}

\bibitem[{{Haehnelt} {et~al.}(1998){Haehnelt}, {Natarajan}, \& {Rees}}]{1998MNRAS.300..817H}
{Haehnelt}, M.~G., {Natarajan}, P., \& {Rees}, M.~J. 1998, Monthly Notices of the Royal Astronomical Society, 300, 817, \dodoi{10.1046/j.1365-8711.1998.01951.x}

\bibitem[{{H{\"a}ring} \& {Rix}(2004)}]{2004ApJ...604L..89H}
{H{\"a}ring}, N., \& {Rix}, H.-W. 2004, The Astrophysical Journal, Letters, 604, L89, \dodoi{10.1086/383567}

\bibitem[{{He} {et~al.}(2024){He}, {Akiyama}, {Enoki}, {Ichikawa}, {Inayoshi}, {Kashikawa}, {Kawaguchi}, {Matsuoka}, {Nagao}, {Onoue}, {Oogi}, {Schulze}, {Toba}, \& {Ueda}}]{2024ApJ...962..152H}
{He}, W., {Akiyama}, M., {Enoki}, M., {et~al.} 2024, \apj, 962, 152, \dodoi{10.3847/1538-4357/ad1518}

\bibitem[{Hearin \& Watson(2013)}]{Hearin2013TheColour}
Hearin, A.~P., \& Watson, D.~F. 2013, Monthly Notices of the Royal Astronomical Society, 435, 1313, \dodoi{10.1093/mnras/stt1374}

\bibitem[{Henriques {et~al.}(2015)Henriques, White, Thomas, Angulo, Guo, Lemson, Springel, \& Overzier}]{Henriques2015}
Henriques, B. M.~B., White, S. D.~M., Thomas, P.~A., {et~al.} 2015, Monthly Notices of the Royal Astronomical Society, 451, 2663, \dodoi{10.1093/mnras/stv705}

\bibitem[{{Hern{\'a}ndez-Y{\'e}venes} {et~al.}(2024){Hern{\'a}ndez-Y{\'e}venes}, {Nagar}, {Arratia}, \& {Jarrett}}]{Hernandez-Yevenes+2024}
{Hern{\'a}ndez-Y{\'e}venes}, J., {Nagar}, N., {Arratia}, V., \& {Jarrett}, T.~H. 2024, \mnras, 531, 4503, \dodoi{10.1093/mnras/stae1372}

\bibitem[{{Hirschmann} {et~al.}(2012){Hirschmann}, {Somerville}, {Naab}, \& {Burkert}}]{2012MNRAS.426..237H}
{Hirschmann}, M., {Somerville}, R.~S., {Naab}, T., \& {Burkert}, A. 2012, Monthly Notices of the Royal Astronomical Society, 426, 237, \dodoi{10.1111/j.1365-2966.2012.21626.x}

\bibitem[{{Hopkins} \& {Hernquist}(2006)}]{2006ApJS..166....1H}
{Hopkins}, P.~F., \& {Hernquist}, L. 2006, \apjs, 166, 1, \dodoi{10.1086/505753}

\bibitem[{{Hopkins} {et~al.}(2005{\natexlab{a}}){Hopkins}, {Hernquist}, {Cox}, {Di Matteo}, {Robertson}, \& {Springel}}]{2005ApJ...630..716H}
{Hopkins}, P.~F., {Hernquist}, L., {Cox}, T.~J., {et~al.} 2005{\natexlab{a}}, The Astrophysical Journal, 630, 716, \dodoi{10.1086/432463}

\bibitem[{{Hopkins} {et~al.}(2007{\natexlab{a}}){Hopkins}, {Hernquist}, {Cox}, {Robertson}, \& {Krause}}]{2007ApJ...669...45H}
{Hopkins}, P.~F., {Hernquist}, L., {Cox}, T.~J., {Robertson}, B., \& {Krause}, E. 2007{\natexlab{a}}, \apj, 669, 45, \dodoi{10.1086/521590}

\bibitem[{{Hopkins} {et~al.}(2005{\natexlab{b}}){Hopkins}, {Hernquist}, {Martini}, {Cox}, {Robertson}, {Di Matteo}, \& {Springel}}]{2005ApJ...625L..71H}
{Hopkins}, P.~F., {Hernquist}, L., {Martini}, P., {et~al.} 2005{\natexlab{b}}, The Astrophysical Journal, Letters, 625, L71, \dodoi{10.1086/431146}

\bibitem[{{Hopkins} {et~al.}(2007{\natexlab{b}}){Hopkins}, {Richards}, \& {Hernquist}}]{Hopkins+2007}
{Hopkins}, P.~F., {Richards}, G.~T., \& {Hernquist}, L. 2007{\natexlab{b}}, \apj, 654, 731, \dodoi{10.1086/509629}

\bibitem[{Hunter(2007)}]{Hunter2007MatplotlibEnvironment}
Hunter, J.~D. 2007, Computing In Science and Engineering, 9

\bibitem[{{Izquierdo-Villalba} {et~al.}(2024){Izquierdo-Villalba}, {Sesana}, {Colpi}, {Spinoso}, {Bonetti}, {Bonoli}, \& {Valiante}}]{Izquierdo-Villalba+2024}
{Izquierdo-Villalba}, D., {Sesana}, A., {Colpi}, M., {et~al.} 2024, \aap, 686, A183, \dodoi{10.1051/0004-6361/202449293}

\bibitem[{{Joshi} {et~al.}(2020){Joshi}, {Pillepich}, {Nelson}, {Marinacci}, {Springel}, {Rodriguez-Gomez}, {Vogelsberger}, \& {Hernquist}}]{2020MNRAS.496.2673J}
{Joshi}, G.~D., {Pillepich}, A., {Nelson}, D., {et~al.} 2020, Monthly Notices of the Royal Astronomical Society, 496, 2673, \dodoi{10.1093/mnras/staa1668}

\bibitem[{{Kashibadze} {et~al.}(2020){Kashibadze}, {Karachentsev}, \& {Karachentseva}}]{2020A&A...635A.135K}
{Kashibadze}, O.~G., {Karachentsev}, I.~D., \& {Karachentseva}, V.~E. 2020, Astronomy \& Astrophysics, 635, A135, \dodoi{10.1051/0004-6361/201936172}

\bibitem[{{Kauffmann} \& {Haehnelt}(2000)}]{2000MNRAS.311..576K}
{Kauffmann}, G., \& {Haehnelt}, M. 2000, Monthly Notices of the Royal Astronomical Society, 311, 576, \dodoi{10.1046/j.1365-8711.2000.03077.x}

\bibitem[{{Kelly} \& {Merloni}(2012)}]{2012AdAst2012E...7K}
{Kelly}, B.~C., \& {Merloni}, A. 2012, Advances in Astronomy, 2012, 970858, \dodoi{10.1155/2012/970858}

\bibitem[{{Kelly} {et~al.}(2011){Kelly}, {Sobolewska}, \& {Siemiginowska}}]{2011ApJ...730...52K}
{Kelly}, B.~C., {Sobolewska}, M., \& {Siemiginowska}, A. 2011, The Astrophysical Journal, 730, 52, \dodoi{10.1088/0004-637X/730/1/52}

\bibitem[{Keres {et~al.}(2003)Keres, Yun, \& Young}]{Keres2003CO2}
Keres, D., Yun, M.~S., \& Young, J.~S. 2003, The Astrophysical Journal, 582, 659, \dodoi{10.1007/978-94-010-0115-1{\_}5}

\bibitem[{{Khandai} {et~al.}(2015){Khandai}, {Di Matteo}, {Croft}, {Wilkins}, {Feng}, {Tucker}, {DeGraf}, \& {Liu}}]{2015MNRAS.450.1349K}
{Khandai}, N., {Di Matteo}, T., {Croft}, R., {et~al.} 2015, Monthly Notices of the Royal Astronomical Society, 450, 1349, \dodoi{10.1093/mnras/stv627}

\bibitem[{{Kim} {et~al.}(2024){Kim}, {Goto}, {Ling}, {Wu}, {Hashimoto}, {Kilerci}, {Ho}, {Uno}, {Wang}, \& {Lin}}]{2024MNRAS.527.5525K}
{Kim}, S.~J., {Goto}, T., {Ling}, C.-T., {et~al.} 2024, \mnras, 527, 5525, \dodoi{10.1093/mnras/stad3499}

\bibitem[{{Klypin} {et~al.}(2002){Klypin}, {Zhao}, \& {Somerville}}]{2002ApJ...573..597K}
{Klypin}, A., {Zhao}, H., \& {Somerville}, R.~S. 2002, The Astrophysical Journal, 573, 597, \dodoi{10.1086/340656}

\bibitem[{{Kormendy} \& {Ho}(2013)}]{2013ARA&A..51..511K}
{Kormendy}, J., \& {Ho}, L.~C. 2013, Annual Review of Astronomy \& Astrophysics, 51, 511, \dodoi{10.1146/annurev-astro-082708-101811}

\bibitem[{{Kormendy} \& {Richstone}(1995)}]{1995ARA&A..33..581K}
{Kormendy}, J., \& {Richstone}, D. 1995, Annual Review of Astronomy \& Astrophysics, 33, 581, \dodoi{10.1146/annurev.aa.33.090195.003053}

\bibitem[{{Kravtsov} {et~al.}(2004){Kravtsov}, {Berlind}, {Wechsler}, {Klypin}, {Gottl{\"o}ber}, {Allgood}, \& {Primack}}]{2004ApJ...609...35K}
{Kravtsov}, A.~V., {Berlind}, A.~A., {Wechsler}, R.~H., {et~al.} 2004, The Astrophysical Journal, 609, 35, \dodoi{10.1086/420959}

\bibitem[{{Kulessa} \& {Lynden-Bell}(1992)}]{1992MNRAS.255..105K}
{Kulessa}, A.~S., \& {Lynden-Bell}, D. 1992, Monthly Notices of the Royal Astronomical Society, 255, 105, \dodoi{10.1093/mnras/255.1.105}

\bibitem[{{Lambrides} {et~al.}(2021){Lambrides}, {Chiaberge}, {Heckman}, {Kirkpatrick}, {Meyer}, {Petric}, {Hall}, {Long}, {Watts}, {Gilli}, {Simons}, {Tchernyshyov}, {Rodriguez-Gomez}, {Vito}, {de la Vega}, {Davis}, {Kocevski}, \& {Norman}}]{2021ApJ...919..129L}
{Lambrides}, E.~L., {Chiaberge}, M., {Heckman}, T., {et~al.} 2021, \apj, 919, 129, \dodoi{10.3847/1538-4357/ac12c8}

\bibitem[{{Lauer} {et~al.}(2007){Lauer}, {Faber}, {Richstone}, {Gebhardt}, {Tremaine}, {Postman}, {Dressler}, {Aller}, {Filippenko}, {Green}, {Ho}, {Kormendy}, {Magorrian}, \& {Pinkney}}]{2007ApJ...662..808L}
{Lauer}, T.~R., {Faber}, S.~M., {Richstone}, D., {et~al.} 2007, The Astrophysical Journal, 662, 808, \dodoi{10.1086/518223}

\bibitem[{{Leja} {et~al.}(2020){Leja}, {Speagle}, {Johnson}, {Conroy}, {van Dokkum}, \& {Franx}}]{2020ApJ...893..111L}
{Leja}, J., {Speagle}, J.~S., {Johnson}, B.~D., {et~al.} 2020, \apj, 893, 111, \dodoi{10.3847/1538-4357/ab7e27}

\bibitem[{{Liepold} \& {Ma}(2024)}]{2024ApJ...971L..29L}
{Liepold}, E.~R., \& {Ma}, C.-P. 2024, \apjl, 971, L29, \dodoi{10.3847/2041-8213/ad66b8}

\bibitem[{{Ling} {et~al.}(2022){Ling}, {Kim}, {Wu}, {Goto}, {Kilerci}, {Hashimoto}, {Lin}, {Wang}, {Ho}, \& {Hsiao}}]{2022MNRAS.517..853L}
{Ling}, C.-T., {Kim}, S.~J., {Wu}, C. K.~W., {et~al.} 2022, \mnras, 517, 853, \dodoi{10.1093/mnras/stac2716}

\bibitem[{{Ling} {et~al.}(2023){Ling}, {Kim}, {Wu}, {Goto}, {Kilerci}, {Hashimoto}, {Lin}, {Wang}, {Ho}, \& {Hsiao}}]{2023MNRAS.522.1138L}
---. 2023, \mnras, 522, 1138, \dodoi{10.1093/mnras/stad918}

\bibitem[{{Little} \& {Tremaine}(1987)}]{1987ApJ...320..493L}
{Little}, B., \& {Tremaine}, S. 1987, The Astrophysical Journal, 320, 493, \dodoi{10.1086/165567}

\bibitem[{{Lodato} \& {Natarajan}(2006)}]{Lodato&Natarajan2006}
{Lodato}, G., \& {Natarajan}, P. 2006, \mnras, 371, 1813, \dodoi{10.1111/j.1365-2966.2006.10801.x}

\bibitem[{{Lodato} \& {Natarajan}(2007)}]{Lodato&Natarajan2007}
---. 2007, \mnras, 377, L64, \dodoi{10.1111/j.1745-3933.2007.00304.x}

\bibitem[{{Ma} {et~al.}(2014){Ma}, {Greene}, {McConnell}, {Janish}, {Blakeslee}, {Thomas}, \& {Murphy}}]{2014ApJ...795..158M}
{Ma}, C.-P., {Greene}, J.~E., {McConnell}, N., {et~al.} 2014, \apj, 795, 158, \dodoi{10.1088/0004-637X/795/2/158}

\bibitem[{{Magorrian} {et~al.}(1998){Magorrian}, {Tremaine}, {Richstone}, {Bender}, {Bower}, {Dressler}, {Faber}, {Gebhardt}, {Green}, {Grillmair}, {Kormendy}, \& {Lauer}}]{1998AJ....115.2285M}
{Magorrian}, J., {Tremaine}, S., {Richstone}, D., {et~al.} 1998, The Astronomical Journal, 115, 2285, \dodoi{10.1086/300353}

\bibitem[{{Maiolino} {et~al.}(2023){Maiolino}, {Scholtz}, {Curtis-Lake}, {Carniani}, {Baker}, {de Graaff}, {Tacchella}, {{\"U}bler}, {D'Eugenio}, {Witstok}, {Curti}, {Arribas}, {Bunker}, {Charlot}, {Chevallard}, {Eisenstein}, {Egami}, {Ji}, {Jones}, {Lyu}, {Rawle}, {Robertson}, {Rujopakarn}, {Perna}, {Sun}, {Venturi}, {Williams}, \& {Willott}}]{Maiolino+2023}
{Maiolino}, R., {Scholtz}, J., {Curtis-Lake}, E., {et~al.} 2023, arXiv e-prints, arXiv:2308.01230, \dodoi{10.48550/arXiv.2308.01230}

\bibitem[{{Marconi} \& {Hunt}(2003)}]{2003ApJ...589L..21M}
{Marconi}, A., \& {Hunt}, L.~K. 2003, The Astrophysical Journal, Letters, 589, L21, \dodoi{10.1086/375804}

\bibitem[{{Marconi} {et~al.}(2004){Marconi}, {Risaliti}, {Gilli}, {Hunt}, {Maiolino}, \& {Salvati}}]{2004MNRAS.351..169M}
{Marconi}, A., {Risaliti}, G., {Gilli}, R., {et~al.} 2004, Monthly Notices of the Royal Astronomical Society, 351, 169, \dodoi{10.1111/j.1365-2966.2004.07765.x}

\bibitem[{{Marian} {et~al.}(2019){Marian}, {Jahnke}, {Mechtley}, {Cohen}, {Husemann}, {Jones}, {Koekemoer}, {Schulze}, {van der Wel}, {Villforth}, \& {Windhorst}}]{2019ApJ...882..141M}
{Marian}, V., {Jahnke}, K., {Mechtley}, M., {et~al.} 2019, \apj, 882, 141, \dodoi{10.3847/1538-4357/ab385b}

\bibitem[{Marinacci {et~al.}(2018)Marinacci, Vogelsberger, Pakmor, Torrey, Springel, Hernquist, Nelson, Weinberger, Pillepich, Naiman, \& Genel}]{Marinacci2018FirstFields}
Marinacci, F., Vogelsberger, M., Pakmor, R., {et~al.} 2018, Monthly Notices of the Royal Astronomical Society, 480, 5113, \dodoi{10.1093/mnras/sty2206}

\bibitem[{{Martizzi} {et~al.}(2020){Martizzi}, {Vogelsberger}, {Torrey}, {Pillepich}, {Hansen}, {Marinacci}, \& {Hernquist}}]{2020MNRAS.491.5747M}
{Martizzi}, D., {Vogelsberger}, M., {Torrey}, P., {et~al.} 2020, Monthly Notices of the Royal Astronomical Society, 491, 5747, \dodoi{10.1093/mnras/stz3418}

\bibitem[{{Marulli} {et~al.}(2007){Marulli}, {Branchini}, {Moscardini}, \& {Volonteri}}]{2007MNRAS.375..649M}
{Marulli}, F., {Branchini}, E., {Moscardini}, L., \& {Volonteri}, M. 2007, Monthly Notices of the Royal Astronomical Society, 375, 649, \dodoi{10.1111/j.1365-2966.2006.11329.x}

\bibitem[{{Mathews}(1978)}]{1978ApJ...219..413M}
{Mathews}, W.~G. 1978, The Astrophysical Journal, 219, 413, \dodoi{10.1086/155794}

\bibitem[{{Matthee} {et~al.}(2024){Matthee}, {Naidu}, {Brammer}, {Chisholm}, {Eilers}, {Goulding}, {Greene}, {Kashino}, {Labbe}, {Lilly}, {Mackenzie}, {Oesch}, {Weibel}, {Wuyts}, {Xiao}, {Bordoloi}, {Bouwens}, {van Dokkum}, {Illingworth}, {Kramarenko}, {Maseda}, {Mason}, {Meyer}, {Nelson}, {Reddy}, {Shivaei}, {Simcoe}, \& {Yue}}]{2024ApJ...963..129M}
{Matthee}, J., {Naidu}, R.~P., {Brammer}, G., {et~al.} 2024, \apj, 963, 129, \dodoi{10.3847/1538-4357/ad2345}

\bibitem[{{McAlpine} {et~al.}(2016){McAlpine}, {Helly}, {Schaller}, {Trayford}, {Qu}, {Furlong}, {Bower}, {Crain}, {Schaye}, {Theuns}, {Dalla Vecchia}, {Frenk}, {McCarthy}, {Jenkins}, {Rosas-Guevara}, {White}, {Baes}, {Camps}, \& {Lemson}}]{2016A&C....15...72M}
{McAlpine}, S., {Helly}, J.~C., {Schaller}, M., {et~al.} 2016, Astronomy and Computing, 15, 72, \dodoi{10.1016/j.ascom.2016.02.004}

\bibitem[{{McConnell} \& {Ma}(2013)}]{2013ApJ...764..184M}
{McConnell}, N.~J., \& {Ma}, C.-P. 2013, The Astrophysical Journal, 764, 184, \dodoi{10.1088/0004-637X/764/2/184}

\bibitem[{{McLure} \& {Dunlop}(2001)}]{2001MNRAS.327..199M}
{McLure}, R.~J., \& {Dunlop}, J.~S. 2001, Monthly Notices of the Royal Astronomical Society, 327, 199, \dodoi{10.1046/j.1365-8711.2001.04709.x}

\bibitem[{{McLure} \& {Dunlop}(2002)}]{2002MNRAS.331..795M}
---. 2002, Monthly Notices of the Royal Astronomical Society, 331, 795, \dodoi{10.1046/j.1365-8711.2002.05236.x}

\bibitem[{{Merloni}(2004)}]{2004MNRAS.353.1035M}
{Merloni}, A. 2004, Monthly Notices of the Royal Astronomical Society, 353, 1035, \dodoi{10.1111/j.1365-2966.2004.08147.x}

\bibitem[{{Merloni} \& {Heinz}(2008)}]{2008MNRAS.388.1011M}
{Merloni}, A., \& {Heinz}, S. 2008, Monthly Notices of the Royal Astronomical Society, 388, 1011, \dodoi{10.1111/j.1365-2966.2008.13472.x}

\bibitem[{{Merritt} \& {Ferrarese}(2001)}]{2001ApJ...547..140M}
{Merritt}, D., \& {Ferrarese}, L. 2001, The Astrophysical Journal, 547, 140, \dodoi{10.1086/318372}

\bibitem[{{Mezcua} {et~al.}(2018){Mezcua}, {Hlavacek-Larrondo}, {Lucey}, {Hogan}, {Edge}, \& {McNamara}}]{Mezcua+2018}
{Mezcua}, M., {Hlavacek-Larrondo}, J., {Lucey}, J.~R., {et~al.} 2018, \mnras, 474, 1342, \dodoi{10.1093/mnras/stx2812}

\bibitem[{{Mezcua} {et~al.}(2024){Mezcua}, {Pacucci}, {Suh}, {Siudek}, \& {Natarajan}}]{2024ApJ...966L..30M}
{Mezcua}, M., {Pacucci}, F., {Suh}, H., {Siudek}, M., \& {Natarajan}, P. 2024, \apjl, 966, L30, \dodoi{10.3847/2041-8213/ad3c2a}

\bibitem[{Mo {et~al.}(1998)Mo, Mao, \& White}]{Mo1998}
Mo, H.~J., Mao, S., \& White, S. D.~M. 1998, Monthly Notices of the Royal Astronomical Society, 295, 319, \dodoi{10.1046/j.1365-8711.1998.01227.x}

\bibitem[{{Moster} {et~al.}(2013){Moster}, {Naab}, \& {White}}]{2013MNRAS.428.3121M}
{Moster}, B.~P., {Naab}, T., \& {White}, S. D.~M. 2013, \mnras, 428, 3121, \dodoi{10.1093/mnras/sts261}

\bibitem[{{Moster} {et~al.}(2010){Moster}, {Somerville}, {Maulbetsch}, {van den Bosch}, {Macci{\`o}}, {Naab}, \& {Oser}}]{2010ApJ...710..903M}
{Moster}, B.~P., {Somerville}, R.~S., {Maulbetsch}, C., {et~al.} 2010, The Astrophysical Journal, 710, 903, \dodoi{10.1088/0004-637X/710/2/903}

\bibitem[{{Moustakas} {et~al.}(2013){Moustakas}, {Coil}, {Aird}, {Blanton}, {Cool}, {Eisenstein}, {Mendez}, {Wong}, {Zhu}, \& {Arnouts}}]{2013ApJ...767...50M}
{Moustakas}, J., {Coil}, A.~L., {Aird}, J., {et~al.} 2013, \apj, 767, 50, \dodoi{10.1088/0004-637X/767/1/50}

\bibitem[{{Murray} {et~al.}(2005){Murray}, {Quataert}, \& {Thompson}}]{2005ApJ...618..569M}
{Murray}, N., {Quataert}, E., \& {Thompson}, T.~A. 2005, The Astrophysical Journal, 618, 569, \dodoi{10.1086/426067}

\bibitem[{{Naab} \& {Ostriker}(2017)}]{2017ARA&A..55...59N}
{Naab}, T., \& {Ostriker}, J.~P. 2017, \araa, 55, 59, \dodoi{10.1146/annurev-astro-081913-040019}

\bibitem[{Naiman {et~al.}(2018)Naiman, Pillepich, Springel, Ramirez-Ruiz, Torrey, Vogelsberger, Pakmor, Nelson, Marinacci, Hernquist, Weinberger, \& Genel}]{Naiman2018FirstEuropium}
Naiman, J.~P., Pillepich, A., Springel, V., {et~al.} 2018, Monthly Notices of the Royal Astronomical Society, 477, 1206, \dodoi{10.1093/mnras/sty618}

\bibitem[{{Natarajan} \& {Treister}(2009)}]{2009MNRAS.393..838N}
{Natarajan}, P., \& {Treister}, E. 2009, Monthly Notices of the Royal Astronomical Society, 393, 838, \dodoi{10.1111/j.1365-2966.2008.13864.x}

\bibitem[{{Natarajan} {et~al.}(2021){Natarajan}, {Tang}, {McGibbon}, {Khochfar}, {Nord}, {Sigurdsson}, {Tricot}, {Cappelluti}, {George}, \& {Hidary}}]{Natarajan+2021}
{Natarajan}, P., {Tang}, K.~S., {McGibbon}, R., {et~al.} 2021, arXiv e-prints, arXiv:2103.13932, \dodoi{10.48550/arXiv.2103.13932}

\bibitem[{Nelson {et~al.}(2015)Nelson, Pillepich, Genel, Vogelsberger, Springel, Torrey, Rodriguez-Gomez, Sijacki, Snyder, Griffen, Marinacci, Blecha, Sales, Xu, \& Hernquist}]{Nelson2015}
Nelson, D., Pillepich, A., Genel, S., {et~al.} 2015, Astronomy and Computing, 13, 12, \dodoi{10.1016/j.ascom.2015.09.003}

\bibitem[{Nelson {et~al.}(2018)Nelson, Pillepich, Springel, Weinberger, Hernquist, Pakmor, Genel, Torrey, Vogelsberger, Kauffmann, Marinacci, \& Naiman}]{Nelson2018FirstBimodality}
Nelson, D., Pillepich, A., Springel, V., {et~al.} 2018, Monthly Notices of the Royal Astronomical Society, 475, 624, \dodoi{10.1093/mnras/stx3040}

\bibitem[{{Nelson} {et~al.}(2019){Nelson}, {Pillepich}, {Springel}, {Pakmor}, {Weinberger}, {Genel}, {Torrey}, {Vogelsberger}, {Marinacci}, \& {Hernquist}}]{2019MNRAS.490.3234N}
{Nelson}, D., {Pillepich}, A., {Springel}, V., {et~al.} 2019, Monthly Notices of the Royal Astronomical Society, 490, 3234, \dodoi{10.1093/mnras/stz2306}

\bibitem[{{Ni} {et~al.}(2022){Ni}, {Di Matteo}, {Bird}, {Croft}, {Feng}, {Chen}, {Tremmel}, {DeGraf}, \& {Li}}]{2022MNRAS.513..670N}
{Ni}, Y., {Di Matteo}, T., {Bird}, S., {et~al.} 2022, \mnras, 513, 670, \dodoi{10.1093/mnras/stac351}

\bibitem[{{Nikolajuk} {et~al.}(2004){Nikolajuk}, {Papadakis}, \& {Czerny}}]{2004MNRAS.350L..26N}
{Nikolajuk}, M., {Papadakis}, I.~E., \& {Czerny}, B. 2004, Monthly Notices of the Royal Astronomical Society, 350, L26, \dodoi{10.1111/j.1365-2966.2004.07829.x}

\bibitem[{{Ntampaka} {et~al.}(2019){Ntampaka}, {ZuHone}, {Eisenstein}, {Nagai}, {Vikhlinin}, {Hernquist}, {Marinacci}, {Nelson}, {Pakmor}, {Pillepich}, {Torrey}, \& {Vogelsberger}}]{2019ApJ...876...82N}
{Ntampaka}, M., {ZuHone}, J., {Eisenstein}, D., {et~al.} 2019, The Astrophysical Journal, 876, 82, \dodoi{10.3847/1538-4357/ab14eb}

\bibitem[{{Oliver} {et~al.}(1997){Oliver}, {Goldschmidt}, {Franceschini}, {Serjeant}, {Efstathiou}, {Verma}, {Gruppioni}, {Eaton}, {Mann}, {Mobasher}, {Pearson}, {Rowan-Robinson}, {Sumner}, {Danese}, {Elbaz}, {Egami}, {Kontizas}, {Lawrence}, {McMahon}, {Norgaard-Nielsen}, {Perez-Fournon}, \& {Gonzalez-Serrano}}]{1997MNRAS.289..471O}
{Oliver}, S.~J., {Goldschmidt}, P., {Franceschini}, A., {et~al.} 1997, \mnras, 289, 471, \dodoi{10.1093/mnras/289.2.471}

\bibitem[{{Pacucci} \& {Loeb}(2020)}]{2020ApJ...895...95P}
{Pacucci}, F., \& {Loeb}, A. 2020, The Astrophysical Journal, 895, 95, \dodoi{10.3847/1538-4357/ab886e}

\bibitem[{{Pacucci} {et~al.}(2023){Pacucci}, {Nguyen}, {Carniani}, {Maiolino}, \& {Fan}}]{2023ApJ...957L...3P}
{Pacucci}, F., {Nguyen}, B., {Carniani}, S., {Maiolino}, R., \& {Fan}, X. 2023, The Astrophysical Journal, Letters, 957, L3, \dodoi{10.3847/2041-8213/ad0158}

\bibitem[{{Parkinson} {et~al.}(2008){Parkinson}, {Cole}, \& {Helly}}]{Parkinson+2008}
{Parkinson}, H., {Cole}, S., \& {Helly}, J. 2008, \mnras, 383, 557, \dodoi{10.1111/j.1365-2966.2007.12517.x}

\bibitem[{{Pearson} {et~al.}(2010){Pearson}, {Oyabu}, {Wada}, {Matsuhara}, {Lee}, {Kim}, {Takagi}, {Goto}, {Im}, {Serjeant}, {Lee}, {Ko}, {White}, \& {Ohyama}}]{2010A&A...514A...8P}
{Pearson}, C.~P., {Oyabu}, S., {Wada}, T., {et~al.} 2010, \aap, 514, A8, \dodoi{10.1051/0004-6361/200913382}

\bibitem[{{Pearson} {et~al.}(2014){Pearson}, {Serjeant}, {Oyabu}, {Matsuhara}, {Wada}, {Goto}, {Takagi}, {Lee}, {Im}, {Ohyama}, {Kim}, \& {Murata}}]{2014MNRAS.444..846P}
{Pearson}, C.~P., {Serjeant}, S., {Oyabu}, S., {et~al.} 2014, \mnras, 444, 846, \dodoi{10.1093/mnras/stu1472}

\bibitem[{{Peeples} {et~al.}(2014){Peeples}, {Werk}, {Tumlinson}, {Oppenheimer}, {Prochaska}, {Katz}, \& {Weinberg}}]{2014ApJ...786...54P}
{Peeples}, M.~S., {Werk}, J.~K., {Tumlinson}, J., {et~al.} 2014, The Astrophysical Journal, 786, 54, \dodoi{10.1088/0004-637X/786/1/54}

\bibitem[{P{\'{e}}rez \& Granger(2007)}]{Perez2007IPythonFor}
P{\'{e}}rez, F., \& Granger, B.~E. 2007, IEEE Journals {\&} Magazines, 9, 21, \dodoi{10.1109/MCSE.2007.53}

\bibitem[{{Pesce} {et~al.}(2021){Pesce}, {Palumbo}, {Narayan}, {Blackburn}, {Doeleman}, {Johnson}, {Ma}, {Nagar}, {Natarajan}, \& {Ricarte}}]{2021ApJ...923..260P}
{Pesce}, D.~W., {Palumbo}, D. C.~M., {Narayan}, R., {et~al.} 2021, The Astrophysical Journal, 923, 260, \dodoi{10.3847/1538-4357/ac2eb5}

\bibitem[{{Piana} {et~al.}(2024){Piana}, {Pu}, \& {Wu}}]{2024MNRAS.530.1732P}
{Piana}, O., {Pu}, H.-Y., \& {Wu}, K. 2024, Monthly Notices of the Royal Astronomical Society, 530, 1732, \dodoi{10.1093/mnras/stae851}

\bibitem[{Pillepich {et~al.}(2018)Pillepich, Nelson, Hernquist, Springel, Pakmor, Torrey, Weinberger, Genel, Naiman, Marinacci, \& Vogelsberger}]{Pillepich2018FirstGalaxies}
Pillepich, A., Nelson, D., Hernquist, L., {et~al.} 2018, Monthly Notices of the Royal Astronomical Society, 475, 648, \dodoi{10.1093/mnras/stx3112}

\bibitem[{{Pillepich} {et~al.}(2018){Pillepich}, {Springel}, {Nelson}, {Genel}, {Naiman}, {Pakmor}, {Hernquist}, {Torrey}, {Vogelsberger}, {Weinberger}, \& {Marinacci}}]{2018MNRAS.473.4077P}
{Pillepich}, A., {Springel}, V., {Nelson}, D., {et~al.} 2018, Monthly Notices of the Royal Astronomical Society, 473, 4077, \dodoi{10.1093/mnras/stx2656}

\bibitem[{{Planck Collaboration} {et~al.}(2016){Planck Collaboration}, {Ade}, {Aghanim}, {Arnaud}, {Ashdown}, {Aumont}, {Baccigalupi}, {Banday}, {Barreiro}, {Bartlett}, {Bartolo}, {Battaner}, {Battye}, {Benabed}, {Beno{\^\i}t}, {Benoit-L{\'e}vy}, {Bernard}, {Bersanelli}, {Bielewicz}, {Bock}, {Bonaldi}, {Bonavera}, {Bond}, {Borrill}, {Bouchet}, {Boulanger}, {Bucher}, {Burigana}, {Butler}, {Calabrese}, {Cardoso}, {Catalano}, {Challinor}, {Chamballu}, {Chary}, {Chiang}, {Chluba}, {Christensen}, {Church}, {Clements}, {Colombi}, {Colombo}, {Combet}, {Coulais}, {Crill}, {Curto}, {Cuttaia}, {Danese}, {Davies}, {Davis}, {de Bernardis}, {de Rosa}, {de Zotti}, {Delabrouille}, {D{\'e}sert}, {Di Valentino}, {Dickinson}, {Diego}, {Dolag}, {Dole}, {Donzelli}, {Dor{\'e}}, {Douspis}, {Ducout}, {Dunkley}, {Dupac}, {Efstathiou}, {Elsner}, {En{\ss}lin}, {Eriksen}, {Farhang}, {Fergusson}, {Finelli}, {Forni}, {Frailis}, {Fraisse}, {Franceschi}, {Frejsel}, {Galeotta}, {Galli}, {Ganga}, {Gauthier}, {Gerbino}, {Ghosh}, {Giard},
  {Giraud-H{\'e}raud}, {Giusarma}, {Gjerl{\o}w}, {Gonz{\'a}lez-Nuevo}, {G{\'o}rski}, {Gratton}, {Gregorio}, {Gruppuso}, {Gudmundsson}, {Hamann}, {Hansen}, {Hanson}, {Harrison}, {Helou}, {Henrot-Versill{\'e}}, {Hern{\'a}ndez-Monteagudo}, {Herranz}, {Hildebrandt}, {Hivon}, {Hobson}, {Holmes}, {Hornstrup}, {Hovest}, {Huang}, {Huffenberger}, {Hurier}, {Jaffe}, {Jaffe}, {Jones}, {Juvela}, {Keih{\"a}nen}, {Keskitalo}, {Kisner}, {Kneissl}, {Knoche}, {Knox}, {Kunz}, {Kurki-Suonio}, {Lagache}, {L{\"a}hteenm{\"a}ki}, {Lamarre}, {Lasenby}, {Lattanzi}, {Lawrence}, {Leahy}, {Leonardi}, {Lesgourgues}, {Levrier}, {Lewis}, {Liguori}, {Lilje}, {Linden-V{\o}rnle}, {L{\'o}pez-Caniego}, {Lubin}, {Mac{\'\i}as-P{\'e}rez}, {Maggio}, {Maino}, {Mandolesi}, {Mangilli}, {Marchini}, {Maris}, {Martin}, {Martinelli}, {Mart{\'\i}nez-Gonz{\'a}lez}, {Masi}, {Matarrese}, {McGehee}, {Meinhold}, {Melchiorri}, {Melin}, {Mendes}, {Mennella}, {Migliaccio}, {Millea}, {Mitra}, {Miville-Desch{\^e}nes}, {Moneti}, {Montier}, {Morgante}, {Mortlock},
  {Moss}, {Munshi}, {Murphy}, {Naselsky}, {Nati}, {Natoli}, {Netterfield}, {N{\o}rgaard-Nielsen}, {Noviello}, {Novikov}, {Novikov}, {Oxborrow}, {Paci}, {Pagano}, {Pajot}, {Paladini}, {Paoletti}, {Partridge}, {Pasian}, {Patanchon}, {Pearson}, {Perdereau}, {Perotto}, {Perrotta}, {Pettorino}, {Piacentini}, {Piat}, {Pierpaoli}, {Pietrobon}, {Plaszczynski}, {Pointecouteau}, {Polenta}, {Popa}, {Pratt}, {Pr{\'e}zeau}, {Prunet}, {Puget}, {Rachen}, {Reach}, {Rebolo}, {Reinecke}, {Remazeilles}, {Renault}, {Renzi}, {Ristorcelli}, {Rocha}, {Rosset}, {Rossetti}, {Roudier}, {Rouill{\'e} d'Orfeuil}, {Rowan-Robinson}, {Rubi{\~n}o-Mart{\'\i}n}, {Rusholme}, {Said}, {Salvatelli}, {Salvati}, {Sandri}, {Santos}, {Savelainen}, {Savini}, {Scott}, {Seiffert}, {Serra}, {Shellard}, {Spencer}, {Spinelli}, {Stolyarov}, {Stompor}, {Sudiwala}, {Sunyaev}, {Sutton}, {Suur-Uski}, {Sygnet}, {Tauber}, {Terenzi}, {Toffolatti}, {Tomasi}, {Tristram}, {Trombetti}, {Tucci}, {Tuovinen}, {T{\"u}rler}, {Umana}, {Valenziano}, {Valiviita}, {Van Tent},
  {Vielva}, {Villa}, {Wade}, {Wandelt}, {Wehus}, {White}, {White}, {Wilkinson}, {Yvon}, {Zacchei}, \& {Zonca}}]{2016A&A...594A..13P}
{Planck Collaboration}, {Ade}, P.~A.~R., {Aghanim}, N., {et~al.} 2016, Astronomy and Astrophysics, 594, A13, \dodoi{10.1051/0004-6361/201525830}

\bibitem[{{Porter} {et~al.}(2014){Porter}, {Somerville}, {Primack}, \& {Johansson}}]{2014MNRAS.444..942P}
{Porter}, L.~A., {Somerville}, R.~S., {Primack}, J.~R., \& {Johansson}, P.~H. 2014, Monthly Notices of the Royal Astronomical Society, 444, 942, \dodoi{10.1093/mnras/stu1434}

\bibitem[{{Press} \& {Schechter}(1974)}]{Press&Schechter1974}
{Press}, W.~H., \& {Schechter}, P. 1974, \apj, 187, 425, \dodoi{10.1086/152650}

\bibitem[{{Ramakrishnan} {et~al.}(2023){Ramakrishnan}, {Nagar}, {Arratia}, {Hern{\'a}ndez-Y{\'e}venes}, {Pesce}, {Nair}, {Bandyopadhyay}, {Medina-Porcile}, {Krichbaum}, {Doeleman}, {Ricarte}, {Fish}, {Blackburn}, {Falcke}, {Bower}, \& {Natarajan}}]{2023Galax..11...15R}
{Ramakrishnan}, V., {Nagar}, N., {Arratia}, V., {et~al.} 2023, Galaxies, 11, 15, \dodoi{10.3390/galaxies11010015}

\bibitem[{{Reines} \& {Volonteri}(2015)}]{Reines&Volonteri2015}
{Reines}, A.~E., \& {Volonteri}, M. 2015, \apj, 813, 82, \dodoi{10.1088/0004-637X/813/2/82}

\bibitem[{{Reynolds}(2021)}]{Reynolds+2021}
{Reynolds}, C.~S. 2021, \araa, 59, 117, \dodoi{10.1146/annurev-astro-112420-035022}

\bibitem[{{Ricarte} \& {Natarajan}(2018)}]{Ricarte+2018a}
{Ricarte}, A., \& {Natarajan}, P. 2018, Monthly Notices of the Royal Astronomical Society, 474, 1995, \dodoi{10.1093/mnras/stx2851}

\bibitem[{{Ricarte} {et~al.}(2023){Ricarte}, {Tiede}, {Emami}, {Tamar}, \& {Natarajan}}]{Ricarte+2023}
{Ricarte}, A., {Tiede}, P., {Emami}, R., {Tamar}, A., \& {Natarajan}, P. 2023, Galaxies, 11, 6, \dodoi{10.3390/galaxies11010006}

\bibitem[{{Ricarte} {et~al.}(2021){Ricarte}, {Tremmel}, {Natarajan}, {Zimmer}, \& {Quinn}}]{Ricarte+2021}
{Ricarte}, A., {Tremmel}, M., {Natarajan}, P., {Zimmer}, C., \& {Quinn}, T. 2021, Monthly Notices of the Royal Astronomical Society, 503, 6098, \dodoi{10.1093/mnras/stab866}

\bibitem[{{Robertson} {et~al.}(2006){Robertson}, {Cox}, {Hernquist}, {Franx}, {Hopkins}, {Martini}, \& {Springel}}]{2006ApJ...641...21R}
{Robertson}, B., {Cox}, T.~J., {Hernquist}, L., {et~al.} 2006, The Astrophysical Journal, 641, 21, \dodoi{10.1086/500360}

\bibitem[{Robitaille {et~al.}(2013)Robitaille, Tollerud, Greenfield, Droettboom, Bray, Aldcroft, Davis, Ginsburg, Price-Whelan, Kerzendorf, Conley, Crighton, Barbary, Muna, Ferguson, Grollier, Parikh, Nair, G{\"{u}}nther, Deil, Woillez, Conseil, Kramer, Turner, Singer, Fox, Weaver, Zabalza, Edwards, Azalee~Bostroem, Burke, Casey, Crawford, Dencheva, Ely, Jenness, Labrie, Lim, Pierfederici, Pontzen, Ptak, Refsdal, Servillat, \& Streicher}]{Robitaille2013Astropy:Astronomy}
Robitaille, T.~P., Tollerud, E.~J., Greenfield, P., {et~al.} 2013, Astronomy and Astrophysics, 558, 33, \dodoi{10.1051/0004-6361/201322068}

\bibitem[{{Rocca-Volmerange} {et~al.}(2007){Rocca-Volmerange}, {de Lapparent}, {Seymour}, \& {Fioc}}]{2007A&A...475..801R}
{Rocca-Volmerange}, B., {de Lapparent}, V., {Seymour}, N., \& {Fioc}, M. 2007, \aap, 475, 801, \dodoi{10.1051/0004-6361:20065217}

\bibitem[{{Rodr{\'\i}guez-Puebla} {et~al.}(2017){Rodr{\'\i}guez-Puebla}, {Primack}, {Avila-Reese}, \& {Faber}}]{2017MNRAS.470..651R}
{Rodr{\'\i}guez-Puebla}, A., {Primack}, J.~R., {Avila-Reese}, V., \& {Faber}, S.~M. 2017, Monthly Notices of the Royal Astronomical Society, 470, 651, \dodoi{10.1093/mnras/stx1172}

\bibitem[{{Saglia} {et~al.}(2016){Saglia}, {Opitsch}, {Erwin}, {Thomas}, {Beifiori}, {Fabricius}, {Mazzalay}, {Nowak}, {Rusli}, \& {Bender}}]{Saglia+2016}
{Saglia}, R.~P., {Opitsch}, M., {Erwin}, P., {et~al.} 2016, \apj, 818, 47, \dodoi{10.3847/0004-637X/818/1/47}

\bibitem[{{Sahu} {et~al.}(2019){Sahu}, {Graham}, \& {Davis}}]{2019ApJ...876..155S}
{Sahu}, N., {Graham}, A.~W., \& {Davis}, B.~L. 2019, The Astrophysical Journal, 876, 155, \dodoi{10.3847/1538-4357/ab0f32}

\bibitem[{{Sales} {et~al.}(2020){Sales}, {Navarro}, {Pe{\~n}afiel}, {Peng}, {Lim}, \& {Hernquist}}]{2020MNRAS.494.1848S}
{Sales}, L.~V., {Navarro}, J.~F., {Pe{\~n}afiel}, L., {et~al.} 2020, Monthly Notices of the Royal Astronomical Society, 494, 1848, \dodoi{10.1093/mnras/staa854}

\bibitem[{{Salucci} {et~al.}(1999){Salucci}, {Szuszkiewicz}, {Monaco}, \& {Danese}}]{1999MNRAS.307..637S}
{Salucci}, P., {Szuszkiewicz}, E., {Monaco}, P., \& {Danese}, L. 1999, Monthly Notices of the Royal Astronomical Society, 307, 637, \dodoi{10.1046/j.1365-8711.1999.02659.x}

\bibitem[{{Sanchez} {et~al.}(2018){Sanchez}, {Bellovary}, {Holley-Bockelmann}, {Tremmel}, {Brooks}, {Governato}, {Quinn}, {Volonteri}, \& {Wadsley}}]{2018ApJ...860...20S}
{Sanchez}, N.~N., {Bellovary}, J.~M., {Holley-Bockelmann}, K., {et~al.} 2018, The Astrophysical Journal, 860, 20, \dodoi{10.3847/1538-4357/aac015}

\bibitem[{{Sato-Polito} {et~al.}(2023){Sato-Polito}, {Zaldarriaga}, \& {Quataert}}]{Sato-Polito+2023}
{Sato-Polito}, G., {Zaldarriaga}, M., \& {Quataert}, E. 2023, arXiv e-prints, arXiv:2312.06756, \dodoi{10.48550/arXiv.2312.06756}

\bibitem[{{Savorgnan} {et~al.}(2016){Savorgnan}, {Graham}, {Marconi}, \& {Sani}}]{2016ApJ...817...21S}
{Savorgnan}, G. A.~D., {Graham}, A.~W., {Marconi}, A., \& {Sani}, E. 2016, The Astrophysical Journal, 817, 21, \dodoi{10.3847/0004-637X/817/1/21}

\bibitem[{{Schutte} {et~al.}(2019){Schutte}, {Reines}, \& {Greene}}]{2019ApJ...887..245S}
{Schutte}, Z., {Reines}, A.~E., \& {Greene}, J.~E. 2019, \apj, 887, 245, \dodoi{10.3847/1538-4357/ab35dd}

\bibitem[{Scott {et~al.}(2013)Scott, Graham, \& Schombert}]{Scott2013THEGALAXIES}
Scott, N., Graham, A.~W., \& Schombert, J. 2013, The Astrophysical Journal, 768, 76, \dodoi{10.1088/0004-637X/768/1/76}

\bibitem[{{Serjeant} {et~al.}(2000){Serjeant}, {Oliver}, {Rowan-Robinson}, {Crockett}, {Missoulis}, {Sumner}, {Gruppioni}, {Mann}, {Eaton}, {Elbaz}, {Clements}, {Baker}, {Efstathiou}, {Cesarsky}, {Danese}, {Franceschini}, {Genzel}, {Lawrence}, {Lemke}, {McMahon}, {Miley}, {Puget}, \& {Rocca-Volmerange}}]{2000MNRAS.316..768S}
{Serjeant}, S., {Oliver}, S., {Rowan-Robinson}, M., {et~al.} 2000, \mnras, 316, 768, \dodoi{10.1046/j.1365-8711.2000.03551.x}

\bibitem[{{Shankar} {et~al.}(2009){Shankar}, {Weinberg}, \& {Miralda-Escud{\'e}}}]{2009ApJ...690...20S}
{Shankar}, F., {Weinberg}, D.~H., \& {Miralda-Escud{\'e}}, J. 2009, The Astrophysical Journal, 690, 20, \dodoi{10.1088/0004-637X/690/1/20}

\bibitem[{{Shankar} {et~al.}(2013){Shankar}, {Weinberg}, \& {Miralda-Escud{\'e}}}]{2013MNRAS.428..421S}
---. 2013, Monthly Notices of the Royal Astronomical Society, 428, 421, \dodoi{10.1093/mnras/sts026}

\bibitem[{{Sharma} {et~al.}(2024){Sharma}, {Choi}, {Somerville}, {Snyder}, {Jhee}, {Kocevski}, {Hirschmann}, {Moster}, {Naab}, {Narayanan}, {Ostriker}, \& {Rosario}}]{2024MNRAS.527.9461S}
{Sharma}, R.~S., {Choi}, E., {Somerville}, R.~S., {et~al.} 2024, \mnras, 527, 9461, \dodoi{10.1093/mnras/stad3836}

\bibitem[{{Shaya} {et~al.}(2017){Shaya}, {Tully}, {Hoffman}, \& {Pomar{\`e}de}}]{2017ApJ...850..207S}
{Shaya}, E.~J., {Tully}, R.~B., {Hoffman}, Y., \& {Pomar{\`e}de}, D. 2017, The Astrophysical Journal, 850, 207, \dodoi{10.3847/1538-4357/aa9525}

\bibitem[{{Shen} {et~al.}(2022{\natexlab{a}}){Shen}, {Eadie}, {Murray}, {Zaritsky}, {Speagle}, {Ting}, {Conroy}, {Cargile}, {Johnson}, {Naidu}, \& {Han}}]{2022ApJ...925....1S}
{Shen}, J., {Eadie}, G.~M., {Murray}, N., {et~al.} 2022{\natexlab{a}}, The Astrophysical Journal, 925, 1, \dodoi{10.3847/1538-4357/ac3a7a}

\bibitem[{{Shen} {et~al.}(2020{\natexlab{a}}){Shen}, {Hopkins}, {Faucher-Gigu{\`e}re}, {Alexander}, {Richards}, {Ross}, \& {Hickox}}]{2020MNRAS.495.3252S}
{Shen}, X., {Hopkins}, P.~F., {Faucher-Gigu{\`e}re}, C.-A., {et~al.} 2020{\natexlab{a}}, \mnras, 495, 3252, \dodoi{10.1093/mnras/staa1381}

\bibitem[{{Shen} {et~al.}(2022{\natexlab{b}}){Shen}, {Vogelsberger}, {Nelson}, {Tacchella}, {Hernquist}, {Springel}, {Marinacci}, \& {Torrey}}]{2022MNRAS.510.5560S}
{Shen}, X., {Vogelsberger}, M., {Nelson}, D., {et~al.} 2022{\natexlab{b}}, Monthly Notices of the Royal Astronomical Society, 510, 5560, \dodoi{10.1093/mnras/stab3794}

\bibitem[{{Shen} {et~al.}(2020{\natexlab{b}}){Shen}, {Vogelsberger}, {Nelson}, {Pillepich}, {Tacchella}, {Marinacci}, {Torrey}, {Hernquist}, \& {Springel}}]{2020MNRAS.495.4747S}
---. 2020{\natexlab{b}}, Monthly Notices of the Royal Astronomical Society, 495, 4747, \dodoi{10.1093/mnras/staa1423}

\bibitem[{{Sijacki} {et~al.}(2015){Sijacki}, {Vogelsberger}, {Genel}, {Springel}, {Torrey}, {Snyder}, {Nelson}, \& {Hernquist}}]{2015MNRAS.452..575S}
{Sijacki}, D., {Vogelsberger}, M., {Genel}, S., {et~al.} 2015, Monthly Notices of the Royal Astronomical Society, 452, 575, \dodoi{10.1093/mnras/stv1340}

\bibitem[{{Silk} {et~al.}(2024){Silk}, {Begelman}, {Norman}, {Nusser}, \& {Wyse}}]{2024ApJ...961L..39S}
{Silk}, J., {Begelman}, M.~C., {Norman}, C., {Nusser}, A., \& {Wyse}, R. F.~G. 2024, \apjl, 961, L39, \dodoi{10.3847/2041-8213/ad1bf0}

\bibitem[{{Silk} \& {Rees}(1998)}]{1998A&A...331L...1S}
{Silk}, J., \& {Rees}, M.~J. 1998, \aap, 331, L1, \dodoi{10.48550/arXiv.astro-ph/9801013}

\bibitem[{{S{\k{a}}dowski}(2009)}]{2009ApJS..183..171S}
{S{\k{a}}dowski}, A. 2009, Astrophysical Journal, Supplement Series, 183, 171, \dodoi{10.1088/0067-0049/183/2/171}

\bibitem[{{S{\k{a}}dowski} {et~al.}(2016){S{\k{a}}dowski}, {Lasota}, {Abramowicz}, \& {Narayan}}]{2016MNRAS.456.3915S}
{S{\k{a}}dowski}, A., {Lasota}, J.-P., {Abramowicz}, M.~A., \& {Narayan}, R. 2016, Monthly Notices of the Royal Astronomical Society, 456, 3915, \dodoi{10.1093/mnras/stv2854}

\bibitem[{{S{\k{a}}dowski} \& {Narayan}(2016)}]{2016MNRAS.456.3929S}
{S{\k{a}}dowski}, A., \& {Narayan}, R. 2016, Monthly Notices of the Royal Astronomical Society, 456, 3929, \dodoi{10.1093/mnras/stv2941}

\bibitem[{{S{\k{a}}dowski} {et~al.}(2014){S{\k{a}}dowski}, {Narayan}, {McKinney}, \& {Tchekhovskoy}}]{2014MNRAS.439..503S}
{S{\k{a}}dowski}, A., {Narayan}, R., {McKinney}, J.~C., \& {Tchekhovskoy}, A. 2014, Monthly Notices of the Royal Astronomical Society, 439, 503, \dodoi{10.1093/mnras/stt2479}

\bibitem[{{Small} \& {Blandford}(1992)}]{1992MNRAS.259..725S}
{Small}, T.~A., \& {Blandford}, R.~D. 1992, Monthly Notices of the Royal Astronomical Society, 259, 725, \dodoi{10.1093/mnras/259.4.725}

\bibitem[{{Somerville} \& {Dav{\'e}}(2015)}]{2015ARA&A..53...51S}
{Somerville}, R.~S., \& {Dav{\'e}}, R. 2015, Annual Review of Astronomy \& Astrophysics, 53, 51, \dodoi{10.1146/annurev-astro-082812-140951}

\bibitem[{{Somerville} {et~al.}(2008){Somerville}, {Hopkins}, {Cox}, {Robertson}, \& {Hernquist}}]{2008MNRAS.391..481S}
{Somerville}, R.~S., {Hopkins}, P.~F., {Cox}, T.~J., {Robertson}, B.~E., \& {Hernquist}, L. 2008, Monthly Notices of the Royal Astronomical Society, 391, 481, \dodoi{10.1111/j.1365-2966.2008.13805.x}

\bibitem[{{Somerville} {et~al.}(2015){Somerville}, {Popping}, \& {Trager}}]{2015MNRAS.453.4337S}
{Somerville}, R.~S., {Popping}, G., \& {Trager}, S.~C. 2015, Monthly Notices of the Royal Astronomical Society, 453, 4337, \dodoi{10.1093/mnras/stv1877}

\bibitem[{{Somerville} \& {Primack}(1999)}]{1999MNRAS.310.1087S}
{Somerville}, R.~S., \& {Primack}, J.~R. 1999, Monthly Notices of the Royal Astronomical Society, 310, 1087, \dodoi{10.1046/j.1365-8711.1999.03032.x}

\bibitem[{Somerville {et~al.}(2001)Somerville, Primack, \& Faber}]{Somerville2001TheGalaxies}
Somerville, R.~S., Primack, J.~R., \& Faber, S.~M. 2001, Monthly Notices of the Royal Astronomical Society, 320, 504

\bibitem[{{Somerville} {et~al.}(2021){Somerville}, {Olsen}, {Yung}, {Pacifici}, {Ferguson}, {Behroozi}, {Osborne}, {Wechsler}, {Pandya}, {Faber}, {Primack}, \& {Dekel}}]{2021MNRAS.502.4858S}
{Somerville}, R.~S., {Olsen}, C., {Yung}, L.~Y.~A., {et~al.} 2021, Monthly Notices of the Royal Astronomical Society, 502, 4858, \dodoi{10.1093/mnras/stab231}

\bibitem[{Spergel {et~al.}(2003)Spergel, Verde, Peiris, Komatsu, Nolta, Bennett, Halpern, Hinshaw, Jarosik, Kogut, Limon, Meyer, Page, Tucker, Weiland, Wollack, \& Wright}]{Spergel2003}
Spergel, D.~N., Verde, L., Peiris, H.~V., {et~al.} 2003, The Astrophysical Journal Supplement Series, 148, 175, \dodoi{10.1086/377226}

\bibitem[{Springel(2005)}]{Springel2005}
Springel, V. 2005, Monthly Notices of the Royal Astronomical Society, 364, 1105, \dodoi{10.1111/j.1365-2966.2005.09655.x}

\bibitem[{Springel {et~al.}(2001)Springel, White, Tormen, \& Kauffmann}]{Springel2001}
Springel, V., White, S.~D., Tormen, G., \& Kauffmann, G. 2001, Monthly Notices of the Royal Astronomical Society, 328, 726, \dodoi{10.1046/j.1365-8711.2001.04912.x}

\bibitem[{Springel {et~al.}(2005)Springel, White, Jenkins, Frenk, Yoshida, Gao, Navarro, Thacker, Croton, Helly, Peacock, Cole, Thomas, Couchman, Evrard, Colberg, \& Pearce}]{Springel2005Nat}
Springel, V., White, S.~D., Jenkins, A., {et~al.} 2005, Nature, 435, 629, \dodoi{10.1038/nature03597}

\bibitem[{Springel {et~al.}(2018)Springel, Pakmor, Pillepich, Weinberger, Nelson, Hernquist, Vogelsberger, Genel, Torrey, Marinacci, \& Naiman}]{Springel2018FirstClustering}
Springel, V., Pakmor, R., Pillepich, A., {et~al.} 2018, Monthly Notices of the Royal Astronomical Society, 475, 676, \dodoi{10.1093/mnras/stx3304}

\bibitem[{Stark {et~al.}(2009)Stark, Mcgaugh, \& Swaters}]{Stark2009AGALAXIES}
Stark, D.~V., Mcgaugh, S.~S., \& Swaters, R.~A. 2009, The Astronomical Journal, 138, 392, \dodoi{10.1088/0004-6256/138/2/392}

\bibitem[{Stevens {et~al.}(2016)Stevens, Croton, \& Mutch}]{Stevens2016}
Stevens, A. R.~H., Croton, D.~J., \& Mutch, S.~J. 2016, Monthly Notices of the Royal Astronomical Society, 461, 859, \dodoi{10.1093/mnras/stw1332}

\bibitem[{Stevens {et~al.}(2018)Stevens, Lagos, Obreschkow, \& Sinha}]{Stevens2018ConnectingSAGE}
Stevens, A. R.~H., Lagos, C. D.~P., Obreschkow, D., \& Sinha, M. 2018, Monthly Notices of the Royal Astronomical Society, 481, 5543, \dodoi{10.1093/MNRAS/STY2650}

\bibitem[{{Stevens} {et~al.}(2024){Stevens}, {Sinha}, {Rohl}, {Sammons}, {Hadzhiyska}, {Hern{\'a}ndez-Aguayo}, \& {Hernquist}}]{2024PASA...41...53S}
{Stevens}, A. R.~H., {Sinha}, M., {Rohl}, A., {et~al.} 2024, \pasa, 41, e053, \dodoi{10.1017/pasa.2024.14}

\bibitem[{{Sturm} \& {Reines}(2024)}]{Sturm&Reines2024}
{Sturm}, M.~R., \& {Reines}, A.~E. 2024, arXiv e-prints, arXiv:2406.06675, \dodoi{10.48550/arXiv.2406.06675}

\bibitem[{{Taylor} {et~al.}(2024){Taylor}, {Finkelstein}, {Kocevski}, {Jeon}, {Bromm}, {Amorin}, {Arrabal Haro}, {Backhaus}, {Bagley}, {Ba{\~n}ados}, {Bhatawdekar}, {Brooks}, {Calabro}, {Chavez Ortiz}, {Cheng}, {Cleri}, {Cole}, {Davis}, {Dickinson}, {Donnan}, {Dunlop}, {Ellis}, {Fernandez}, {Fontana}, {Fujimoto}, {Giavalisco}, {Grazian}, {Guo}, {Hathi}, {Holwerda}, {Hirschmann}, {Inayoshi}, {Kartaltepe}, {Khusanova}, {Koekemoer}, {Kokorev}, {Larson}, {Leung}, {Lucas}, {McLeod}, {Napolitano}, {Onoue}, {Pacucci}, {Papovich}, {P{\'e}rez-Gonz{\'a}lez}, {Pirzkal}, {Somerville}, {Trump}, {Wilkins}, {Yung}, \& {Zhang}}]{2024arXiv240906772T}
{Taylor}, A.~J., {Finkelstein}, S.~L., {Kocevski}, D.~D., {et~al.} 2024, arXiv e-prints, arXiv:2409.06772, \dodoi{10.48550/arXiv.2409.06772}

\bibitem[{{Terrazas} {et~al.}(2020){Terrazas}, {Bell}, {Pillepich}, {Nelson}, {Somerville}, {Genel}, {Weinberger}, {Habouzit}, {Li}, {Hernquist}, \& {Vogelsberger}}]{2020MNRAS.493.1888T}
{Terrazas}, B.~A., {Bell}, E.~F., {Pillepich}, A., {et~al.} 2020, Monthly Notices of the Royal Astronomical Society, 493, 1888, \dodoi{10.1093/mnras/staa374}

\bibitem[{{Toomre}(1964)}]{Toomre1964}
{Toomre}, A. 1964, The Astrophysical Journal, 139, 1217, \dodoi{10.1086/147861}

\bibitem[{{Tremaine} {et~al.}(2002){Tremaine}, {Gebhardt}, {Bender}, {Bower}, {Dressler}, {Faber}, {Filippenko}, {Green}, {Grillmair}, {Ho}, {Kormendy}, {Lauer}, {Magorrian}, {Pinkney}, \& {Richstone}}]{2002ApJ...574..740T}
{Tremaine}, S., {Gebhardt}, K., {Bender}, R., {et~al.} 2002, The Astrophysical Journal, 574, 740, \dodoi{10.1086/341002}

\bibitem[{Tremonti {et~al.}(2004)Tremonti, Heckman, Kauffmann, Brinchmann, Charlot, White, Seibert, Peng, Schlegel, Uomoto, Fukugita, \& Brinkmann}]{Tremonti2004TheSurvey}
Tremonti, C.~A., Heckman, T.~M., Kauffmann, G., {et~al.} 2004, The Astrophysical Journal, 613, 898, \dodoi{10.1086/423264}

\bibitem[{{Tundo} {et~al.}(2007){Tundo}, {Bernardi}, {Hyde}, {Sheth}, \& {Pizzella}}]{2007ApJ...663...53T}
{Tundo}, E., {Bernardi}, M., {Hyde}, J.~B., {Sheth}, R.~K., \& {Pizzella}, A. 2007, The Astrophysical Journal, 663, 53, \dodoi{10.1086/518225}

\bibitem[{{Ueda} {et~al.}(2014){Ueda}, {Akiyama}, {Hasinger}, {Miyaji}, \& {Watson}}]{Ueda+2014}
{Ueda}, Y., {Akiyama}, M., {Hasinger}, G., {Miyaji}, T., \& {Watson}, M.~G. 2014, \apj, 786, 104, \dodoi{10.1088/0004-637X/786/2/104}

\bibitem[{Van Der~Walt {et~al.}(2011)Van Der~Walt, Colbert, \& Varoquaux}]{VanDerWalt2011TheComputation}
Van Der~Walt, S., Colbert, S.~C., \& Varoquaux, G. 2011, Computing in Science and Engineering, 13, 22, \dodoi{10.1109/MCSE.2011.37}

\bibitem[{{Vestergaard} \& {Peterson}(2006)}]{2006ApJ...641..689V}
{Vestergaard}, M., \& {Peterson}, B.~M. 2006, The Astrophysical Journal, 641, 689, \dodoi{10.1086/500572}

\bibitem[{Virtanen {et~al.}(2020)Virtanen, Gommers, Oliphant, Haberland, Reddy, Cournapeau, Burovski, Peterson, Weckesser, Bright, {van der Walt}, Brett, Wilson, Millman, Mayorov, Nelson, Jones, Kern, Larson, Carey, Polat, Feng, Moore, {VanderPlas}, Laxalde, Perktold, Cimrman, Henriksen, Quintero, Harris, Archibald, Ribeiro, Pedregosa, {van Mulbregt}, \& {SciPy 1.0 Contributors}}]{2020SciPy-NMeth}
Virtanen, P., Gommers, R., Oliphant, T.~E., {et~al.} 2020, Nature Methods, 17, 261, \dodoi{10.1038/s41592-019-0686-2}

\bibitem[{{Vito} {et~al.}(2018){Vito}, {Brandt}, {Yang}, {Gilli}, {Luo}, {Vignali}, {Xue}, {Comastri}, {Koekemoer}, {Lehmer}, {Liu}, {Paolillo}, {Ranalli}, {Schneider}, {Shemmer}, {Volonteri}, \& {Wang}}]{2018MNRAS.473.2378V}
{Vito}, F., {Brandt}, W.~N., {Yang}, G., {et~al.} 2018, \mnras, 473, 2378, \dodoi{10.1093/mnras/stx2486}

\bibitem[{{Vogelsberger} {et~al.}(2014{\natexlab{a}}){Vogelsberger}, {Genel}, {Springel}, {Torrey}, {Sijacki}, {Xu}, {Snyder}, {Bird}, {Nelson}, \& {Hernquist}}]{2014Natur.509..177V}
{Vogelsberger}, M., {Genel}, S., {Springel}, V., {et~al.} 2014{\natexlab{a}}, Nature, 509, 177, \dodoi{10.1038/nature13316}

\bibitem[{{Vogelsberger} {et~al.}(2014{\natexlab{b}}){Vogelsberger}, {Genel}, {Springel}, {Torrey}, {Sijacki}, {Xu}, {Snyder}, {Nelson}, \& {Hernquist}}]{2014MNRAS.444.1518V}
---. 2014{\natexlab{b}}, Monthly Notices of the Royal Astronomical Society, 444, 1518, \dodoi{10.1093/mnras/stu1536}

\bibitem[{{Vogelsberger} {et~al.}(2020){Vogelsberger}, {Nelson}, {Pillepich}, {Shen}, {Marinacci}, {Springel}, {Pakmor}, {Tacchella}, {Weinberger}, {Torrey}, \& {Hernquist}}]{2020MNRAS.492.5167V}
{Vogelsberger}, M., {Nelson}, D., {Pillepich}, A., {et~al.} 2020, Monthly Notices of the Royal Astronomical Society, 492, 5167, \dodoi{10.1093/mnras/staa137}

\bibitem[{{Volonteri} {et~al.}(2016){Volonteri}, {Dubois}, {Pichon}, \& {Devriendt}}]{2016MNRAS.460.2979V}
{Volonteri}, M., {Dubois}, Y., {Pichon}, C., \& {Devriendt}, J. 2016, \mnras, 460, 2979, \dodoi{10.1093/mnras/stw1123}

\bibitem[{{Volonteri} {et~al.}(2003){Volonteri}, {Haardt}, \& {Madau}}]{2003ApJ...582..559V}
{Volonteri}, M., {Haardt}, F., \& {Madau}, P. 2003, The Astrophysical Journal, 582, 559, \dodoi{10.1086/344675}

\bibitem[{{Wandel} {et~al.}(1999){Wandel}, {Peterson}, \& {Malkan}}]{1999ApJ...526..579W}
{Wandel}, A., {Peterson}, B.~M., \& {Malkan}, M.~A. 1999, The Astrophysical Journal, 526, 579, \dodoi{10.1086/308017}

\bibitem[{{Wechsler} \& {Tinker}(2018)}]{2018ARA&A..56..435W}
{Wechsler}, R.~H., \& {Tinker}, J.~L. 2018, \araa, 56, 435, \dodoi{10.1146/annurev-astro-081817-051756}

\bibitem[{{Weinberger} {et~al.}(2017){Weinberger}, {Springel}, {Hernquist}, {Pillepich}, {Marinacci}, {Pakmor}, {Nelson}, {Genel}, {Vogelsberger}, {Naiman}, \& {Torrey}}]{2017MNRAS.465.3291W}
{Weinberger}, R., {Springel}, V., {Hernquist}, L., {et~al.} 2017, Monthly Notices of the Royal Astronomical Society, 465, 3291, \dodoi{10.1093/mnras/stw2944}

\bibitem[{{Weinberger} {et~al.}(2018){Weinberger}, {Springel}, {Pakmor}, Nelson, Genel, Pillepich, Vogelsberger, Marinacci, Naiman, Torrey, \& Hernquist}]{10.1093/mnras/sty1733}
{Weinberger}, R., {Springel}, V., {Pakmor}, R., {et~al.} 2018, Monthly Notices of the Royal Astronomical Society, 479, 4056, \dodoi{10.1093/mnras/sty1733}

\bibitem[{White \& Rees(1978)}]{White1978}
White, S. D.~M., \& Rees, M.~J. 1978, Monthly Notices of the Royal Astronomical Society, 183, 341

\bibitem[{{Wolf} {et~al.}(2023){Wolf}, {Nandra}, {Salvato}, {Buchner}, {Onoue}, {Liu}, {Arcodia}, {Merloni}, {Ciroi}, {Di Mille}, {Burwitz}, {Brusa}, {Ishimoto}, {Kashikawa}, {Matsuoka}, {Urrutia}, \& {Waddell}}]{2023A&A...669A.127W}
{Wolf}, J., {Nandra}, K., {Salvato}, M., {et~al.} 2023, \aap, 669, A127, \dodoi{10.1051/0004-6361/202244688}

\bibitem[{{Wright} {et~al.}(2010){Wright}, {Eisenhardt}, {Mainzer}, {Ressler}, {Cutri}, {Jarrett}, {Kirkpatrick}, {Padgett}, {McMillan}, {Skrutskie}, {Stanford}, {Cohen}, {Walker}, {Mather}, {Leisawitz}, {Gautier}, {McLean}, {Benford}, {Lonsdale}, {Blain}, {Mendez}, {Irace}, {Duval}, {Liu}, {Royer}, {Heinrichsen}, {Howard}, {Shannon}, {Kendall}, {Walsh}, {Larsen}, {Cardon}, {Schick}, {Schwalm}, {Abid}, {Fabinsky}, {Naes}, \& {Tsai}}]{2010AJ....140.1868W}
{Wright}, E.~L., {Eisenhardt}, P. R.~M., {Mainzer}, A.~K., {et~al.} 2010, The Astronomical Journal, 140, 1868, \dodoi{10.1088/0004-6256/140/6/1868}

\bibitem[{{Wu} {et~al.}(2023){Wu}, {Ling}, {Goto}, {Kim}, {Hashimoto}, {Kilerci}, {Lin}, {Wang}, {Uno}, {Ho}, \& {Hsiao}}]{2023MNRAS.523.5187W}
{Wu}, C. K.~W., {Ling}, C.-T., {Goto}, T., {et~al.} 2023, \mnras, 523, 5187, \dodoi{10.1093/mnras/stad1769}

\bibitem[{{Wyithe} \& {Loeb}(2003)}]{2003ApJ...595..614W}
{Wyithe}, J. S.~B., \& {Loeb}, A. 2003, The Astrophysical Journal, 595, 614, \dodoi{10.1086/377475}

\bibitem[{{Yang} {et~al.}(2023){Yang}, {Caputi}, {Papovich}, {Arrabal Haro}, {Bagley}, {Behroozi}, {Bell}, {Bisigello}, {Buat}, {Burgarella}, {Cheng}, {Cleri}, {Dav{\'e}}, {Dickinson}, {Elbaz}, {Ferguson}, {Finkelstein}, {Grogin}, {Hathi}, {Hirschmann}, {Holwerda}, {Huertas-Company}, {Hutchison}, {Iani}, {Kartaltepe}, {Kirkpatrick}, {Kocevski}, {Koekemoer}, {Kokorev}, {Larson}, {Lucas}, {P{\'e}rez-Gonz{\'a}lez}, {Rinaldi}, {Shen}, {Trump}, {de la Vega}, {Yung}, \& {Zavala}}]{2023ApJ...950L...5Y}
{Yang}, G., {Caputi}, K.~I., {Papovich}, C., {et~al.} 2023, \apjl, 950, L5, \dodoi{10.3847/2041-8213/acd639}

\bibitem[{{Yang} {et~al.}(2003){Yang}, {Mo}, \& {van den Bosch}}]{2003MNRAS.339.1057Y}
{Yang}, X., {Mo}, H.~J., \& {van den Bosch}, F.~C. 2003, Monthly Notices of the Royal Astronomical Society, 339, 1057, \dodoi{10.1046/j.1365-8711.2003.06254.x}

\bibitem[{{Yao} {et~al.}(2023){Yao}, {Ravi}, {Gezari}, {van Velzen}, {Lu}, {Schulze}, {Somalwar}, {Kulkarni}, {Hammerstein}, {Nicholl}, {Graham}, {Perley}, {Cenko}, {Stein}, {Ricarte}, {Chadayammuri}, {Quataert}, {Bellm}, {Bloom}, {Dekany}, {Drake}, {Groom}, {Mahabal}, {Prince}, {Riddle}, {Rusholme}, {Sharma}, {Sollerman}, \& {Yan}}]{2023ApJ...955L...6Y}
{Yao}, Y., {Ravi}, V., {Gezari}, S., {et~al.} 2023, \apjl, 955, L6, \dodoi{10.3847/2041-8213/acf216}

\bibitem[{{York} {et~al.}(2000){York}, {Adelman}, {Anderson}, {Anderson}, {Annis}, {Bahcall}, {Bakken}, {Barkhouser}, {Bastian}, {Berman}, {Boroski}, {Bracker}, {Briegel}, {Briggs}, {Brinkmann}, {Brunner}, {Burles}, {Carey}, {Carr}, {Castander}, {Chen}, {Colestock}, {Connolly}, {Crocker}, {Csabai}, {Czarapata}, {Davis}, {Doi}, {Dombeck}, {Eisenstein}, {Ellman}, {Elms}, {Evans}, {Fan}, {Federwitz}, {Fiscelli}, {Friedman}, {Frieman}, {Fukugita}, {Gillespie}, {Gunn}, {Gurbani}, {de Haas}, {Haldeman}, {Harris}, {Hayes}, {Heckman}, {Hennessy}, {Hindsley}, {Holm}, {Holmgren}, {Huang}, {Hull}, {Husby}, {Ichikawa}, {Ichikawa}, {Ivezi{\'c}}, {Kent}, {Kim}, {Kinney}, {Klaene}, {Kleinman}, {Kleinman}, {Knapp}, {Korienek}, {Kron}, {Kunszt}, {Lamb}, {Lee}, {Leger}, {Limmongkol}, {Lindenmeyer}, {Long}, {Loomis}, {Loveday}, {Lucinio}, {Lupton}, {MacKinnon}, {Mannery}, {Mantsch}, {Margon}, {McGehee}, {McKay}, {Meiksin}, {Merelli}, {Monet}, {Munn}, {Narayanan}, {Nash}, {Neilsen}, {Neswold}, {Newberg}, {Nichol}, {Nicinski},
  {Nonino}, {Okada}, {Okamura}, {Ostriker}, {Owen}, {Pauls}, {Peoples}, {Peterson}, {Petravick}, {Pier}, {Pope}, {Pordes}, {Prosapio}, {Rechenmacher}, {Quinn}, {Richards}, {Richmond}, {Rivetta}, {Rockosi}, {Ruthmansdorfer}, {Sandford}, {Schlegel}, {Schneider}, {Sekiguchi}, {Sergey}, {Shimasaku}, {Siegmund}, {Smee}, {Smith}, {Snedden}, {Stone}, {Stoughton}, {Strauss}, {Stubbs}, {SubbaRao}, {Szalay}, {Szapudi}, {Szokoly}, {Thakar}, {Tremonti}, {Tucker}, {Uomoto}, {Vanden Berk}, {Vogeley}, {Waddell}, {Wang}, {Watanabe}, {Weinberg}, {Yanny}, {Yasuda}, \& {SDSS Collaboration}}]{2000AJ....120.1579Y}
{York}, D.~G., {Adelman}, J., {Anderson}, John~E., J., {et~al.} 2000, The Astronomical Journal, 120, 1579, \dodoi{10.1086/301513}

\bibitem[{{Yung} {et~al.}(2021){Yung}, {Somerville}, {Finkelstein}, {Hirschmann}, {Dav{\'e}}, {Popping}, {Gardner}, \& {Venkatesan}}]{2021MNRAS.508.2706Y}
{Yung}, L.~Y.~A., {Somerville}, R.~S., {Finkelstein}, S.~L., {et~al.} 2021, Monthly Notices of the Royal Astronomical Society, 508, 2706, \dodoi{10.1093/mnras/stab2761}

\bibitem[{{Yung} {et~al.}(2019){Yung}, {Somerville}, {Finkelstein}, {Popping}, \& {Dav{\'e}}}]{2019MNRAS.483.2983Y}
{Yung}, L.~Y.~A., {Somerville}, R.~S., {Finkelstein}, S.~L., {Popping}, G., \& {Dav{\'e}}, R. 2019, Monthly Notices of the Royal Astronomical Society, 483, 2983, \dodoi{10.1093/mnras/sty3241}

\bibitem[{{Yung} {et~al.}(2022){Yung}, {Somerville}, {Ferguson}, {Finkelstein}, {Gardner}, {Dav{\'e}}, {Bagley}, {Popping}, \& {Behroozi}}]{2022MNRAS.515.5416Y}
{Yung}, L.~Y.~A., {Somerville}, R.~S., {Ferguson}, H.~C., {et~al.} 2022, Monthly Notices of the Royal Astronomical Society, 515, 5416, \dodoi{10.1093/mnras/stac2139}

\bibitem[{{Yung} {et~al.}(2023){Yung}, {Somerville}, {Finkelstein}, {Behroozi}, {Dav{\'e}}, {Ferguson}, {Gardner}, {Popping}, {Malhotra}, {Papovich}, {Rhoads}, {Bagley}, {Hirschmann}, \& {Koekemoer}}]{2023MNRAS.519.1578Y}
{Yung}, L.~Y.~A., {Somerville}, R.~S., {Finkelstein}, S.~L., {et~al.} 2023, Monthly Notices of the Royal Astronomical Society, 519, 1578, \dodoi{10.1093/mnras/stac3595}

\bibitem[{{Zanisi} {et~al.}(2020){Zanisi}, {Shankar}, {Lapi}, {Menci}, {Bernardi}, {Duckworth}, {Huertas-Company}, {Grylls}, \& {Salucci}}]{2020MNRAS.492.1671Z}
{Zanisi}, L., {Shankar}, F., {Lapi}, A., {et~al.} 2020, Monthly Notices of the Royal Astronomical Society, 492, 1671, \dodoi{10.1093/mnras/stz3516}

\bibitem[{{Zhang} {et~al.}(2023){Zhang}, {Behroozi}, {Volonteri}, {Silk}, {Fan}, {Hopkins}, {Yang}, \& {Aird}}]{2023MNRAS.518.2123Z}
{Zhang}, H., {Behroozi}, P., {Volonteri}, M., {et~al.} 2023, Monthly Notices of the Royal Astronomical Society, 518, 2123, \dodoi{10.1093/mnras/stac2633}

\bibitem[{{Zhou} {et~al.}(2010){Zhou}, {Zhang}, {Wang}, \& {Zhu}}]{2010ApJ...710...16Z}
{Zhou}, X.-L., {Zhang}, S.-N., {Wang}, D.-X., \& {Zhu}, L. 2010, The Astrophysical Journal, 710, 16, \dodoi{10.1088/0004-637X/710/1/16}

\bibitem[{Zwaan {et~al.}(2005)Zwaan, Meyer, Staveley-Smith, \& Webster}]{Zwaan2005TheGalaxies}
Zwaan, M.~A., Meyer, M.~J., Staveley-Smith, L., \& Webster, R.~L. 2005, Monthly Notices of the Royal Astronomical Society, 359, L30, \dodoi{10.1111/j.1745-3933.2005.00029.x}

\end{thebibliography}

\appendix
 
   

\section{Model calibration}
\label{app:calibration}
Calibrating models prior to comparison is crucial, as any differences might lead to inconsistencies between models. {\sc Dark Sage} employs prescriptions containing free parameters, whose values are determined through model calibration. Specifically, in this version of {\sc Dark Sage}, only eight free parameters are allowed to vary during the calibration process. This calibration is manually conducted using a set of observational constraints, including the stellar mass function \citep{Baldry2008}, H{\sc i} \citep{Zwaan2005TheGalaxies} and H2 mass functions \citep{Keres2003CO2}, the H{\sc i}--stellar mass scaling relation \citep{Brown2015TheGalaxies}, the black hole--bulge mass relation \citep{Scott2013THEGALAXIES}, the Baryonic Tully--Fisher relation \citep{Stark2009AGALAXIES}, the galaxy mass--metallicity relation \citep{Tremonti2004TheSurvey}, and the mean cosmic star formation density--redshift relation \citep{Somerville2001TheGalaxies}. All of these with the exception of the mean cosmic star formation density--redshift relation are adopted from $z=0$. Further details can be found in appendix A of \citet{Stevens2016}.

\citetalias{Ricarte+2018a} explores several models with 3-5 tuned parameters. Here, we use the power-law model, which assigns to SMBHs a random Eddington ratio drawn from a power-law distribution. This model uses 5 parameters tuned to reproduce the $z=0.1$ AGN bolometric luminosity function \citep{Hopkins+2007,Ueda+2014} and the local M-$\sigma$ relation \citep{Saglia+2016}. To obtain realistic scatter for the black hole mass and luminosity functions, \citetalias{Ricarte+2018a} calibrates to the local SMBH mass function, by manually adjusting the scatter to achieve a satisfactory alignment with acceptable mass function estimates from \citet{2009ApJ...690...20S} (see \citetalias{Ricarte+2018a} Appendix B).

The {\sc Santa Cruz} SAM contains around 25 free parameters  that are calibrated by hand to match a set of observations at $z=0$. These observations include: the stellar mass function \citep{2013MNRAS.436..697B}, the
ratio of stellar mass-to-halo mass as a function of halo mass \citep{2017MNRAS.470..651R}, the cold
gas fraction (defined as $f_{\rm cold} \equiv (M_{\rm HI} + M_{\rm H2} )/M_{*})$ as a function of stellar mass for disk-dominated galaxies (defined as having bulge-to-total stellar mass ratios B/T $<$ 0.4) \citep{2014ApJ...786...54P, 2018RMxAA..54..443C}, the stellar metallicity as a function of stellar mass \citep{2005MNRAS.362...41G}, and the black hole mass-bulge mass relation \citep{2013ApJ...764..184M} (see \citet{2019MNRAS.483.2983Y} Appendix B and \citet{2022MNRAS.517.6091G}).

The {\sc IllustrisTNG} simulation is tuned to match the stellar mass function, the stellar-to-halo mass relation, the total gas mass content within the virial radius
(r500) of massive groups, the stellar mass—stellar size, and the black hole–galaxy mass relations all at $z=0$, in addition to the cosmic star formation rate density up to $z=10$ (see \citet{2018MNRAS.473.4077P}). 

{\sc TRINITY} contains fifty six parameters calibrated by an MCMC to fit the observed stellar mass function from $z=0-8$, the galaxy quenched fractions from $z=0-4$, cosmic star formation rates from $z=0-10$, specific star formation rates from $z=0-9$, galaxy UV luminosity functions from $z=9-10$, quasar luminosity functions from $z=0-5$, quasar probability distribution functions from $z=0-2.5$, active SMBH mass function from $z=0-5$, SMBH mass-bulge mass relation at $z=0$, and the SMBH mass distribution of bright quasars from $z=5.8-6.5$ (see the observational references used in Tables 4-10 of \citet{2023MNRAS.518.2123Z}).

As stated here, {\sc TRINITY} is the only model that is tuned to follow scaling relations at high-redshift, whereas the {\sc Santa Cruz} SAM and, \citetalias{Ricarte+2018a} are calibrated to $z=0$ observations. {\sc Dark Sage} and {\sc IllustrisTNG} are calibrated to match the observed cosmic star formation history. While there is some overlap in the calibration quantities used for the SAMs and {\sc IllustrisTNG}, each method also incorporates distinct observational data not used by the other. Furthermore, the specific observational studies chosen for calibration differ between techniques in some cases. Although quantifying differences in calibration procedures is ideal, it is an expensive task that requires further work beyond this paper. Nonetheless, current model predictions that match a set of observations still yield valuable results.

\section{Eddington ratio distributions binned by halo mass} \label{BHgrowth_histdist}

\begin{figure*}[t]
\centering
\includegraphics[width=\columnwidth, clip]{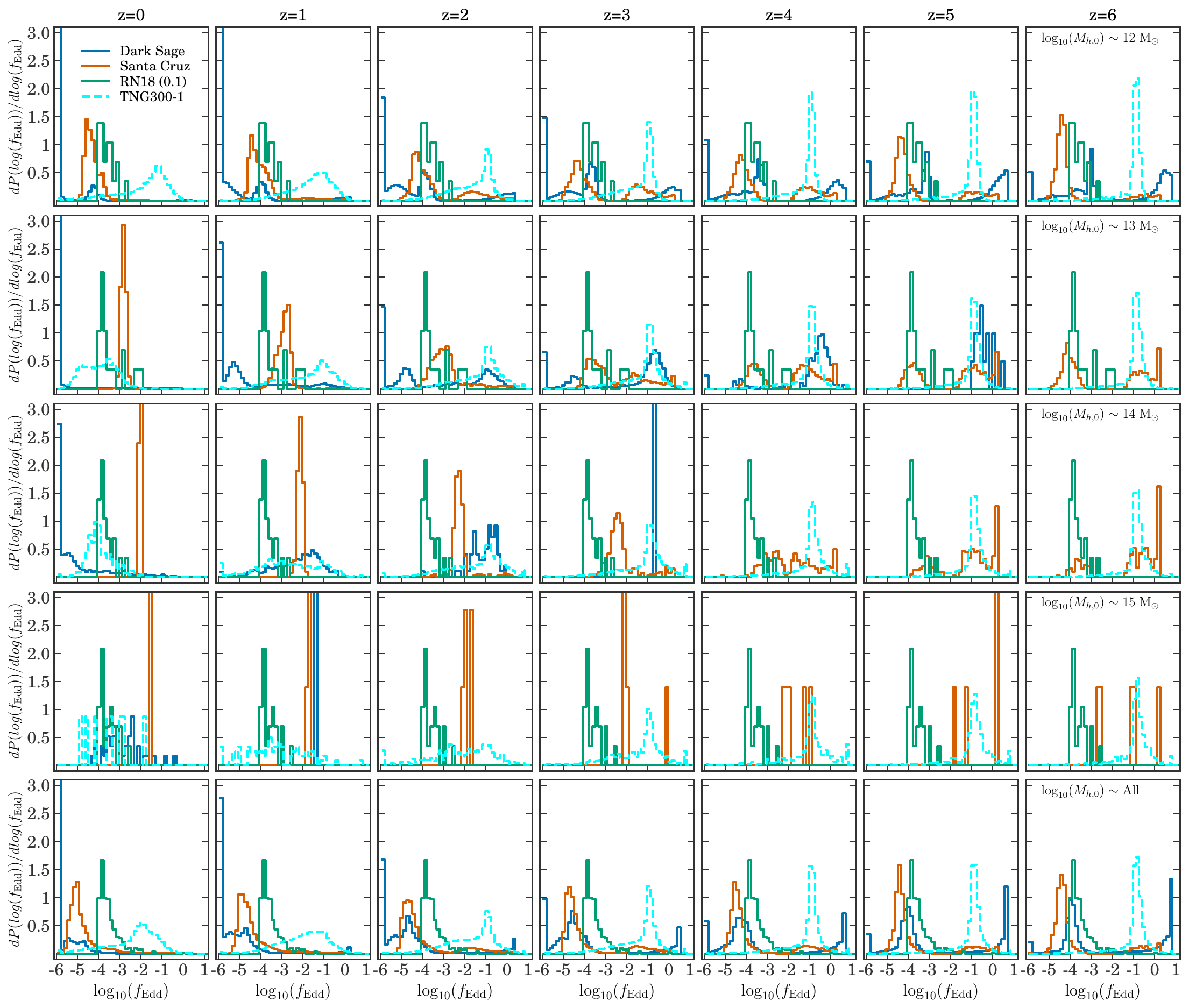}
\caption{Histograms of the Eddington ratio distributions are shown for {\sc Dark Sage} (solid dark blue line), the {\sc Santa Cruz} SAM (solid brown line), and \citetalias{Ricarte+2018a} (solid green line with $p_{merge}$=0.1). The \citetalias{Ricarte+2018a} model with $p_{merge}$=1.0 has similar Eddington ratio distributions to the $p_{merge}$=0.1 model, so only one is presented. These models are compared to {\sc TNG300-1} (dashed cyan line). The grey shaded region denote valus of $\log_{10}(f_\mathrm{Edd}) \leq -6$. Each column shows varying redshifts from $z=0$ (first column) up to $z=6$ (last column). Each row shows different halo mass bins from $\sim 10^{12}\, \mathrm{M}_{\odot} {\rm (first row)}$ to $\sim 10^{15}\, \mathrm{M}_{\odot}$ (fourth row). The last row shows all halo masses combined. The data displayed in the figures show the largest progenitor galaxy for a root $\log_{10}(M_{h}) \sim 12-15$ at $z=0$. All models show a diversity of Eddington ratio distributions across halo masses and redshifts.}\label{fig:Eddingtondistributinos_halobins}

\end{figure*}

Figure \ref{fig:Eddingtondistributinos_halobins} shows the Eddington ratio distributions similar to figure \ref{fig:Eddington_ratio}, but here, we break down all galaxies into bins of halo masses. At $z=0$, {\sc Dark Sage} displays a trimodal distribution with peaks at $\log_{10}(f_\mathrm{Edd}) \sim -6$, $-5.5$, and $-4.5$. The peak at $\log_{10}(f_\mathrm{Edd}) \sim -6$ is primarily contributed by galaxies with halo masses around $M_h \sim 10^{12-13}, \mathrm{M}{\odot}$. The peak at $\log{10}(f_\mathrm{Edd}) \sim -5.5$ is driven by galaxies with $M_h \sim 10^{14}, \mathrm{M}{\odot}$, while the peak at $\log{10}(f_\mathrm{Edd}) \sim -4.5$ corresponds to black holes in low-mass halos. The {\sc Santa Cruz} SAM displays a narrow distribution around $\log{10}(f_\mathrm{Edd}) \sim -4$ at $z=0$, which is heavily composed of black holes in low-mass halos. Interestingly, at $z>3$, black holes in galaxies with $M_h \sim 10^{12-14}, \mathrm{M}{\odot}$ exhibit a bimodal Eddington ratio distribution. The first peak ranges from $\log{10}(f_\mathrm{Edd}) \sim -4$ to $-3$, while the second peak spans from $\log_{10}(f_\mathrm{Edd}) \sim -1$ to $1$. We find that the first peak dominates at all redshifts and halo masses. The \citetalias{Ricarte+2018a} model shows consistent Eddington distributions across redshift and halo mass. {\sc TNG300-1} shows a broad distribution that peaks around $\log_{10}(f_\mathrm{Edd}) \sim -2$. Black holes in $M_h \sim 10^{12}, \mathrm{M}{\odot}$ drive such high $\log_{10}(f_\mathrm{Edd})$. At fixed $z$, this peak in the Eddington ratio distribution is consistent across all halo mass bins (see figure \ref{BHgrowth_histdist}). 

\end{document}